\documentclass[manuscript,nonacm]{acmart} 
\AtBeginDocument{%
  \providecommand\BibTeX{{%
    \normalfont B\kern-0.5em{\scshape i\kern-0.25em b}\kern-0.8em\TeX}}}

\setcopyright{acmcopyright}
\copyrightyear{2018}
\acmYear{2018}
\acmDOI{XXXXXXX.XXXXXXX}

\acmConference[]{}{}{}
\acmBooktitle{} 
\acmPrice{15.00}
\acmISBN{978-1-4503-XXXX-X/18/06}

\usepackage{wrapfig}
\usepackage{url}
\usepackage{graphicx}
\usepackage{xcolor}
\usepackage[normalem]{ulem}
\usepackage[title]{appendix}
\usepackage{array}
\usepackage{longtable}
\definecolor{tealblue}{rgb}{0.0, 0.5, 0.5}
\newcommand{\soutc}[1]{}
\newcommand{\souts}[1]{}
\newcommand{\blu}[1]{\textcolor{black}{#1}}
\newcommand{\blues}[1]{\textcolor{black}{#1}}
\begin{document}

\begin{teaserfigure}
\centering
  \includegraphics[width=1.05\textwidth]{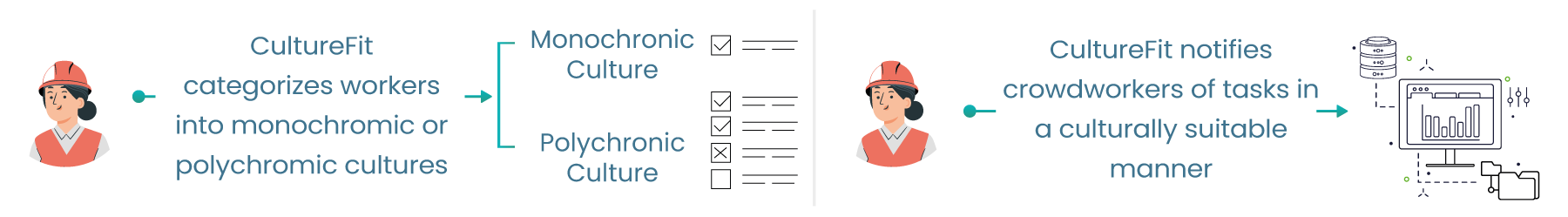}
  \caption{\blu{Overview of CultureFit's Functionality.}}
  \label{fig:teaser}
\end{teaserfigure}
\title{\blues{A Culturally-Aware Tool for Crowdworkers: Leveraging Chronemics to Support Diverse Work Styles}}



\author{Carlos Toxtli}
\affiliation{%
  \institution{Clemson University}
  \city{Clemson}
  \country{USA}
}

\author{Christopher Curtis}
\affiliation{%
  \institution{Northeastern University}
  \city{Boston}
  \country{USA}
}

\author{Saiph Savage}
\affiliation{%
  \institution{Northeastern University \& Universidad Nacional Autonoma de Mexico (UNAM)}
  \city{Boston}
  \country{USA \& Mexico}
}






\renewcommand{\shortauthors}{Toxtli, et al.}

\begin{abstract}
Crowdsourcing markets are expanding worldwide, but often feature standardized interfaces that ignore the cultural diversity of their workers, negatively impacting their well-being and productivity. To transform these workplace dynamics, this paper proposes creating culturally-aware workplace tools, specifically designed to adapt to the cultural dimensions of monochronic and polychronic work styles. 
We illustrate this approach with "CultureFit," a tool that we engineered based on extensive research in Chronemics and culture theories. To study and evaluate our tool in the real world, we conducted a field experiment with 55 workers from 24 different countries. Our field experiment revealed that CultureFit significantly improved the earnings of workers from cultural backgrounds often overlooked in design. Our study is among the pioneering efforts to examine culturally aware digital labor interventions. It also provides access to a dataset with over two million data points on culture and digital work, which can be leveraged for future research in this emerging field. The paper concludes by discussing the importance and future possibilities of incorporating cultural insights into the design of tools for digital labor.

\end{abstract}

\maketitle

\section{Introduction}
\blu{Crowdworkers significantly enhance AI services \cite{gaur2016effects, jung2012improving, krishna2017visual}, yet often face poor working conditions \cite{gray2019ghost}, particularly non-US/European workers whose cultural backgrounds differ from the intended design of crowdsourcing platforms \cite{chen2021desperate, posada2022coloniality, toxtli2021quantifying}. This issue usually stems from the assumption that crowdworkers are a homogeneous group \cite{gordon2022jury}, neglecting their diverse cultural backgrounds \cite{ kapania2023hunt}. Moreover, a notable trend in design has emerged advocating for minimizing cultural impact in work interfaces, aiming for global uniformity in their design rather than customizing these systems to accommodate cultural nuances \cite{norman2012does,yaaqoubi2020practitioners, norman2013design}. Consequently, many work interfaces have strived for uniform standards, and have ignored worker diversity \cite{hill2004cross,jones1983transaction,irani2015cultural}.}

\soutc{Crowdworkers play a crucial role in enabling  AI to deliver better services \cite{gaur2016effects, jung2012improving, krishna2017visual}. However, crowdworkers have also reported unsatisfactory working conditions \cite{gray2019ghost}, especially for crowdworkers who come from outside the US and Europe (i.e., they might come from regions with cultural backgrounds that may differ from those for which the crowdsourcing platform was originally designed) \cite{chen2021desperate, posada2022coloniality, toxtli2021quantifying}. Part of the problem might arise because the current paradigm assumes that crowdworkers are a large
homogeneous group, and hence workers' culture or general backgrounds are not considered \cite{gordon2022jury, kapania2023hunt}. Another concerning issue is that numerous researchers and practitioners have advocated for minimizing the impact of culture in online work interfaces \cite{norman2012does,yaaqoubi2020practitioners}. They contend that modern office activities should follow worldwide standards to render uniform all work transactions. They have advocated for the design of interfaces and systems tailored to particular work activities rather than workers' cultural peculiarities \cite{norman2013design}. As a result, many work interfaces aim to follow uniform standards independent of the diverse cultural background of the workers \cite{hill2004cross,jones1983transaction,irani2015cultural}. }

\blu{However, interfaces often reflect the cultural biases of their designers \cite{brejcha2015cross}, inadvertently embedding their cultural norms \cite{reinecke2011improving,rai2017editor,sun2012cross}. This can lead to designs that unintentionally require "outside workers" to adapt or modify their behaviors \cite{sun2012cross,meyer2014culture}, potentially hindering their success and effectiveness in their jobs \cite{irani2010postcolonial,casilli2017global,hall1989understanding,hall1971paradox}. A solution can be to create culturally aware tools for crowdworkers, yet research into integrating culture theory into such designs remains limited \cite{sayago2023cultures,lascau2019monotasking,mansouri2024does}. Further research is crucial to assess these systems' effectiveness and their potential benefits for crowdworkers from varied cultural backgrounds.}

\soutc{However, it is important to recognize that such interfaces are not culture-free; rather, they often reflect the cultural work norms of their designers \cite{irani2010postcolonial}. This unintentionally imposes the expectation for "outside workers" to conform and adapt their practices, which can ultimately hinder their success  \cite{casilli2017global,hall1989understanding,hall1971paradox}.
One way to potentially address this problem is to create culturally aware tools and platforms for crowdworkers. However, research in applying culture theory ideas into the design of crowdworker interfaces has been limited \cite{lascau2019monotasking}. Further investigations are essential to examine the performance of these systems and understand the potential benefits they can bring to workers from diverse cultural backgrounds.}

\blu{To address this knowledge gap, we focus on designing a tool that aims to enhance crowdworkers' experiences by incorporating cultural considerations. Drawing from culture theory, we apply Chronemics—a discipline used in various social sciences like Organizational Psychology and Anthropology—to inform our interface design \cite{ballard2004communication, hall1989understanding, bluedorn1999polychronicity, robinson1990handbook}. Chronemics helps us distinguish between "monochronic" and "polychronic" work practices, which influence how different cultures manage time and tasks \cite{hall1971paradox, nardon2009culture}. For instance, monochronic cultures (usually involving people from the United States, Germany, Scandinavia) focus on sequential task handling, whereas polychronic cultures (usually involving people from Latin America, Africa, South Asia) prioritize multitasking and social interactions \cite{cotte1999juggling,trompenaars2011riding}.}

\blu{For this purpose, we implemented these concepts into CultureFit\footnote{Plugin link:\blues{\url{https://github.com/ai4he/toloka-cultural-assistant}}}, a tool that dynamically adjusts its interface to notify workers of potential tasks according to ``Monochronic'' or ``Polychronic'' settings, thereby catering to the diverse cultural preferences of crowdworkers. Figures~\ref{fig:teaser} and \ref{fig:notifications} illustrate CultureFit and show its interface in use. CultureFit equips polychronic workers with a notification system that extends beyond the crowdsourcing platform, enabling them to consider work opportunities while involved in various other activities, such as using desktop applications or browsing social media. This flexibility, supported by culture theory \cite{bluedorn2002human}, leverages the multitasking preferences of polychronic workers. Conversely, the system notifies monochronic workers of available tasks only when they are on the crowdsourcing platform and have completed their current tasks, minimizing distractions and enhancing focus. This approach, which helps maintain their preferred work schedules, aligns with the preferences culture theory identifies for monochronic workers \cite{inbook, hall1971paradox}. By acknowledging workers' cultural differences, our tool proposes a novel design space that shifts the adaptation burden from workers to the interface itself, enabling digital workers from diverse backgrounds to potentially thrive in a space that respects and integrates their cultural work patterns \cite{gray2019ghost,alvarado2021decolonial}. }

\blu{To evaluate the effectiveness of our novel tool, we conducted an IRB-approved field study to investigate its potential to improve the experiences of crowdworkers. We employed a between-subjects design for our study, where workers who identified with either polychronic or monochronic cultural traits used CultureFit, while polychronic and monochronic workers in the control groups did not. Fig. \ref{fig:experimento} presents an overview of our field experiment. Together, we were able to recruit crowdworkers spanning 24 countries across regions including the United States, Africa, Latin America, Europe, and South and Southeast Asia. Workers in our study completed over 2,300 tasks for 158 requesters on the Toloka crowdsourcing platform, generating more than two million anonymized data points with information about workers' digital labor and cultural traits.  We plan to publicly release this dataset upon the paper's publication. Notably, through our results we found that polychronic workers—often disadvantaged in crowd work \cite{chen2020,gupta2014turk,toxtli2021quantifying,lascau2019monotasking}—saw a 258\% wage increase when using CultureFit.} 

\soutc{Our IRB-approved field study studies whether such a tool can yield substantial benefits in wages and user experience for workers. The study employed a between-subjects design, where real-world workers with either polychronic or monochronic cultural traits utilized our \soutc{recommendation} system. 
These workers spanned 24 countries and across various regions, including: The United States, Africa, Latin America, Europe as well as South and South East Asia.
Over 2,300 tasks were completed by these workers for more than 158 requesters on the Toloka crowdsourcing platform, generating over two million anonymized data points on workers and their labor. 
Both monochronic and polychronic workers reported significant enhancements in user experience while using CultureFit. Our study also uncovered that polychronic workers, a population traditionally hurt by crowd work \cite{chen2020,gupta2014turk,toxtli2021quantifying,lascau2019monotasking}, significantly increased their wages, experiencing a 258\% wage increase using CultureFit. 
Our post survey also uncovered that these workers felt CultureFit used their strengths significantly more.}

\blu{In this paper, we contribuite: 1) a culturally aware crowd work notification tool; 2) a case study on applying culture theories to crowd work tools; 3) a two-week field experiment showcasing how CultureFit enhances wages for polychronic crowdworkers; 4) design recommendations for incorporating culture theories into crowd work tools; 5) over two million anonymized data points on digital labor and culture, to be publicly available post-publication, for future research studying cultural impacts on computer-supported collaborative work systems.}

\soutc{In this paper, we contribute: 1) a culturally aware recommender system for crowdworkers; 2) a case study in applying culturally sensitive theories to mechanisms for recommending labor;  3) a two-week field experiment showcasing that CultureFit improves the wages and perceptions of polychronic crowdworkers; 4) design recommendations for integrating cultural theories into crowd work tools; 5) over two million anonymized telemetry data points for researchers to study the impact of culture on digital labor which will be publicly available upon publication.}

\subsubsection*{Positionality Statement}
As researchers in the HCI community, our commitment lies in designing and evaluating technology that embraces diverse perspectives and experiences. We acknowledge the problematic nature of crowd work, recognizing the unfavorable conditions it imposes on workers, which pose significant challenges to their well-being \cite{martin2014being}. Extensive research has shown that crowd work disproportionately impacts workers outside the United States and Europe, resulting in significantly lower earnings for these individuals \cite{gupta2014turk}. Consequently, our research aims to foster inclusive technologies that foster positive outcomes for all. Our diverse team includes authors from Latin America and the United States, featuring individuals with indigenous heritage and a leader in diversity, equity, and inclusion at their institution. Among us, two authors practice ``Polychronic'' work styles, while one adheres to a ``Monochronic'' approach.

\soutc{Our objective is to create inclusive, accessible, and equitable technology that serves all communities. By contributing to the development of technology that truly supports everyone, we aspire to make a positive impact. We firmly believe that by prioritizing inclusivity in our technological designs, we can mitigate biases and work towards a more just and equitable world.}

\section{Related Work}
Our research connects with the following key areas of prior literature: 

\subsection{Universal Design.}
Vast work in the field of universal design has questioned the usefulness of culturally-informed interfaces \cite{norman2013design,steinfeld2012universal,norman2016living}. 
While universal design is aware of the clashes that exist due to the lack of cultural awareness in design, it also argues that people will eventually adjust to the non-culturally-aware interfaces \cite{kose1998barrier,norman2013design}. 
This view-point has allowed universal design to make products and provide ``equivalent experiences'' for a wider range of possible users than addressing the specific accessibility or inclusive goals of smaller targeted populations \cite{heylighen2014nature,story2001principles,steinfeld2012universal}.
Universal design has thus positioned itself as a ``culturally-neutral'' design framework \cite{hardy2019design,hardy2018rural}. 
Norman and the researchers who follow his work have argued that designs become more approachable to a wider global audience upon the removal of individualistic and cultural human elements \cite{norman1991cognitive}. 
Norman argues that people will adapt to activities in ways not necessarily innate, supporting the prevailing importance of designing for ``activity over culture'' \cite{norman2013design}. 

Our research considers that while it might be true that people will eventually adapt to interfaces not tailored to their culture \cite{norman2013design}, the adaptation period can generate harm \cite{sikkema1987design, rau2015cross}, e.g., cost people their livelihood \cite{gray2019ghost}. Prior work has also established that there is a link between cross-cultural differences and outcomes like mental health and job performance \cite{inbook}. Current research on homogeneous designs also indicates that prevailing design norms may indeed lead to reduced success for some crowdworkers \cite{hill2004cross,jones1983transaction,irani2015cultural}. Additionally, there is concern that crowdworkers from cultures underrepresented in design frequently have to engage in higher rates of unpaid labor, leading to poorer work experiences \cite{inbook,article,newlands2021crowdwork, toxtli2021quantifying,gupta2014turk,chen2020}.
Consequently, we advocate for culturally-aware design.

\subsection{Chronemics \blues{and Different Cultural Dimensions within Time Management}.}
A key goal of designing culturally aware interfaces is determining what long-term aspects of a culture are important to consider \cite{quesenbery2011global,marcus2006cross}. 
Hall and Hall \cite{hall1990understanding,hall1989dance}, argued that key cultural variables to consider are people's concepts of time. This falls under the study of Chronemics, which  is the study of time perceptions, and \emph{"... includes time orientation, understanding and organisation, the use of and reaction to time pressures, the innate and learned awareness of time..."} \cite{reynolds2007encyclopedia}. 
This theory extends to the social and cultural level, and can be used describe work patterns across cultures and societies \cite{hall1990understanding}. Hall and Hall identified two main cultural concepts of time: (1) monochronic, and (2) polychronic \cite{hall1990understanding}. 
They established that cultures with a monochronic time use, view time as linear  \cite{hall1990understanding}. This helps these individuals focus on doing one thing at a time, as time becomes something that can be scheduled and compartmentalized \cite{meredith2017project,hall1990understanding}. According to the theory, Monochronic workers typically prioritize schedules highly \cite{trompenaars2011riding}, often valuing them above building social relationships \cite{bluedorn1992many}. They also prefer to limit multi-tasking and minimize distractions  \cite{lascau2019monotasking,trompenaars2011riding}.

By contrast, polychronic cultures view time as something occurring within the context of multiple events and the constant involvement of other people \cite{hall1990understanding,hall1989dance}. 
In these cultures, it is more important to complete human transactions than to adhere to schedules \cite{trompenaars2011riding}. Polychronic workers thus tend to value receiving unpredictable alerts, particularly those related to relationship building \cite{trompenaars2011riding}, as well as favor flexible, spontaneous, and concurrent work schedules \cite{hall1989beyond, hall1989dance, lee1999time}, and prefer to work while engaging in other activities \cite{inbook, article}. We use insights from this culture theory to design better tools for global crowdworkers. 

Our research is also inspired by previous studies on designing culturally aware systems \cite{sun2012cross,reinecke2010culturally}, as well as recent HCI research that has analyzed the monochronic and polychronic characteristics of crowdworkers, providing targeted design recommendations \cite{lascau2022crowdworkers,lascau2019monotasking}. Overall, we utilize this prior work to guide our tool design.

\subsection{Culture and Crowd Work.} Crowdsourcing platforms allow non-experts to find new jobs \cite{gray2019ghost}. These platforms are usually also not tied to a particular geographical region \cite{hardy2019designing,flores2020challenges}. \soutc{Together, this has helped crowdsourcing markets to be labeled a solution for facilitating the financial recovery of regions suffering from economic hardship \cite{flores2020challenges,hanrahan2020reciprocal}} The results are that  \soutc{more and} more people from different parts of the world are exploring crowd work as a viable job option \cite{kaziunas2019precarious,flores2020challenges}. \soutc{especially where local opportunities are limited.} The growing global nature of crowd work has led HCI researchers to study the demographics of crowdworkers \cite{hara2019worker,ross2010crowdworkers}, helping to shed light on the different cultures present in crowd work and the problems they face \cite{moreschi2020brazilian,newlands2021crowdwork,martin2014being,flores2020challenges,hanrahan2020reciprocal}. On the other hand, recent research has identified that many crowdworkers have a tendency to follow polychronic work patterns \cite{lascau2022crowdworkers}. Workers with this type of cultural background have also tended to experience more hardships \cite{chen2020,martin2014being,manrai1995effects,gong2009national,hooker2003working}, and have been joining crowdsourcing platforms in increasing numbers \cite{moreschi2020brazilian,newlands2021crowdwork,flores2020challenges}. This prior work serves as motivation for our research, as it highlights the need to create tools that can support workers with diverse cultural backgrounds.

\subsection{The Tooling Add-on Movement.} Crowdsourcing platforms have recently been augmented with a suite of tools to improve them for both workers and requesters \cite{savage2021research,Hanrahan2015,Savage2024,Do2024,DelosSantos2023}. Such a tooling movement has focused on creating change for workers and requesters without needing buy-in from the crowdsourcing platforms themselves \cite{savage2021research,kittur2013future}. Most of these tools have taken the form of ``web add-ons'' (plugins) that provide additional functionalities to the platforms  \cite{williams2019perpetual,sannon2019privacy,rivera2021want}. For instance, Turkopticon enhances the functionality of the crowdsourcing platform of Amazon Mechanical Turk by enabling workers to rate requesters \cite{irani2013turkopticon,silberman2015operating}. This ``add-on'' has helped workers fight power imbalances and information asymmetries \cite{irani2013turkopticon,sannon2019privacy,Kingsley2015}. \soutc{Related work has studied tools that integrate different workflows, avatars, or self-reflection mechanisms to improve workers' labor and experiences \cite{pradhan2022search,hettiachchi2021challenge,qiu2021using,edixhoven2021improving,braylan2021aggregating,rahman2019constructing,kasunic2019turker}. Similarly, other tools have focused on guiding requesters to improve the instructions for their tasks \cite{nouri2021iclarify,bragg2018sprout,manam2018wingit}.} We take inspiration from this to now augment crowdsourcing platforms with culturally aware interfaces.

\subsection{{Worker Community System Designs.}}
\blu{A number of data-driven tools have facilitated collaboration and community-building among workers \cite{pradhan2022search,hettiachchi2021challenge,qiu2021using,edixhoven2021improving,braylan2021aggregating,rahman2019constructing,kasunic2019turker}, including crowdworkers. These tools have improved how workers collaborate and help foster a sense of unity. For example, Coworker.org, an NGO dedicated to empowering workers and developing power-building strategies in the modern economy \cite{fiorentino2019petitions}, has developed a calculator tool that skillfully analyzes the wage data of platform workers \cite{ShiptTransparencyCalculator}. This tool clarifies payment practices and encourages mutual support among workers, enabling them to challenge unfair pay practices from requesters. Other tools have been designed to help workers collaborate with each other to identify potential wage theft and unpaid work \cite{toxtli2021quantifying,calacci2022bargaining,WageTheftIntro2023,platform2020wage}, as well as flag safety risks \cite{alimahomed2021surveilling,ShromaMonitoring2023}. Tools like Turkopticon and Turkerview are specifically designed to facilitate crowdworker collaboration by allowing them to share information about requesters, ultimately aiding in the identification of more favorable job opportunities \cite{irani2013turkopticon,savage2020becoming}. Additionally, online forums like Turker Nation have played a crucial role in cultivating a community and promoting collaborative learning among crowdworkers, enhancing their collective knowledge and support network \cite{zyskowski2018crowded,martin2014being,wang2017community}. These platforms and tools have begun to foster mutual support for workplace challenges. Our paper explores how to enhance support for crowdworkers by recognizing their cultural identities.}

\begin{wrapfigure}{r}{0.465\textwidth} 
  \centering
  \includegraphics[width=0.44\textwidth]{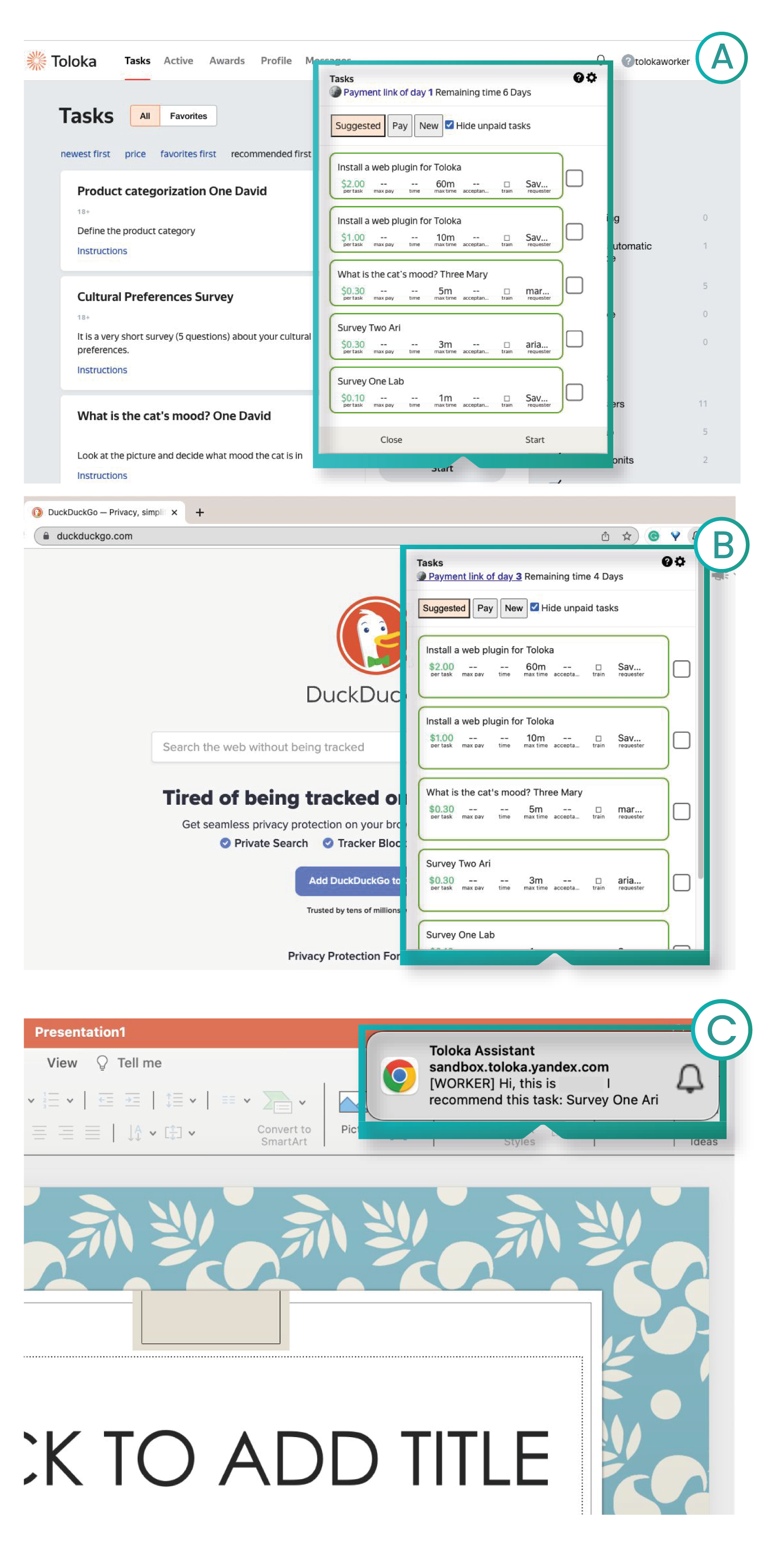} 
  \caption{Screenshots of CultureFit adapting its notification interface to workers' culture. For monochronic workers: (a) CultureFit notifies about tasks on Toloka. For polychronic workers: CultureFit notifies about tasks on Toloka while: (b) browsing other sites; or (c) engaged in other computer activities.}
  \label{fig:notifications}
\end{wrapfigure}

\section{CultureFit} 

\blu{CultureFit is a tool designed to improve the experiences of crowdworkers by tailoring job notifications to their cultural backgrounds. Drawing from the ``Tooling Add-on Movement'' in crowd work \cite{williams2019perpetual,sannon2019privacy,rivera2021want,irani2013turkopticon,silberman2015operating}, CultureFit functions autonomously as a Chrome plugin, operating independently without requiring official support from crowdsourcing platforms such as Toloka. Fig. \ref{fig:teaser} presents an overview of CultureFit, and Fig. \ref{fig:notifications} presents screenshots of it.} \blu{Grounded in culture theory \cite{hall1959silent,hall1989understanding}, our tool is designed to accommodate cultural variations in time perception, specifically monochronic and polychronic cultural orientations. CultureFit leverages these distinctions to tailor notifications about available tasks to workers \cite{lewis2010cultures,ballard2004communication,huang2010effects,mccrickard2003chewar}.} 

\blu{It can be important to observe that by notifying workers about tasks on Toloka, CultureFit can help to streamline the process of finding work \cite{gray2019ghost}. This can help to minimize the unpaid labor of task searching—one of the most burdensome types of unpaid labor in crowd work \cite{toxtli2021quantifying}. Overall, CultureFit is a crowd work tool that streamlines the discovery of relevant tasks through culturally aware notifications, thereby reducing workers' need to navigate extensive lists. Unlike other tools aimed at aiding workers in finding tasks \cite{savage2020becoming,wijenayake2023combining,de2023independiente,xie2023understanding}, CultureFit distinctively integrates cultural sensitivity into its notification design.}

\blu{Another thing to observe is that CultureFit has a task recommendation algorithm, which determines which tasks workers are notified about. For more details on how this recommendation algorithm works, please see our Appendix. Note that while interesting, this recommendation algorithm is not our main research contribution and is hence not detailed here.} 

\blu{CultureFit enhances cultural awareness in crowd work through two culturally sensitive notifications:}

\soutc{is an intelligent tool that recommends labor to crowdworkers. It exists as a web plugin that recommends tasks from the Toloka crowdsourcing platform}. \soutc{Based on culture theory \cite{hall1959silent,hall1989understanding}, the tool was made with the understanding that there are cultural differences in workers' time perceptions. Ballard et al has previously studied how the length, diversity, and feedback around tasks meaningfully affect time perceptions \cite{ballard2004communication}, indeed this is the driving principle behind Chronemics. We leverage this insight in how the tool adapts its interface to people with monochronic and polychronic work traits. 
We design within the Toloka platform, leveraging its diverse, global workforce and their unique cultural perspectives \cite{toloka_2023}. To enhance our tool design, we connected with research studying the impact of cultural time differences on interface design, specifically, we noted research which studied notification systems for both Mono and Poly chronic workers \cite{huang2010effects,mccrickard2003chewar}. Here we applied the same content-based recommender system in both the monochronic, and ploychronic populations. (To address the cold-start problem our model was trained off of data initially collected from pilot studies in line with prior work \cite{savage2020becoming}). For all workers the recommendations within the extension's popup are sorted according to the content-based model's ranking of them, this is in contrast with the chronological, newest-to-oldest, display that Toloka employed at the time of the study. However, there are substantial differences in the interfaces which we detail below. }

\subsection{Polychronic \soutc{Recommendation} \blu{Work Notification }Interface.}
CultureFit caters to workers with polychronic \blu{cultural} traits, who, according to the theory, favor flexible, spontaneous, and concurrent work schedules \cite{hall1989beyond, hall1989dance, lee1999time}. \soutc{To support the preferences of polychronic workers,} \blu{Thus,} our tool \soutc{recommends work opportunistically} \blu{opportunistically notifies polycronic workers about job opportunities} across the worker's computer operating system, web browser, or  crowdsourcing marketplace. In this way, the tool is able to reach workers while they are doing different activities. Fig. \ref{fig:notifications} presents how CultureFit can \soutc{recommend} \blu{notify workers about} labor \blu{opportunities} within different contexts, and also in a peripheral manner, to facilitate opportunistic work schedules \blu{and multi-tasking} (something related work recommends for polychronic workers \cite{lascau2019monotasking}). 
\blu{CultureFit's Polychronic Interface also alerts workers to social media updates, emails, and messages from requesters and fellow crowdworkers. This feature, informed by culture theory \cite{hall1989dance,lee1999time}, aims to support the building of social connections, which is important to polychronic cultures, who prioritize meaningful human interactions and relationships.}

\soutc{CultureFit also offers the option to receive outside notifications, such as updates from social media, as well as internal notifications, such as alerts when they receive a message from a requester or a co-worker. This feature takes into account that polychronic workers tend to value receiving unpredictable alerts, particularly those related to relationship building.} \soutc{By enabling these types of notifications, CultureFit aims to better accommodate the multitasking and social preferences of polychronic crowdworkers}.

\subsection{Monochronic \blu{Notification} \soutc{Recommendation} Interface.} In the case of workers who \soutc{are determined to} follow Monochronic labor patterns \cite{hall1959silent}, CultureFit \soutc{recommends} \blu{notifies about} tasks in a way that will allow them to manage their work days in a focused and efficient manner. \soutc{Given that, according to theory, for these kinds of workers schedules are usually of high importance and often seen as even more important than ``building social relationships'' \cite{bluedorn1992many}.} For this purpose, CultureFit \soutc{recommends tasks } \blu{provides task notifications} only when workers are on the Toloka platform (Fig. \ref{fig:notifications}.a), considering that these workers, according to the theory, prefer to limit multi-tasking \cite{bluedorn1992many}. CultureFit \soutc{recommends} \blu{also makes sure to notify workers about} tasks only after the worker\blu{s} \soutc{has} \blu{have} finished the labor they are currently doing, and \soutc{recommends} \blu{only notifies them about} tasks that will fit within the worker's established schedule. For this purpose, the tool favors \blu{notifying about} tasks that can be completed within the worker's schedule over tasks that might require working over time, even if the tasks match the worker's preferred type or come from their favorite requesters (to predict the amount of time a task will take, we use techniques from prior work \cite{saito2019turkscanner,savage2020becoming}). \soutc{CultureFit's design follows recommendations from prior work, e.g., having mechanisms that minimize distractions \cite{lascau2019monotasking}.} \blu{For monochronic workers, CultureFit also strategically limits notifications from social media, emails, and messages from requesters on the crowdsourcing platform. This approach aligns with research findings that emphasize monochronic workers' preferences for fewer interruptions, and a more focused approach to social connections and work tasks \cite{lascau2019monotasking,bluedorn1992many}.}

\section{User Scenarios} \blu{We present user scenarios on how monochronic and polychronic workers would use CultureFit. We aim to enhance understanding of the context in which CultureFit is used, especially within the Toloka crowdsourcong platform.}  
\subsection{\blu{User Scenario for Monochronic Crowdworker: Bob}}
\blu{Bob is a dedicated monochronic crowdworker who prefers to work in a structured manner. He likes to focus on one task at a time and finds it distracting to sift through multiple tasks on traditional crowdsourcing platforms. Bob's typical workday is well-planned, with specific times allocated for different activities, including work, meals, and leisure.}

{Before CultureFit,} \blu{Bob would log into Toloka every morning spending a significant amount of time searching through tasks to find the ones that fit his schedule and expertise \cite{Kingsley2024}. This process was time-consuming and led to frustration, as Bob felt that he could have used this time to actually work on tasks. He also found it challenging to resist the temptation of non-work-related notifications (e.g., activities on social media), which occasionally disrupted his focus.}

After adopting CultureFit, \blu{Bob now receives notifications about tasks on Toloka that are aligned with his schedule and work preferences. This change has improved his work efficiency and satisfaction. He no longer needs to manually search for tasks each morning. Instead, Bob receives a curated list of tasks right before his designated work time, allowing him to dive straight into focused work. CultureFit also blocks non-work-related notifications during his work hours, helping him maintain his concentration.} \blu{Bob appreciates how CultureFit understands his monochronic work style and tailors the notification system to enhance his focus. This allows him to be more productive and achieve a better work-life balance, as he can now dedicate his planned work time more effectively and enjoy his leisure time without worrying about missing out on tasks.}

\subsection{\blu{User Scenario for Polychronic Crowdworker: Alejandro}}
\blu{Alejandro is a dynamic polychronic crowdworker who thrives on multitasking. He enjoys working on multiple tasks simultaneously and often juggles work with other activities like socializing or hobbies. Alejandro prefers a flexible work environment where he can switch between tasks as his interest and energy dictate.}

\blu{Before CultureFit,} \blu{Alejandro found traditional crowdsourcing platforms somewhat restrictive and inefficient for his working style. He would often miss out on social happenings or interesting tasks because he was too engrossed in another activity. Alejandro wanted a system that would automatically notify him of new opportunities, eliminating the need to continuously monitor the crowdsourcing platform or sift through lengthy lists of available tasks.}

\blu{After adopting CultureFit, Alejandro's work experience has been transformed as the tool provides notifications that support his multitasking nature. Now, while engaged in one task on Toloka, he receives alerts for other tasks that match his interests, allowing him to seamlessly transition between jobs without losing momentum. Additionally, CultureFit alerts him to social happenings and networking opportunities, ensuring he does not miss out on valuable human-to-human experiences.} \blu{Alejandro appreciates the tool's flexibility, particularly its ability to recognize his polychronic tendencies and offer a work experience tailored to his dynamic style. With CultureFit, he feels more connected to work opportunities and social engagements, enhancing both his professional and personal life.}

{\bf \soutc{Work Recommendation Algorithm.}}

\section{Methods}
\begin{figure}[h]
\includegraphics[width=1.0\linewidth]{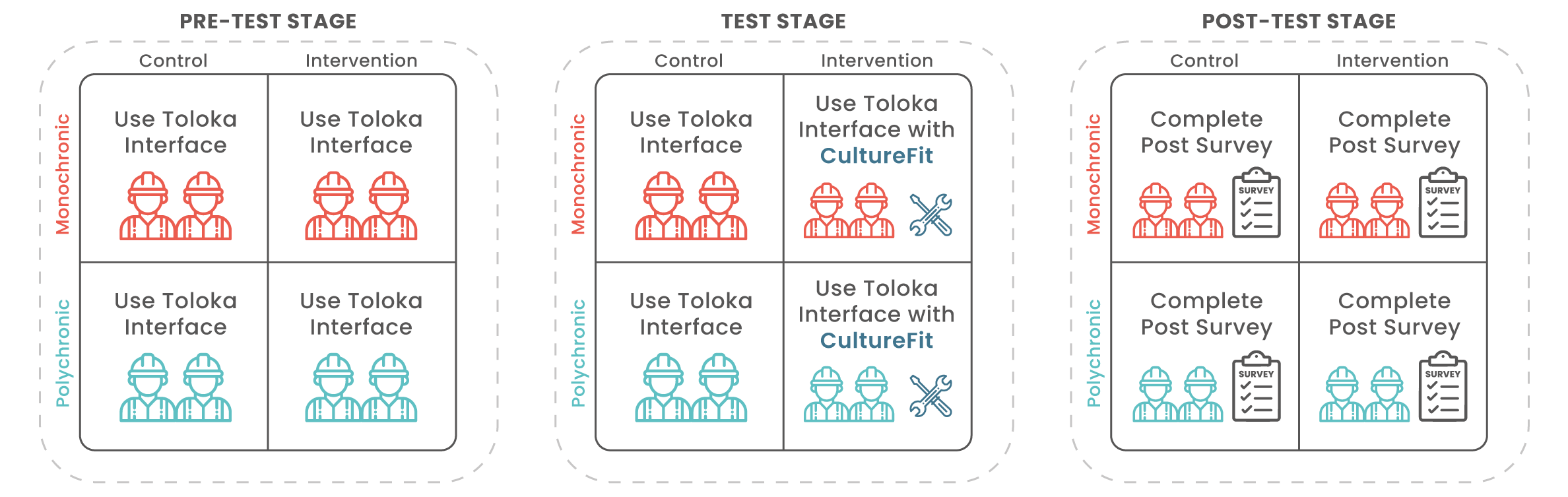}
\caption{Overview of our study that involved a 2x2 between subject study with three different stages.}
\label{fig:experimento}
\end{figure}
\blu{Our IRB-approved field experiment focused on comparing the experiences of crowdworkers who utilized CultureFit and those who did not. This helped us better grasp the impact of culturally aware tools on crowdworkers. We thus implemented a between-subjects design, which included control groups that did not use CultureFit and intervention groups that did. These groups were further divided into sub-groups based on the cultural orientations of the workers, categorizing them as either monochronic or polychronic. We had overall the following groups:} 
\blu{
\begin{enumerate}
    \item Monochronic workers using CultureFit on Toloka (intervention group).
    \item Polychronic workers using CultureFit on Toloka (intervention group).
    \item Monochronic workers following their usual work routine on Toloka WITHOUT CultureFit (control group).
    \item Polychronic workers following their usual work routine on Toloka WITHOUT CultureFit (control group).
\end{enumerate}}

\blu{Having these control and intervention groups, laid the groundwork for a comparative study of workers' experiences. For example, it facilitates an investigation into the experiences of monochronic workers who engaged with CultureFit as opposed to monochronic workers in the control group without access. It also allowed us to compare the experiences of monochronic and polychronic workers with access to CultureFit. Note that both control and intervention groups worked on the existing Toloka interface, but only the intervention groups accessed also the CultureFit web-plugin.}

\blu{Note that for our study, we also implemented a ``randomized control-group pre-test/post-test" design where we divided our study into three phases: pre-test, test, and post-test \cite{christensen2011research}. The test phase signifies the period when workers in the intervention groups begin utilizing CultureFit. The pre-test and post-test phases are designed to capture the state of affairs before and after the application of CultureFit, respectively. Fig. \ref{fig:experimento} presents an overview of this experiment design.} \blu{It is also important to highlight that given the potential market fluctuations in crowdsourcing markets  \cite{bloomfield2019quantitative,savage2020becoming,chiang2018crowd}, merely comparing data from before and after our tool was used may not accurately reflect our tool's effectiveness. This is due to the possibility that any observed improvements in workers' wages might be attributed to general market trends rather than our tool specifically. This is also why we implemented both control and intervention conditions alongside the ``before-and-after'' stages. By comparing worker groups with and without access to our tool (intervention and control conditions), and by conducting before and after analyses of each condition, we aimed to more effectively isolate the actual impact of CultureFit. Overall, this methodological approach helps us discern our tool's real world contribution to workers' experiences amidst the dynamic nature of the Toloka crowdsourcing market.}

\soutc{We conducted an IRB-approved field experiment to study how culture theory might improve the labor outcomes of workers. Similar to prior work, we measured labor outcomes based on workers' wages, their digital traces, and their perceptions. Conducting the experiment in the real world allows us to investigate the natural usages of our system, and better understand the strengths and weaknesses of culture-aware tools in crowd work. Throughout the study we were committed to complete transparency with the participants regarding the data collected from them, and the nature of the recommendations they would receive. In line with prior work in addition to providing this information participants also had the option to drop from the study and still receive compensation.} 

\soutc{For our field experiment, we considered the following four conditions: 1.Crowdworkers with monochronic cultural traits do their labor with CultureFit; 2. Crowdworkers with polychronic cultural traits do their labor with CultureFit; 3. Monochronic crowdworkers do their labor as normal (control); and 4. Polychronic crowdworkers do their labor as normal (control). We used a ``randomized control-group pre-test test'' design for our field experiment \cite{christensen2011research}. This setup is similar to a between-subjects study, but also includes measurements taken before and after the intervention \cite{bloomfield2019quantitative}. Such a setup helps us to better understand how our tool leads to changes in labor outcomes. We thus split our experiment into stages:}

\subsection{Pre-Test Stage.} \blu{We initiated our study with a 7-day ``Pre-Test Stage" applicable to all workers. As mentioned above, this pre-test phase helped us establish a baseline understanding of the wages, perceptions, and digital behaviors among crowdworkers across our four conditions. The duration and structure of this stage, as well as subsequent stages in our study, were informed by prior research \cite{chiang2018crowd,savage2020becoming,toxtli2021quantifying}. To establish our baseline, we guided all workers through a structured process:}

\begin{enumerate}

     \item \blu{\textbf{Pre-Survey Completion}: All workers completed a pre-survey that we crafted to gather information about their demographics, crowd work experiences, and to classify them as either polychronic or monochronic workers. Our Appendix presents our pre-survey.}
     
    \item \blu{\textbf{Web-plugin Installation}: All workers installed a web-plugin equipped with telemetry tracking, which we designed in accordance with prior work \cite{toxtli2021quantifying,saito2019turkscanner}. The web-plugin helps us to monitor and quantify workers' digital behavior (e.g., hourly wages and task completion rates). We allowed the uninstallation of our web-plugin at any time and offered financial rewards for study participation. After installing our web-plugin and finishing the pre-survey, all workers received \$2 USD, an amount set to surpass the US federal minimum wage of \$7.25/hour for a task under 10 minutes \cite{silberman2018responsible}. Workers also earned an extra \$0.5 daily for simply keeping our web-plugin installed, totaling \$3.5 for the 7-day Pre-Test Stage.}

    \item \blu{\textbf{Do Crowd Work}: Participants did crowd work as usual for 7 days. (e.g., they engaged in activities such as data labeling, completing surveys, communicating with requesters, or searching for tasks \cite{toxtli2021quantifying}). This allowed us to observe their baseline.}
\end{enumerate}

\subsection{Test Stage.} 
\blu{During our 7-day Test Stage, workers in the intervention group gained access to CultureFit. CultureFit was activated via the web-plugin previously installed. Crowdworkers using CultureFit began receiving culturally-aware notifications about potential tasks. These workers were encouraged to use CultureFit at their discretion. Meanwhile, the control groups continued their usual crowd work activities, mirroring the Pre-Test Stage. Note that having these control and intervention groups, as well as the different stages, helped us to identify market fluctuations that could influence the results we observed with CultureFit.}

\soutc{We had a 7-day Test Stage to study how workers' experiences and wages changed when using our tool. The duration is consistent with prior work, and was decided on, as opposed to a longer time frame, to help against the known difficulty of engaging crowdworkers in long form studies \cite{huang2016there} which would challenge the feasibility of the study. Workers in the CultureFit conditions received culturally-aware recommendations and those in control did their work as normal without receiving recommendations (they just kept the tool installed for the telemetry tracking). Akin to previous work that had conducted field experiments on crowdsourcing platforms \cite{toxtli2021quantifying}, we paid participants another \$0.5 for each day they used our tool during this stage. On average, we paid participants a total of \$3.5 for using our tool. In all conditions, we continued paying participants each day they kept our tool installed. Finally, to avoid this remuneration interfering with our study, we paid these bonuses at the end of our experiment and did not include our remuneration when calculating workers' wages. Across conditions, we kept the telemetry tracking ``on'' to study the change in workers' wages and digital behavior}

\subsection{\blu{{\bf Post-Test Stage.}}}\blu{At the experiment's end, workers in all conditions completed a post-survey about their experiences during the Test Stage, earning \$1.0 for participation. Overall, participants received \$10 for the entire study. They were instructed to uninstall CultureFit, with the plugin automatically deactivating if not manually removed.}

\soutc{Workers in all conditions were asked to complete a post-survey at the end of the experiment. The survey provided us with information about the experiences and perceptions of workers during the Test Stage. We paid participants another \$1.0 for completing the survey. In total, the participants received a total of \$10 for the entire field study. Participants were asked to uninstall CultureFit at this final stage. The plugin was fully deactivated in case the participants missed uninstalling it. }

\soutc{{\bf Field Experiment: Quantitative Data Analysis.} Across conditions, we analyzed the quantitative data collected from our telemetry tracking using the Kruskal-Wallis Omnibus test with the Dunn-Bonferroni post-hoc test. These statistical tests help us to identify conditions with significant differences. We also computed the Mann-Whitney U test for independent group pairwise comparisons. The test allows us to compare the behavior of monochronic versus polychronic workers. Further, we used the Wilcoxon signed-rank test to identify the effect between stages for dependent groups. Particularly, workers' improvements between the Pre-Test and the Test Stages. We use the tie correction for the pairwise tests. Finally, we also studied participants' Likert scale responses in our surveys. Here, we computed the median values of the five-level Likert questions for each condition.}

\soutc{\bf Field Experiment: Qualitative Analysis.} 

\subsection{Data Analysis of Workers' Survey Data. } \blu{We conducted both quantitative and qualitative analyses of participant survey responses. We analyzed their Likert scale answers from surveys, calculating the median of the five-level questions for each condition. We also used affinity diagramming to study open-ended responses from pre- and post-surveys, identifying common themes \cite{lucero2015using,harboe2015real}. This involved color-coding responses for our four experiment conditions and organizing them by question. We further categorized responses using intra-question affinity diagramming, with the authors independently coding data and collaboratively developing three thematic codes.}

\soutc{We analyzed workers' responses to the open-ended questions in the pre-survey and post-survey using affinity diagramming (also known as affinity mapping). This enabled us to identify themes across participant responses. We selected affinity diagramming because it efficiently organizes self-reported information into distinct clusters. 
To begin, raw responses were color-coded to match the four experiment conditions (monochronic/CultureFit, polychronic/CultureFit, monochronic/Control, and polychronic/Control) and organized according to the question prompt. Next, intra-question affinity diagramming was used to categorize responses to each question individually and generate consolidated insights across the survey questions. 
As a group, the authors first independently coded the data and together developed a set of three thematic codes that were applied to the survey responses.}

\subsection{\bf Participants.} \blu{Similar to prior work \cite{lascau2019monotasking}, we used the Toloka crowdsourcing platform to recruit participants. We posted a detailed task description outlining all study requirements and steps, including completion of surveys, web-plugin installation, and data collection. Workers were also informed that they could drop out of the study at any time and would be compensated for the duration of their participation. Workers interested in participating were required to complete a pre-survey questionnaire upon accepting the study terms. This pre-survey used the Multitasking Preference Inventory (MPI) to evaluate workers' polychronic or monochronic tendencies \cite{poposki2010multitasking}. The MPI, a 14-item self-assessment tool, is more effective than earlier scales like the Inventory of Polychronic Values (IPV) in distinguishing individual and cultural work styles \cite{bluedorn1999polychronicity}. It includes statements rated on a 5-point scale, assessing multitasking behavior and task-switching preferences  \cite{poposki2010multitasking}. Scores indicate a preference for either monochronic or polychronic work habits. We categorized workers into monochronic or polychronic groups based on their stated preferences.} We recruited 55 participants (23 monochronic, 32 polychronic) from a diverse pool across 24 countries, including the US, Middle-East, Central and South America, various African countries, South and Southeast Asia, and Eastern and Western Europe. Note that a power analysis for a dichotomous endpoint-one-sample study indicated the need for at least 9 participants per condition, assuming a $\alpha$ (Type I error) of 0.2, $\beta$ (Type II error) of 0.05, an anticipated incidence of 0.5, and a power of 0.8 ($1-\beta$). 

We then grouped participants into four conditions:

\begin{itemize}
  \item {\bf Monochronic/CultureFit (M/CF).} 12 workers from monochronic cultures used CultureFit.
  \item {\bf Polychronic/CultureFit (P/CF).} 17 workers from polychronic cultures used CultureFit.
  \item {\bf Monochronic/Control (M/S).} 11 workers from monochronic cultures, who conducted their work as (s)tandard.
  \item {\bf Polychronic/Control (P/S).} 15 workers from polychronic cultures, who conducted their work as (s)tandard.
\end{itemize}

\section{Results}
\blu{We originally recruited 126 Toloka workers, of whom 55 remained throughout our two-week field experiment. This retention rate is typical for real-world experiments conducted on crowdsourcing platforms \cite{huang2016there,hara2018data,toxtli2021quantifying,chiang2018crowd,savage2020becoming}. This paper focuses on the 55 workers who completed our study throughout its duration. However, section 6.7 compares these 55 participants with those who dropped out to identify any tendencies or traits possibly influencing dropout rates in real-world longitudinal studies. }

\subsection{Results: Overview}
\blu{Our study included 55 workers—23 monochronic and 32 polychronic, with a median age of 30—which is a typical sample size for real world deployments \cite{huang2016there,toxtli2021quantifying,chiang2018crowd,savage2020becoming}. We collected over two million telemetry logs on Toloka, tracking workers' mouse clicks, scrolls, keyboard activity, and page transitions, along with wage and task data, and interactions with our tool's notifications. Over two weeks, workers engaged with 2,303 tasks, with monochronic workers interacting with a median of 89 tasks ($\mu=156$, $\sigma=193$), and polychronic workers with a median of 83 tasks ($\mu=255$, $\sigma=587$).} 

\begin{figure}[h]
\includegraphics[width=1.0\linewidth]{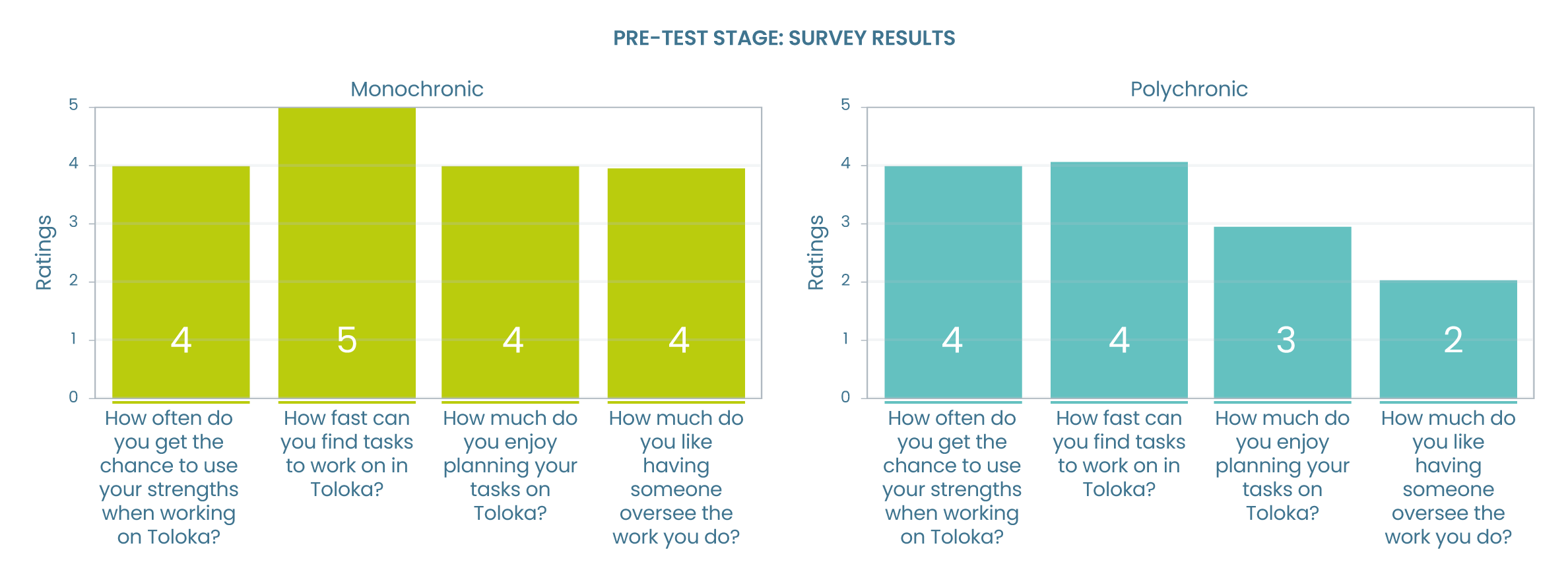}
\caption{\blu{Pre-survey Overview: Monochronic vs. Polychronic Workers}}
\label{fig:presurvey}
\end{figure}

\begin{table}[t]
{\begin{tabular}{lllll}
\toprule
{ \textbf{\small Question}} & 
{ \textbf{ \small Monochronic}} &
{ \textbf{\small Polychronic}} \\ \midrule
{ {\small \blu{How often do you get the chance to use your strengths when working on Toloka?}}}         & 
{ 4 ($\mu=4.0$, $\sigma=1.0$)} & 
{ 4 ($\mu=3.8$, $\sigma=1.1$)} \\
{ {\small \blu{How fast can you find tasks to work on in Toloka?}}}        & 
{ 5 ($\mu=4.4$, $\sigma=0.6$)} & 
{ 4 ($\mu=3.6$, $\sigma=1.2$)} \\
{ {\small \blu{How much do you enjoy planning your tasks on Toloka?}}}         & 
{ 4 ($\mu=3.8$, $\sigma=1.0$)} & 
{ 3 ($\mu=3.2$, $\sigma=0.9$)} \\
{ {\small \blu{How much do you like having someone oversee the work you do?}}} & 
{ 4 ($\mu=3.6$, $\sigma=1.2$)} & 
{ 2 ($\mu=2.0$, $\sigma=1.1$)} \\\bottomrule
\end{tabular}}
\caption{\label{tab:presurveydata}\blu{Overview of Pre-survey responses.}}
\end{table}

\subsection{\blu{Pre-Test Stage: Survey Results.}} 
\blu{Fig. \ref{fig:presurvey} and Table \ref{tab:presurveydata} present our pre-test survey results. We investigated differences in views between monochronic and polychronic workers by computing median Likert scale values and applying the Mann-Whitney U Test. Results revealed no significant differences in workers' perceived use of their strengths in their jobs ($H=233$, $p=0.32$). However, significant cultural differences emerged in workers' preferences for planning. In specific,  monochronic workers showed a stronger preference than polychronic workers for planning their work on Toloka ($U=394$, $p=0.07$). Monochronic workers also preferred closer supervision, similarly marking a significant difference in supervision preferences ($U=170$, $p=0.001$). These findings are consistent with the known cultural preferences of monochronic individuals, who typically favor structured schedules and more formal work relationships \cite{trompenaars2004managing,hall1989beyond}. Conversely, under half of the monochronic workers (10 out of 23) reported using web forums (e.g., Reddit, Quora, Facebook Groups) for work assistance, without favoring any specific platform. Meanwhile, two-thirds of polychronic participants used web forums, notably favoring Facebook. Only 10\% of monochronic workers used job assistance tools, with no polychronic workers doing so.}

\begin{figure}[h]
\includegraphics[width=1.0\linewidth]{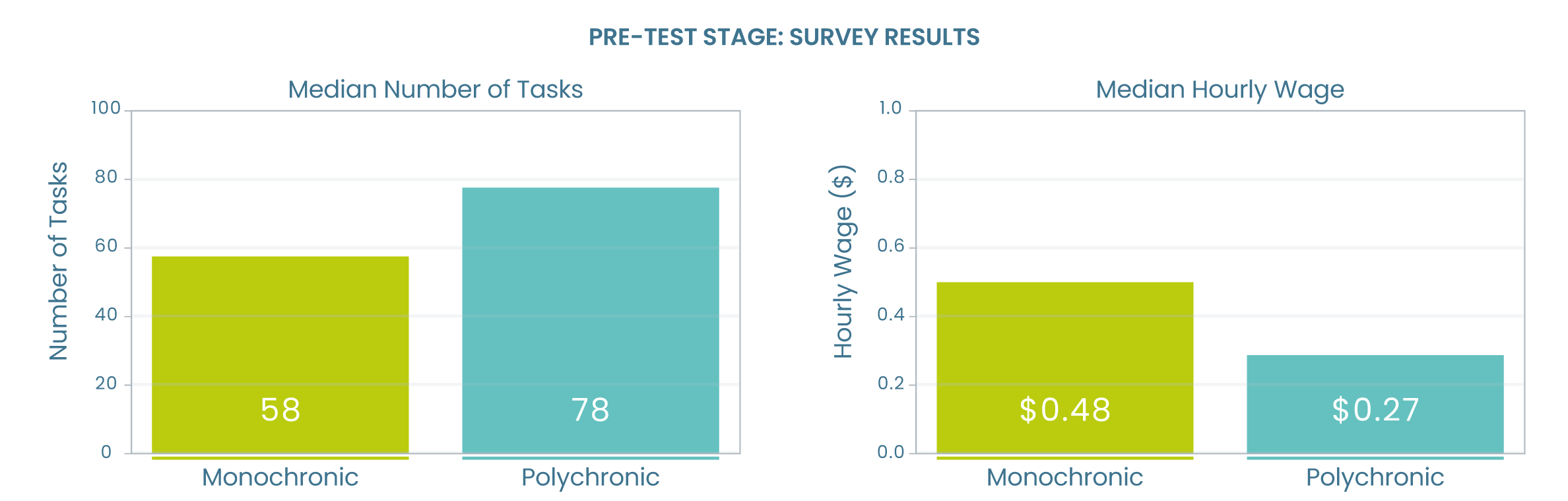}
\caption{Pre-Test Telemetry Log Summary for workers of different cultural groups.}
\label{fig:preTestLog}
\end{figure}

\begin{table}[h]
{\begin{tabular}{lllll}
\toprule
{ \textbf{ \blu{Variable}}} & 
{ \textbf{ \blu{Monochronic}}} &
{ \textbf{ \blu{Polychronic}}} \\ \midrule
{ \blu{Median Number of Tasks}}         & 
{ \blu{58 tasks ($\mu=85$, $\sigma=84$)}} & 
{ \blu{78 tasks ($\mu=89$, $\sigma=76$)}} \\
{ \blu{Median Hourly Wage}}        & 
{\blu{\$0.48 ($\mu=\$1.05$, $\sigma=\$1.40$)}} & 
{ \blu{\$0.27 ($\mu=\$1.51$, $\sigma=\$2.46$)}} \\\bottomrule
\end{tabular}}
\caption{\label{tab:telePre} \blu{Telemetry Log Results for the Pre-Test Stage.}}
\end{table}

\subsection{\blu{Pre-Test Stage: Telemetry Logs Results.}} 
\blu{We also collected data on the wages and number of tasks that workers in our pre-test stage completed using our web plugins with Telemetry tracking, following the same methodologies from prior work \cite{savage2020becoming,toxtli2021quantifying}. Table \ref{tab:telePre} and Fig. \ref{fig:preTestLog} presents a summary of these results. Monochronic workers finished fewer tasks but earned slightly higher hourly wages than polychronic workers. Next, we wanted to study if these differences were significant. We thus first computed for each cultural group the Shapiro-Wilk test on their distribution of tasks and their distribution of wages. Across groups, this test indicated that we were working with non-normal distributions (p-value < .05). Based on this, to evaluate potential significant differences in the task and wage distributions between monochronic and polychronic workers, we decided to employ the Kruskal-Wallis Omnibus test. This non-parametric test, designed for comparing median values across independent distributions, allowed us to first analyze the disparities in task distributions between monochronic and polychronic workers. Subsequently, we applied the same technique to assess differences in the wage distributions of polychronic and monochronic workers. Our findings showed no significant differences in neither the number of tasks that monochronic and polychronic workers completed ($H=7$, $p=0.06$) nor in their wages ($H=4$, $p=0.25$).}

\soutc{{ \bf Pre-Test Stage Results: Telemetry Logs.} With our telemetry logs, we detected that during the week-long Pre-Test stage, monochronic workers completed a median of 58 tasks ($\mu=85$, $\sigma=84$), and had a median hourly wage of \$0.48 ($\mu=\$1.05$, $\sigma=\$1.40$).  At this stage, we did not observe significant differences between the number of tasks completed nor the wages of the monochronic workers who were in the control condition and those in the CultureFit condition. Similarly, polychronic workers completed a median of 78 tasks ($\mu=89$, $\sigma=76$), and had a median hourly wage of \$0.27 ($\mu=\$1.51$, $\sigma=\$2.46$). Across conditions, we also did not identify any significant differences in the wages nor number of completed tasks of the polychronic workers.  Overall, despite minor differences between the digital behaviors of the monochronic and polychronic workers across different conditions, we did not see in the Pre-Test stage significant difference in either the number of tasks completed by workers ($H=7$, $p=0.06$), nor in their wages ($H=4$, $p=0.25$). As discussed in the Methods section, we calculated statistical significant differences via the Kruskal-Wallis Omnibus test. We also calculated workers' wages based on prior work. It is important to note that the median number of tasks performed by Toloka workers in our study is lower than the median performed by crowdworkers in Amazon Mechanical Turk. }

\subsection{\blu{Test Stage: Telemetry Logs Results.}}
\blu{Our goal in the Test Stage was to investigate the effects of embedding culture theory within crowd work tools. This includes examining shifts in workers'  wages and digital behaviors as indicators of change.}
\begin{table}[ht]
{\begin{tabular}{lll}\toprule
{ \textbf{Condition}} & 
{ \textbf{Pre-Test}} & 
{ \textbf{Test}} \\\midrule
{ Monochronic/CultureFit}  & 
{ \$0.34 ($\mu=\$0.60$, $\sigma=\$0.92$)} & 
{ \$0.39 ($\mu=\$1.65$, $\sigma=\$1.70$)}  \\
{ Polychronic/CultureFit} & 
{ \$0.16 ($\mu=\$1.27$, $\sigma=\$1.78$)} & 
{ \$0.68 ($\mu=\$1.31$, $\sigma=\$1.63$)}  \\
{ Monochronic/Control} & 
{ \$0.62 ($\mu=\$1.50$, $\sigma=\$1.89$)} & 
{ \$0.69 ($\mu=\$1.65$, $\sigma=\$1.70$)}  \\
{ Polychronic/Control}  & 
{ \$0.39 ($\mu=\$1.76$, $\sigma=\$3.14$)} & 
{ \$1.04 ($\mu=\$1.89$, $\sigma=\$2.04$)} \\\bottomrule
\end{tabular}}
\caption{\label{tab:earnings}Workers' median earnings during the Pre-Test and Test stages across the different conditions.}
\end{table}

\begin{figure}[h]
  \begin{center}
    \includegraphics[width=0.6\textwidth]{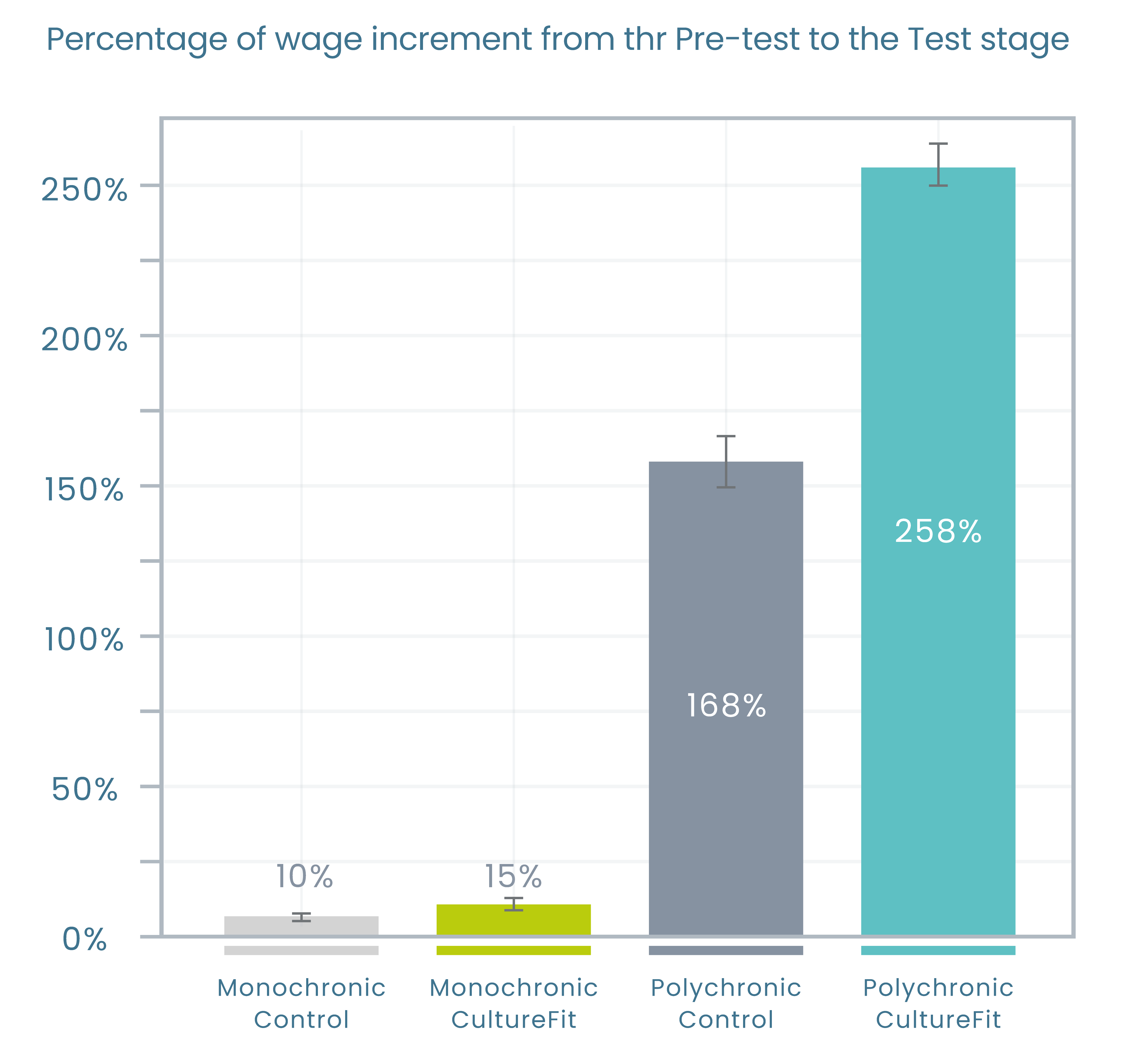}
  \end{center}
  \caption{Overview of workers' increase in wages from the Pre-Test Stage to the Test Stage across conditions.  \blu{Error bars represent 95\% confidence intervals.} CultureFit significantly increased the wages of polychronic workers.}
  \label{fig:logs-income}
\end{figure}

\begin{figure}[h]
  \begin{center}
    \includegraphics[width=.8\linewidth]{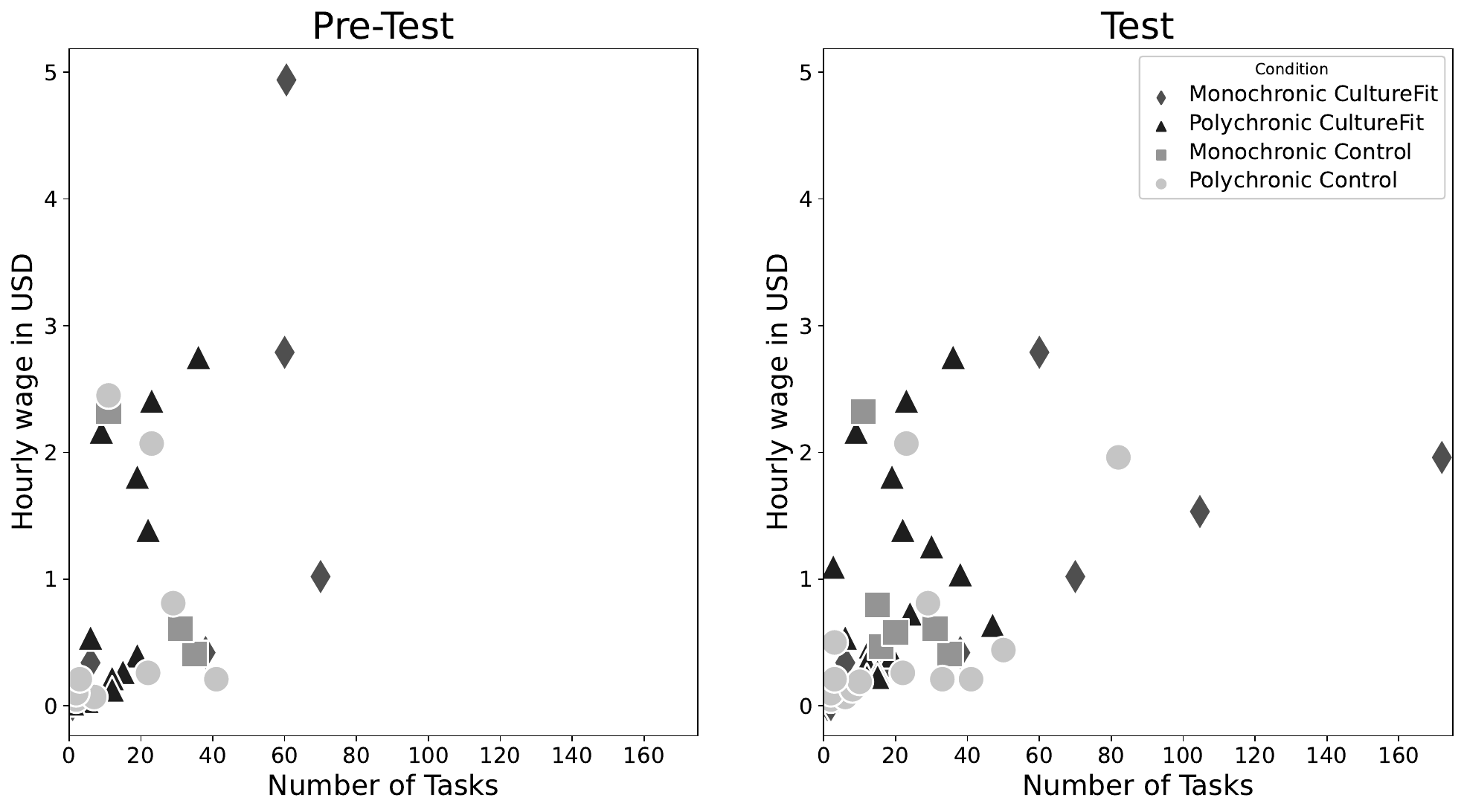}
  \end{center}
  \caption{Overview per condition of the number of tasks each worker completed (X-axis) and the total wages they received (Y-axis).} 
  \label{fig:scatterPlot}
  \vspace{-4pt}
\end{figure}

\subsubsection{\blu{Hourly Wages.}} \blu{To assess the potential impact of CultureFit on workers' wages, we first established the baseline median hourly wage for each condition (control and CultureFit) during the Pre-Test phase. We then compared this baseline to workers'  median hourly wage during the Test Stage. It is important to note that we calculated workers' hourly wages using the same methodologies as previous studies \cite{savage2020becoming,toxtli2021quantifying}. The results, detailed in Table \ref{tab:earnings}, show workers' median hourly wages across the four different conditions and stages. Figure \ref{fig:logs-income} illustrates the percentage increase in the median hourly wages of workers between the Pre-Test and Test Stages. To calculate this percentage change, we subtracted the median Pre-Test wage from the median Test wage and then divided the result by the median Pre-Test wage. This highlights how median hourly wages evolved across the different conditions during our study.}

\blu{Our results uncovered that the wages for workers increased during the Test stage across all conditions. Next, we studied whether this increase was significant. For this purpose, we first conducted the Shapiro-Wilk test on the wage distributions across conditions during the Pre-Test and during the Test stages, finding non-normal distributions (p-value < .05). Based on this, we decided to use the Kruskal-Wallis Omnibus test to assess if the wage changes observed in each condition were significant. We found that workers in the polychronic CultureFit condition did change significantly their wages ($Z=11$, $p=0.01$). The median wages of these workers increased 258\%, going from \$0.19 to \$0.68 USD. For the other conditions, we did not observe any significant wage increases.} \blu{For instance, polychronic workers in the control condition saw median wages rise from \$0.39 to \$1.04 USD, indicating an increase of 168\%. But, the change was not significant $(Z=9$, \(p=0.48\)). This lack of significance is likely due to high wage variability among polychronic workers in the control group (\(\sigma=2.04\)), indicating inconsistent wage increases across workers. Similarly, monochronic workers using CultureFit experienced a 15\% wage increase, from \$0.62 to \$0.69 USD per hour, but this was also not significant (\(Z=5\), \(p=0.31\)). Monochronic workers in the control condition had a 10\% wage increase, which was again not statistically significant (\(Z=4\), \(p=0.43\)). These non-significant variations in wages between the pre-test and test stages could just reflect natural fluctuations in the crowdsourcing market \cite{savage2020becoming}.}

\subsubsection{\blu{Analyzing Shifts in Digital Work Patterns.}} \blu{To further study the changes our tool might have created in workers' digital traces, we created scatter plots from the Pre-Test and Test stages (Fig. \ref{fig:scatterPlot}), plotting each worker's task count (X-axis) against their earnings (Y-axis). In these scatter plots, each point represents a worker that is color and figure coded according to their condition. Comparing Fig. \ref{fig:scatterPlot}'s Pre-Test scatter plot to its Test one, we see two key trends: an upward shift indicating higher hourly wages and a rightward shift reflecting increased task completion. The upward trend is most notable in the Polychronic CultureFit group where workers experienced significant wage increases.}

\subsection{\blu{Post-Test Stage: Post-Survey Quantitative Results.}}
\begin{figure}[h]
  \begin{center}
    \includegraphics[width=.8\linewidth]{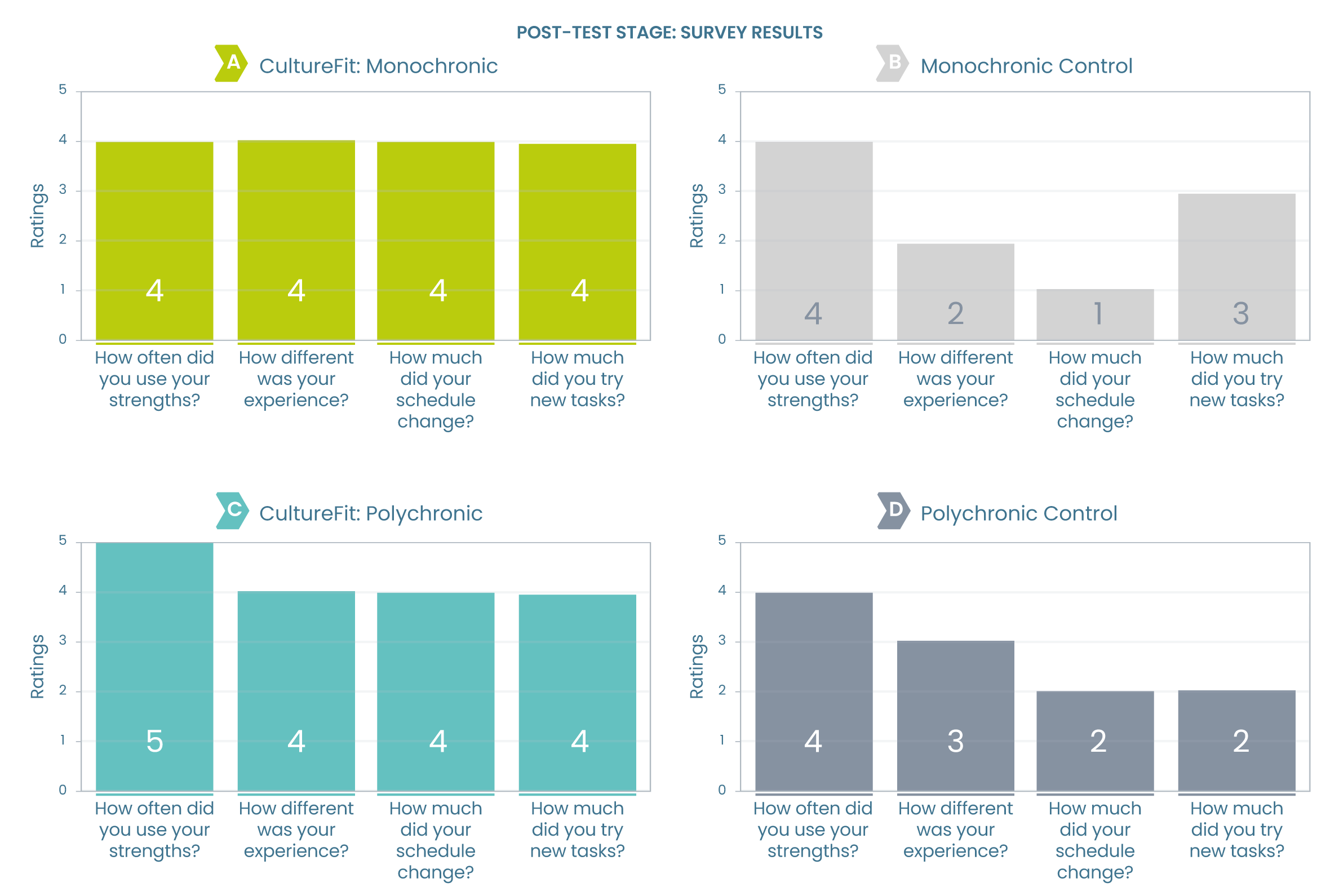}
  \end{center}
  \caption{\blu{Median Post-Survey Response Summary.}} 
  \label{fig:postSurvey}
  \vspace{-4pt}
\end{figure}
\blu{After the Test Stage, our tool ceased operations, halting notifications and behavior tracking, and redirected workers to a post-survey. Fig \ref{fig:postSurvey} and Table \ref{tab:postsurveydata} shows the median perceptions of workers from the post-survey under different conditions. We studied whether CultureFit usage led to significantly different perceptions compared to non-users, i.e., we analyzed the responses between workers utilizing CultureFit (Figs. \ref{fig:postSurvey}.A and \ref{fig:postSurvey}.C) and those who did not (Figs. \ref{fig:postSurvey}.B and \ref{fig:postSurvey}.D).  For this purpose, we first computed the Shapiro-Wilk test, which indicated non-normal distributions for both groups across survey questions (\(p < .05\)). As a result, we utilized the Kruskal-Wallis test to determine whether significant differences existed among the responses from these groups.} 
\begin{table}[ht]
\centering\resizebox{\columnwidth}{!}{\begin{tabular}{lllll}\toprule
{ \textbf{\LARGE Question}} & 
{ \textbf{\LARGE Monochronic/CultureFit}} &
{ \textbf{\LARGE Polychronic/CultureFit}} & {\large \textbf{\LARGE Monochronic/Control}} &
{ \textbf{\LARGE Polychronic/Control}}\\ \midrule
{ \blu{\LARGE How often did you get to use your strengths while working on Toloka this week?}}          & 
{\LARGE 4 ($\mu=4.0$, $\sigma=0.8$)} & 
{\LARGE 5 ($\mu=4.5$, $\sigma=0.8$)} & 
{\LARGE 4 ($\mu=4.0$, $\sigma=0.8$)} & 
{\LARGE 4 ($\mu=4.1$, $\sigma=0.8$)} \\
{ \blu{\LARGE How different did your experience feel while working on Toloka this week?}}      & 
{\LARGE 4 ($\mu=3.9$, $\sigma=0.9$)} & 
{\LARGE 4 ($\mu=3.6$, $\sigma=1.3$)} & 
{\LARGE 2 ($\mu=3.0$, $\sigma=1.5$)} & 
{\LARGE 3 ($\mu=2.9$, $\sigma=1.3$)} \\
{\blu{\LARGE How much has your work schedule on Toloka changed this week compared to previous weeks?}}    & 
{\LARGE 4 ($\mu=3.4$, $\sigma=1.2$)} & 
{\LARGE 4 ($\mu=3.2$, $\sigma=1.5$)} & 
{\LARGE 1 ($\mu=2.3$, $\sigma=1.8$)} & 
{\LARGE 2 ($\mu=2.4$, $\sigma=1.5$)} \\
{\blu{\LARGE How much did you get to try out new tasks on Toloka this week?}}    & 
{\LARGE 4 ($\mu=3.7$, $\sigma=0.7$)} & 
{\LARGE 4 ($\mu=3.2$, $\sigma=1.4$)} & 
{\LARGE 3 ($\mu=3.1$, $\sigma=1.6$)} & 
{\LARGE 2 ($\mu=2.6$, $\sigma=1.2$)} \\\bottomrule
\end{tabular}}
\caption{\label{tab:postsurveydata}Overview of the median Post-survey responses.}
\end{table}
Our analysis uncovered significant differences between CultureFit users (Figs. \ref{fig:postSurvey}.A and \ref{fig:postSurvey}.C) and non-users (Figs. \ref{fig:postSurvey}.B and \ref{fig:postSurvey}.D) in reporting changes to their experiences (\(U=210, p=0.03\)) and schedules (\(U=209, p=0.03\)) during the Test-Stage. 

We also investigated CultureFit's impact on enhancing workers' perceptions of utilizing their strengths. To quantify this change, we utilized the Wilcoxon signed-rank Z test, comparing perceptions from the Pre-Test to the Test stages. This test evaluates the impact of an intervention on a participant group by comparing the variations in their responses before and after the intervention. It does not require the differences between paired observations to adhere to a normal distribution. Polychronic workers in the CultureFit condition had a significant shift in how much they felt that they utilized their strengths (Fig \ref{fig:postSurvey}.C), moving from a median of 4 (\(\mu=3.8\), \(\sigma=1.1\)) in the Pre-Test to 5 (\(\mu=4.5\), \(\sigma=0.8\)) (\(Z=3\), \(p=0.01\)) in the Test stage. We did not observe significant changes in other conditions (Fig \ref{fig:postSurvey}.A, B, and D).

\subsection{\blu{Post-Test Stage: Post-Survey Qualitative Results.}}
\blu{We analyzed workers' open-ended survey responses, identifying common themes:}

\subsubsection{\blu{Culturally-Aware Interfaces and Improved Work Practices.}} Workers using CultureFit reported that the tool brought about positive changes in their work behavior: \emph{``Despite it [CultureFit] changing my behavior, it was for the best '' [P/CF~10].} Workers using CultureFit felt that it transformed their work habits for the better by providing a better understanding of time dynamics in the crowdsourcing market, which enabled them to manage their work schedule more effectively: \emph{``It [the tool] gave me a better sense of how much time I should work. The tool allowed me to stay on top of what jobs were in Toloka and what was new. So I was able to schedule my time more effectively.'' [M/CF~7].} Further, it was interesting to observe that workers considered that despite the tools changing their work practices, they did not feel that the tool interfered with their work: \emph{``I love how the plugin doesn't really interfere and affect your work'' [P/CF~4].} It is important to note that the belief of CultureFit leading to better work practices occurred for both monochronic and polychronic workers. Overall, workers from both cultures perceived a positive change in their work processes when using the tool. Polychronic workers were pleased with the positive change that CultureFit had on their work, and did not express that the tool negatively interfered with their work. Meanwhile, participants from monochronic cultures also noticed a positive change. The tool helped them to be more mindful of the time spent using the platform.

\subsubsection{\blu{Culture and Social Features in Work Tools.}} Polychronic workers differed from monochronic workers in the features they desired in future tools, regardless of the conditions they were. Polychronic workers valued having tools with ``social features''. In specific, polychronic workers mentioned that they preferred applications that facilitated communication with the crowdsourcing platform, requesters, and other workers. Also, polychronic workers with access to CultureFit expressed that they valued that the interface helped them to connect properly with requesters and other workers: \emph{``The tool provides a friendly link between the client [requesters] and the service provider [workers], that interests me [P/CF~8].''} \blu{Note that the "friendly link" mentioned by workers primarily refers to notifications about messages from requesters on Toloka—notifications that Toloka's interface design did not prioritize or effectively alert workers to.} Further, polychronic workers in the control mentioned that in future tools they would like to see the integration of social features: \emph{``Plugins should also be tools that increase communication between clients [requesters] and providers [workers]. The ability to communicate more [with requesters] and hear back about whether or not my efforts are appreciated would be helpful [P/C~9].''} Finally, monochronic workers did not mention wanting social features. This might be because culture theory argues that such features are not as important in monochronic cultures \cite{bouncken2004cultural,ursu2021exploring}.

\subsubsection{\blu{Positive Experiences with Culturally-Aware Notification Tools.}} Participants in the CultureFit conditions reported that our tool led them to have more positive and engaging labor experiences. This result resembles what prior research has found about crowdworkers and their positive experiences with tooling \cite{williams2019perpetual,kaplan2018striving}. The positive sentiment with our tool led some to engage in more crowd work. As one participant expressed, \emph{``Since installing the plugin, I felt more inclined to check out Toloka on a daily basis, and by doing that, I found a few more tasks that I usually wouldn't have seen, this meant I worked on a few more tasks, and I enjoyed these [M/C~5].''}

\subsection{\blu{Analyzing the Crowdworker Dropouts}}
\begin{figure}[h]
   \begin{center}
     \includegraphics[width=.9\linewidth]{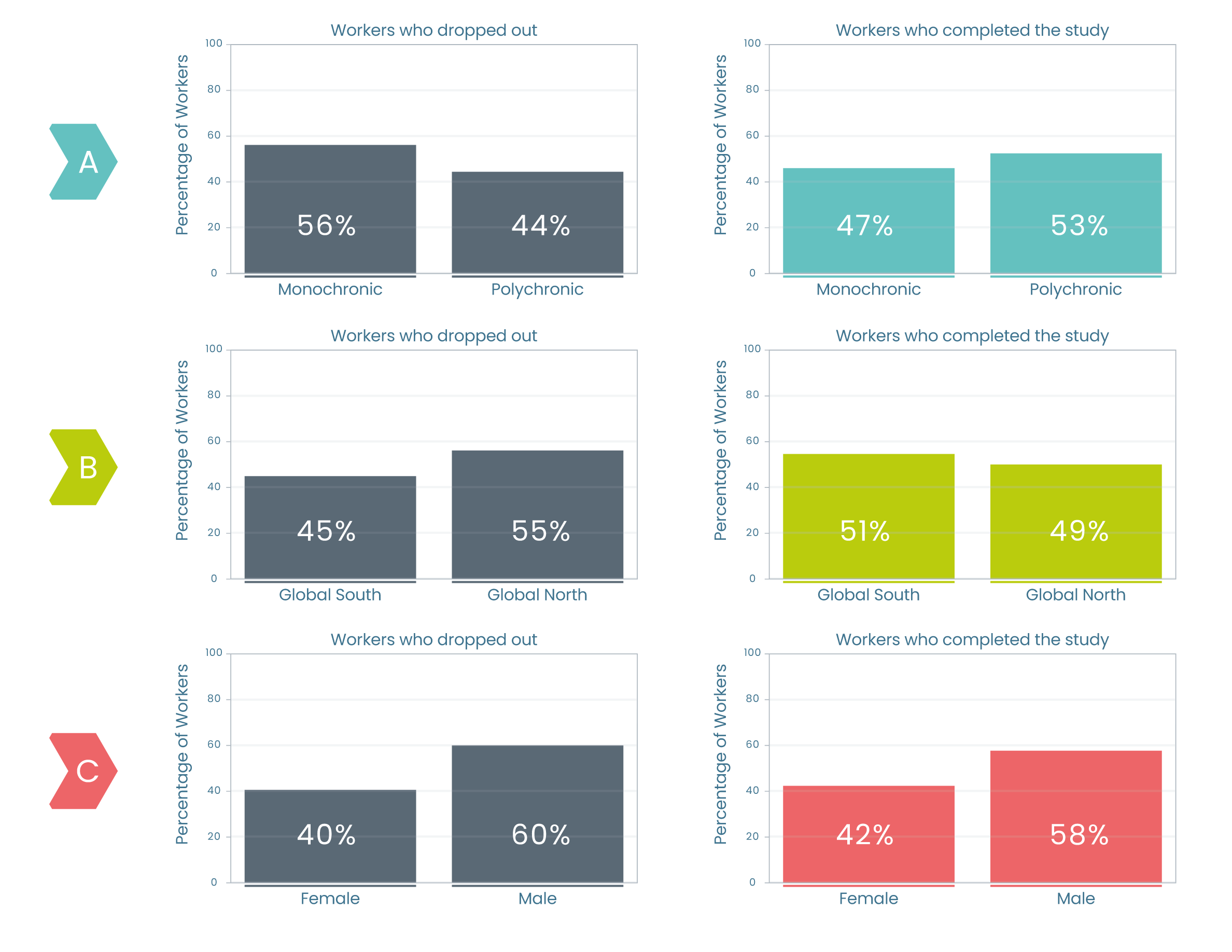}
   \end{center}
   \caption{\blu{Characteristics of ``persisting'' crowdworkers and those who ``dropped out''}.} 
   \label{fig:participantsPlot}
   \vspace{-4pt}
 \end{figure}

\blu{This paper focuses on the 55 crowdworkers who completed our study. However, we also examined the characteristics of the "dropouts" to gain insights into the dynamics of conducting real-world experiments with new crowd work tool designs. Figure \ref{fig:participantsPlot} presents different traits contrasting ``dropouts'' with those who stayed.} 

\blu{Fig. \ref{fig:participantsPlot}.A shows the completion and dropout rates among workers with different cultural traits, with nearly equal numbers of monochronic (56\%) and polychronic (44\%) workers dropping out. Next, we investigated if there were significant differences in cultural traits between workers who dropped out and those who remained in the study. A Chi-square test of independence revealed that there was no significant association between the cultural traits of workers who dropped out or those who continued participating in the study, $X^2$ (1, N=77)=1.91, p=0.16.}

\blu{In Figure \ref{fig:participantsPlot}.B, we categorized workers' countries of origin into global south or north based on existing literature \cite{wallerstein2020world,stiglitz2004globalization,GlobalSouth2023}, and displayed the distribution through bar plots showing the percentage of workers from each region. A Chi-square test of independence indicated no significant link between workers' geographic regions of origin and their choices to either discontinue or persist in the study, $X^2$ (34, N=77)=32.66, p=0.53.} 

\blu{In Fig. \ref{fig:participantsPlot}.C, we display the gender distributions for workers who dropped out versus those who stayed in the study. A Chi-square test of independence found no significant relationship between the gender of workers and those who dropped out or remained in the study, $X^2$ (1, N=77)=0.005, p=0.94.}

\blu{Overall, these findings from the Chi-square tests can help to dispel the notion of survivorship bias linked to the demographics of workers. Next, we examined if reductions in wages or the number of tasks completed might have prompted workers to exit our study, questioning if our tool could have negatively affected workers' productivity and led to their departure. However, our analysis showed that all dropouts occurred within the pre-test stage (usually also within the first day), before workers received any notifications from our tool. Therefore, dropouts were not due to a decline in productivity from using CultureFit. Future studies could benefit from interviewing the dropouts to understand why some crowdworkers persist in longitudinal studies while others choose to leave.}

\section{Discussion}
Our research pioneers ways to start integrating \soutc{certain} cultural dimensions into \blu{the design of tools for crowdworkers} \soutc{tool design, in order to better meet the diverse preferences and backgrounds of crowdworkers}. Such direction is vital to the field of \soutc{design} \blu{CSCW}, as it opens up new opportunities for creating \soutc{tools and} \blu{computational} artifacts \blu{that can support} \soutc{for} the global workforce that exists on crowdsourcing platforms \cite{fox2020worker,potocka2022crowdwork,varanasi2022feeling,de2015rise,difallah2018demographics}. Through our real-world field experiments, we \blu{started to see} \soutc{discovered} the \blu{potential} benefits of having a culturally-sensitive tool, particularly for workers with polychronic traits \cite{difallah2018demographics,morden1999models,lee2011global,lascau2019monotasking}. This \blu{can be important} \soutc{is significant}, as previous studies had shown that these workers are often disproportionately affected by the challenges of crowd work \cite{gray2019ghost,chen2020,toxtli2021quantifying,lascau2019monotasking}.  

\subsection{\blu{Understanding the Integration of Cultural Insights in Crowd Work Tool Design}} 
\blu{Inspired by culture theory \cite{ballard2004communication, hall1989understanding, bluedorn1999polychronicity, robinson1990handbook}, we created a culturally aware tool tailored to both polychronic and monochronic crowdworkers. Based on the theory, we anticipated that our tool would enhance productivity and satisfaction for both groups. We expected polychronic workers would find value in our tool's notifications designed to enhance multitasking and adapt to flexible schedules \cite{hall1989understanding}. Conversely, monochronic workers would appreciate our tool's notifications that promote structured and sequential task management, supporting their preference for orderly progress \cite{hall1989understanding,bluedorn1999polychronicity,hall1971paradox, nardon2009culture}. In summary, we expected that our tool's culturally aware notifications would help both groups thrive within crowd work.}

\blu{However, our results were unexpected. Initially, in the Pre-Test phase of our field experiment, both polychronic and monochronic workers demonstrated similar performance in terms of tasks completed and hourly wages. Yet, when they began using our tool during the Test phase, a notable divergence occurred: polychronic workers saw a significant increase in their wages, while monochronic workers experienced no such improvement (see Fig. \ref{fig:logs-income}). Consequently, it became clear that our culturally aware tool did not universally enhance labor outcomes. } 

\blu{For example, our scatter plot, illustrated in Figure \ref{fig:scatterPlot}, reveals that most monochronic workers in the Test stage are clustered at the lower regions of both the X and Y axes, mirroring their positioning in the control condition and in the Pre-Test phase. This similarity further showcases that our tool did not impact the economic outcomes or the number of tasks monochronic workers completed. In  contrast, Figure \ref{fig:scatterPlot} in the Test stage shows a noticeable shift for polychronic workers. Initially concentrated around the origin point (0,0) in the Pre-Test phase, these workers now appear more prominently distributed across higher values of both the X and Y axes in the Test visualization. This distribution shift indicates that they completed more tasks and received higher wages, suggesting that our tool effectively aligned with their multitasking abilities and preferences for engaging in diverse and simultaneous activities.}

\subsubsection{\blu{Why might we see these results?}} \blu{The differential impact that CultureFit had to monochornic and polychronic workers could be due to the design bias in workplace technologies \cite{hall1959silent,slack2005culture,marcus2000crosscurrents}, which are often unconsciously tailored to monochronic preferences — organized, sequential, and linear task presentations \cite{trompenaars1993riding,bigelow1998cultural}. This inherent design focus may mean that monochronic workers do not experience as significant a change with the introduction of a culturally aware tool because the tool aligns closely with the existing cultural bias in tool design, which already favors monochronic workers. Conversely, polychronic workers, who are less catered to by standard designs \cite{norman2013design,slack2005culture,hall1959silent}, likely experience more noticeable benefits when a tool is finally adapted to fit their cultural work style. Essentially, while existing tools support monochronic workers well \cite{hall1959silent}, our culturally aware tool likely fills an important gap for polychronic workers whose natural work tendencies are often forgotten in design \cite{rau2012cross,hofstede2001culture}.}

\blu{We believe that the disparity in tool effectiveness revealed by our results underscores the need for designing culturally aware tools, especially for populations traditionally overlooked in the design process. Culturally aware tools are likely to have greater impact on them. We could thus envision future research focusing on designing tools that cater specifically to the culture and needs of rural crowdworkers in the United States, a group often forgotten in mainstream tool design \cite{flores2020challenges,khovanskaya2017reworking,carr2009hollowing}. Creating tools that resonate with the culture of rural areas could be more impactful than continuing to design primarily for urban workers, who might already have access to a wide array of specialized tools \cite{duncan2015worlds,chambers2014rural}. Similarly, it may be more beneficial to design tools that are adapted to the cultures of the Global South rather than continuing to focus predominantly on the Global North \cite{collier2008bottom,prahalad2008fortune}, where there is already an abundance of systems tailored to local needs. This approach can not only promote inclusivity, but can also maximize the potential impact of the computational artifacts by being tailored to cultures that are traditionally underrepresented in design.}

\subsection{\blu{Understanding the Feasibility and Challenges of Culturally-Aware Tools in Crowd Work}} 
\blu{Next, we discuss the challenges and feasibility of our proposed system design.}
 
\emph{Feasibility.} Prior work has provided important design recommendations for crowdworkers with polychronic or monochronic traits \cite{lascau2019monotasking,Lasacau2024}. However, several of these recommendations focus on completely re-designing crowdsourcing platforms \cite{gaikwad2015daemo}, or forcing workers and requesters to change their behaviors \cite{gaikwad2016boomerang,toxtli2020reputation,brabham2012crowdsourcing}. However, pressing people and platforms to change is not always feasible \cite{Cheon2024}. In our design of CultureFit, we focused on designing a tool that could co-exist within existing crowdsourcing platforms, and could automatically adapt to the cultural background of the workers (without neither forcing workers, requesters, nor platforms to have to make significant changes). Our approach can thus make it more feasible to start to create culturally-aware \blu{crowd work tools.} \soutc{opportunities} \blu{Overall, in system design, we acknowledge the advantages of implementing a "tool add-on" architecture that complements existing work environments without major disruption \cite{Newbold2022,Brumby2013}. Future tools for digital labor platforms should embrace this approach to enhance their feasibility.} \blu{Note that feasibility also entails ensuring equal access to tools, which is important for preventing economic disparities among crowdworkers. \cite{williams2019perpetual,gray2019ghost}. To foster equitable tool access, we aim to opensource CultureFit, engage with worker collectives, and employ proven global tool access and adoption strategies \cite{kraemer2009one,angel2015participatory}.}

\soutc{The proliferation of culturally aware tools may also affect how much workers follow the recommendations from such tools. It is likely that monochronic workers did not significantly increase their wages because CultureFit's design for monochronic cultures was similar to other tools that these workers already used.  However, the fact that providing polychronic workers with a culturally-aware tool did significantly improve their wages, without needing any modifications to the crowdsourcing platform  suggests that designing culturally-aware tools can be impactful, especially for traditionally excluded populations. Equal tool access, crucial to prevent economic disparities among crowdworkers, should be considered alongside design \cite{williams2019perpetual,gray2019ghost}. To promote equitable distribution and tool adaptation, we plan to open-source CultureFit and collaborate with worker collectives, incorporating research on global tool access and adoption \cite{kraemer2009one,angel2015participatory}.}

\subsubsection*{Challenges.} There are the challenges that can arise with our culturally-aware tool: Requesters from monochronic cultures may be unhappy with polychronic workers who utilize CultureFit for multitasking. This dissatisfaction can stem from cultural differences, as monochronic requesters could perceive multitasking negatively \cite{lee1999time, bluedorn1992many}. To tackle this issue, we propose keeping workers' practices hidden from requesters and only showing them the final outcomes of their work. However, a complication arises from the fact that certain digital labor platforms now employ "surveillance mechanisms" allowing requesters and platforms to monitor workers' actions and determine payment based on their work practices \cite{upwork_2022}. This surveillance assumes that only one work practice is correct, e.g., one that excludes multitasking. To address this challenge, we envision implementing CultureFit on the requesters' side as well. This would provide guidance to requesters on embracing the cultural diversity of their workforce \cite{Kim2023}.

\subsection{\blu{Designing the Future of Culturally-Aware Tools for Crowd Work}}
\blu{Our paper presents findings that we hope will inspire the development of novel crowd work tools. Next, we outline new design directions inspired by our findings and previous research:}

\subsubsection{\blu{Designing Crowd Work Tools for Play and Socialization}}
\blu{Crowd work often isolates workers \cite{gray2019ghost}, limiting their social connections \cite{yao2021together}. Our post-study revealed that some workers from polychronic cultures appreciated CultureFit's socialization notifications, emphasizing the significance of enabling social interactions at work for them. Future crowd work tools could improve socialization by integrating CSCW research on friendship interfaces \cite{schwanda2012see,erickson2000social}, or research on designing work interfaces for "play for play's sake \cite{kasunic2019crowd,blythe2004funology,monk2002funology,zichermann2020gamification}," akin to having ping pong tables in the workplace \cite{friedman2014best,becker1990total}. These tools could better cater to diverse cultural preferences, thereby enhancing worker engagement.} 

\subsubsection{\blu{Designing Crowd Work Tools for Multiple Goals}}
\blu{Our scatter plot (Fig. \ref{fig:scatterPlot}) showed that some of the monochronic workers who used our tool increased their task completion, as indicated by more advanced points along the X-axis. It is likely that our culturally aware tool helped these workers concentrate better and complete tasks without interruption \cite{Janssen2019}, which explains the uptick in activity. However, the plot also showed no corresponding rise along the Y-axis, which measured hourly wages. Overall, our findings indicate that although CultureFit is culturally aware, it did not significantly increase earnings for monochronic workers. In the future, research could delve into creating tools that not only promote cultural awareness but also directly focus on boosting earnings or assist workers in achieving their different goals. Incorporating insights from previous studies focused on diverse objectives—such as wage increases, skill development, and creative pursuits \cite{chiang2018crowd,savage2020becoming,suzuki2016atelier,Shaer2024} —can help researchers develop a more comprehensive approach to designing tools for crowd work.
This approach would enhance task efficiency and support workers' broader career development, offering more inclusive support systems \cite{Kun2023}.}

\subsubsection{\blu{Designing Tools for Cultural Understanding in Crowd Work}}
\blu{A novel direction for culturally aware crowd work tools could involve integrating a real-time cultural exchange platform into the workflow. This feature would allow workers from diverse backgrounds to interactively share cultural insights and practices, thereby enriching the work environment and fostering mutual understanding. For example, workers could engage in brief, timed exchanges sharing cultural greetings, traditions, or personal experiences before starting collaborative tasks. This would deepen understanding and appreciation among workers. Additionally, integrating AI could help overcome language barriers and merge cultural connectivity with productivity, enhancing worker engagement and broadening global perspectives. This future research could benefit from connections with prior CSCW work in ``Cross-Cultural Communication Studies'' \cite{yuan2013understanding,li2023improving,setlock2004taking,lindtner2012cultural,bjorn2016practice}, ``Educational Technology Research'' \cite{roberts2005computer,conole2012designing,hmelo2013international,stahl2006group}, and ``Social Computing'' \cite{christakis2009connected,ciborra1997groupware,meiselwitz2020social}. These fields could guide the tool's design to handle diverse communication styles and cultural norms \cite{hofstede2005cultures}, make the platform more engaging and educational with immersive technologies \cite{gregory2016learning}, facilitate social interactions after or during work \cite{Mella2024,kasunic2019turker}, and even create a supportive worker community with effective moderation tools \cite{kraut2012building,Seering2019,Seering2018}.}


\subsection*{Limitations and Future Work.} The study's findings were limited by the methodology, population, and platform used. 
Future research should consider field experiments which focus on more granular comparisons - such as comparing polychronic workers in Latin America and India. This would compliment recent research which includes investigations into individual regions and niches, such as a new study profiling the working conditions of data anotators in India \cite{10.1145/3491102.3502121}. \soutc{Understanding these cultural differences would benefit crowd work research.} 
Furthermore, the study showed that workers changed their digital behaviors when using the tool, but it remains unclear how they would adapt and use it over the long term. However, there are known challenges regarding the difficulties of performing long term studies with crowdworkers \cite{huang2016there}, additionally, given the novel nature of the work there is little to no established digital infrastructure to support a longitudinal study at this time. This leaves long-term studies and development in the means to do so as valuable future work.
\soutc{Although the current design with CultureFit has not received worker complaints about interruptions or feeling disrupted, the possibility of similar designs having overbearing notifications causing issues for polychronic workers should be considered. Examining these boundaries and how task nature affects also workers' preferences would be valuable.} \blu{Within our limitations we also have to recognize the challenge of ``demand characteristics'' — subtle signals in experiments that can shape participants' behavior to match perceived expectations. Demand characteristics might have affected participants' reported satisfaction with our tool. However, we took several steps to minimize their influence. We carefully used neutral language in our instructions and avoided leading questions to try to ensure that responses were genuine. To reduce the sense of being studied, we also designed our tool as a lightweight web plugin integrated seamlessly into workers' usual digital labor platform. Additionally, we established a control group to help discern the true impact of these characteristics. Moreover, we guaranteed anonymity and confidentiality for all responses, encouraging participants to share their honest feedback without reservation}. \blues{While our study demonstrates the potential benefits of incorporating cultural awareness into crowd work tools, particularly concerning cultural dimensions of time management styles, we acknowledge that this represents only one facet of cultural diversity. Future work could explore how other cultural dimensions, such as power distance or individualism vs. collectivism \cite{hofstede2001culture,hofstede2005cultures}, might be integrated into culturally-aware crowd work tool design.} Despite its limitations, we hope to inspire researchers to incorporate cultural dimensions into their crowd work designs.

\section{Conclusion}
\blu{Crowdsourcing markets often feature standardized interfaces that do not accommodate the cultural diversity of workers, which can adversely affect their well-being and productivity \cite{posada2022coloniality,hofstede2005cultures,meyer2014culture,sun2012cross,marcus2006cross,Roffarello2023,Ricaurte2019}. \blues{Our research studies how considering the cultural dimensions of monochronic and polychronic work styles can positively transform crowdworkers' experiences.} Our paper proposes the creation of culturally customized workplace systems, exemplified by our tool "CultureFit," designed based on Chronemics and culture theories. Through a field experiment involving 55 workers from 24 countries, we found that CultureFit significantly enhanced earnings for culturally diverse workers frequently overlooked in design \cite{hall1959silent,hofstede1984cultural,posada2022coloniality}. Moreover, we will introduce a novel dataset on culture and digital work, laying the groundwork for further research in the area. Overall, our findings underline the significance of incorporating cultural insights into digital labor tool design,  and providing insights for future research directions.}\\

{\bf Acknowledgments.} This research was supported by an NSF CAREER grant. Special thanks to the workers who participated in this research, as well as Dmitry Ustalov, Anastasia Lucas, Eeshani Mondal for their invaluable feedback.

\bibliographystyle{ACM-Reference-Format}
\bibliography{main}


\begin{thebibliography}{197}


\ifx \showCODEN    \undefined \def \showCODEN     #1{\unskip}     \fi
\ifx \showDOI      \undefined \def \showDOI       #1{#1}\fi
\ifx \showISBNx    \undefined \def \showISBNx     #1{\unskip}     \fi
\ifx \showISBNxiii \undefined \def \showISBNxiii  #1{\unskip}     \fi
\ifx \showISSN     \undefined \def \showISSN      #1{\unskip}     \fi
\ifx \showLCCN     \undefined \def \showLCCN      #1{\unskip}     \fi
\ifx \shownote     \undefined \def \shownote      #1{#1}          \fi
\ifx \showarticletitle \undefined \def \showarticletitle #1{#1}   \fi
\ifx \showURL      \undefined \def \showURL       {\relax}        \fi
\providecommand\bibfield[2]{#2}
\providecommand\bibinfo[2]{#2}
\providecommand\natexlab[1]{#1}
\providecommand\showeprint[2][]{arXiv:#2}

\bibitem[Glo(2023)]%
        {GlobalSouth2023}
 \bibinfo{year}{2023}\natexlab{}.
\newblock \bibinfo{title}{Global North and Global South}.
\newblock \bibinfo{howpublished}{\url{https://en.wikipedia.org/wiki/Global_North_and_Global_South}}.
\newblock
\newblock
\shownote{Accessed: 2023-04-27}.


\bibitem[Alimahomed-Wilson and Reese(2021)]%
        {alimahomed2021surveilling}
\bibfield{author}{\bibinfo{person}{Jake Alimahomed-Wilson} {and} \bibinfo{person}{Ellen Reese}.} \bibinfo{year}{2021}\natexlab{}.
\newblock \showarticletitle{Surveilling Amazon’s warehouse workers: racism, retaliation, and worker resistance amid the pandemic}.
\newblock \bibinfo{journal}{\emph{Work in the Global Economy}} \bibinfo{volume}{1}, \bibinfo{number}{1-2} (\bibinfo{year}{2021}), \bibinfo{pages}{55--73}.
\newblock


\bibitem[Alvarado~Garcia et~al\mbox{.}(2021)]%
        {alvarado2021decolonial}
\bibfield{author}{\bibinfo{person}{Adriana Alvarado~Garcia}, \bibinfo{person}{Juan~F Maestre}, \bibinfo{person}{Manuhuia Barcham}, \bibinfo{person}{Marilyn Iriarte}, \bibinfo{person}{Marisol Wong-Villacres}, \bibinfo{person}{Oscar~A Lemus}, \bibinfo{person}{Palak Dudani}, \bibinfo{person}{Pedro Reynolds-Cu{\'e}llar}, \bibinfo{person}{Ruotong Wang}, {and} \bibinfo{person}{Teresa Cerratto~Pargman}.} \bibinfo{year}{2021}\natexlab{}.
\newblock \showarticletitle{Decolonial pathways: Our manifesto for a decolonizing agenda in hci research and design}. In \bibinfo{booktitle}{\emph{Extended Abstracts of the 2021 CHI Conference on Human Factors in Computing Systems}}. \bibinfo{pages}{1--9}.
\newblock


\bibitem[{\'A}ngel et~al\mbox{.}(2015)]%
        {angel2015participatory}
\bibfield{author}{\bibinfo{person}{Walter {\'A}ngel}, \bibinfo{person}{Saiph Savage}, {and} \bibinfo{person}{Nataly Moreno}.} \bibinfo{year}{2015}\natexlab{}.
\newblock \showarticletitle{Participatory stoves: Designing renewable energy technologies for the rural sector}. In \bibinfo{booktitle}{\emph{Proceedings of the 18th ACM Conference Companion on Computer Supported Cooperative Work \& Social Computing}}. \bibinfo{pages}{259--262}.
\newblock


\bibitem[Baeza-Yates et~al\mbox{.}(1999)]%
        {baeza1999modern}
\bibfield{author}{\bibinfo{person}{Ricardo Baeza-Yates}, \bibinfo{person}{Berthier Ribeiro-Neto}, {et~al\mbox{.}}} \bibinfo{year}{1999}\natexlab{}.
\newblock \bibinfo{booktitle}{\emph{Modern information retrieval}}. Vol.~\bibinfo{volume}{463}.
\newblock \bibinfo{publisher}{ACM press New York}.
\newblock


\bibitem[Ballard and Seibold(2004)]%
        {ballard2004communication}
\bibfield{author}{\bibinfo{person}{Dawna~I Ballard} {and} \bibinfo{person}{David~R Seibold}.} \bibinfo{year}{2004}\natexlab{}.
\newblock \showarticletitle{Communication-related organizational structures and work group temporal experiences: the effects of coordination method, technology type, and feedback cycle on members' construals and enactments of time}.
\newblock \bibinfo{journal}{\emph{Communication Monographs}} \bibinfo{volume}{71}, \bibinfo{number}{1} (\bibinfo{year}{2004}), \bibinfo{pages}{1--27}.
\newblock


\bibitem[Becker(1990)]%
        {becker1990total}
\bibfield{author}{\bibinfo{person}{Franklin~D Becker}.} \bibinfo{year}{1990}\natexlab{}.
\newblock \showarticletitle{The total workplace: Facilities management and the elastic organization}.
\newblock \bibinfo{journal}{\emph{(No Title)}} (\bibinfo{year}{1990}).
\newblock


\bibitem[Bigelow(1998)]%
        {bigelow1998cultural}
\bibfield{author}{\bibinfo{person}{H Bigelow}.} \bibinfo{year}{1998}\natexlab{}.
\newblock \showarticletitle{Cultural Differences in the Workplace}.
\newblock \bibinfo{journal}{\emph{Unpublished manuscript, RMIT University, Victoria, Australia}} (\bibinfo{year}{1998}).
\newblock


\bibitem[Bj{\o}rn et~al\mbox{.}(2016)]%
        {bjorn2016practice}
\bibfield{author}{\bibinfo{person}{Pernille Bj{\o}rn}, \bibinfo{person}{Luigina Ciolfi}, \bibinfo{person}{Mark Ackerman}, \bibinfo{person}{Geraldine Fitzpatrick}, {and} \bibinfo{person}{Volker Wulf}.} \bibinfo{year}{2016}\natexlab{}.
\newblock \showarticletitle{Practice-based CSCW Research: ECSCW bridging across the Atlantic}. In \bibinfo{booktitle}{\emph{Proceedings of the 19th ACM Conference on Computer Supported Cooperative Work and Social Computing Companion}}. \bibinfo{pages}{210--220}.
\newblock


\bibitem[Bloomfield and Fisher(2019)]%
        {bloomfield2019quantitative}
\bibfield{author}{\bibinfo{person}{Jacqueline Bloomfield} {and} \bibinfo{person}{Murray~J Fisher}.} \bibinfo{year}{2019}\natexlab{}.
\newblock \showarticletitle{Quantitative research design}.
\newblock \bibinfo{journal}{\emph{Journal of the Australasian Rehabilitation Nurses Association}} \bibinfo{volume}{22}, \bibinfo{number}{2} (\bibinfo{year}{2019}), \bibinfo{pages}{27--30}.
\newblock


\bibitem[Bluedorn(2002)]%
        {bluedorn2002human}
\bibfield{author}{\bibinfo{person}{Allen~C Bluedorn}.} \bibinfo{year}{2002}\natexlab{}.
\newblock \bibinfo{booktitle}{\emph{The human organization of time: Temporal realities and experience}}.
\newblock \bibinfo{publisher}{Stanford University Press}.
\newblock


\bibitem[Bluedorn et~al\mbox{.}(1999)]%
        {bluedorn1999polychronicity}
\bibfield{author}{\bibinfo{person}{Allen~C Bluedorn}, \bibinfo{person}{Thomas~J Kalliath}, \bibinfo{person}{Michael~J Strube}, {and} \bibinfo{person}{Gregg~D Martin}.} \bibinfo{year}{1999}\natexlab{}.
\newblock \showarticletitle{Polychronicity and the Inventory of Polychronic Values (IPV): The development of an instrument to measure a fundamental dimension of organizational culture}.
\newblock \bibinfo{journal}{\emph{Journal of managerial psychology}} \bibinfo{volume}{14}, \bibinfo{number}{3/4} (\bibinfo{year}{1999}), \bibinfo{pages}{205--231}.
\newblock


\bibitem[Bluedorn et~al\mbox{.}(1992)]%
        {bluedorn1992many}
\bibfield{author}{\bibinfo{person}{Allen~C Bluedorn}, \bibinfo{person}{Carol~Felker Kaufman}, {and} \bibinfo{person}{Paul~M Lane}.} \bibinfo{year}{1992}\natexlab{}.
\newblock \showarticletitle{How many things do you like to do at once? An introduction to monochronic and polychronic time}.
\newblock \bibinfo{journal}{\emph{Academy of Management Perspectives}} \bibinfo{volume}{6}, \bibinfo{number}{4} (\bibinfo{year}{1992}), \bibinfo{pages}{17--26}.
\newblock


\bibitem[Blythe and Monk(2004)]%
        {blythe2004funology}
\bibfield{author}{\bibinfo{person}{Mark Blythe} {and} \bibinfo{person}{Andrew Monk}.} \bibinfo{year}{2004}\natexlab{}.
\newblock \showarticletitle{Funology 2}.
\newblock  (\bibinfo{year}{2004}).
\newblock


\bibitem[Bouncken(2004)]%
        {bouncken2004cultural}
\bibfield{author}{\bibinfo{person}{Ricarda~B Bouncken}.} \bibinfo{year}{2004}\natexlab{}.
\newblock \showarticletitle{Cultural diversity in entrepreneurial teams: Findings of new ventures in Germany}.
\newblock \bibinfo{journal}{\emph{Creativity and Innovation Management}} \bibinfo{volume}{13}, \bibinfo{number}{4} (\bibinfo{year}{2004}), \bibinfo{pages}{240--253}.
\newblock


\bibitem[Brabham(2012)]%
        {brabham2012crowdsourcing}
\bibfield{author}{\bibinfo{person}{Daren~C Brabham}.} \bibinfo{year}{2012}\natexlab{}.
\newblock \showarticletitle{Crowdsourcing: A model for leveraging online communities}.
\newblock In \bibinfo{booktitle}{\emph{The participatory cultures handbook}}. \bibinfo{publisher}{Routledge}, \bibinfo{pages}{120--129}.
\newblock


\bibitem[Braylan and Lease(2021)]%
        {braylan2021aggregating}
\bibfield{author}{\bibinfo{person}{Alexander Braylan} {and} \bibinfo{person}{Matthew Lease}.} \bibinfo{year}{2021}\natexlab{}.
\newblock \showarticletitle{Aggregating Complex Annotations via Merging and Matching}. In \bibinfo{booktitle}{\emph{Proceedings of the 27th ACM SIGKDD Conference on Knowledge Discovery \& Data Mining}}. \bibinfo{pages}{86--94}.
\newblock


\bibitem[Brejcha(2015)]%
        {brejcha2015cross}
\bibfield{author}{\bibinfo{person}{Jan Brejcha}.} \bibinfo{year}{2015}\natexlab{}.
\newblock \bibinfo{booktitle}{\emph{Cross-cultural human-computer interaction and user experience design: a semiotic perspective}}.
\newblock \bibinfo{publisher}{CRC Press}.
\newblock


\bibitem[Brumby et~al\mbox{.}(2013)]%
        {Brumby2013}
\bibfield{author}{\bibinfo{person}{Duncan~P. Brumby}, \bibinfo{person}{Anna~L. Cox}, \bibinfo{person}{Jonathan Back}, {and} \bibinfo{person}{Sandy~JJ Gould}.} \bibinfo{year}{2013}\natexlab{}.
\newblock \showarticletitle{Recovering from an interruption: Investigating speed accuracy trade offs in task resumption behavior}.
\newblock \bibinfo{journal}{\emph{Journal of Experimental Psychology: Applied}} \bibinfo{volume}{19}, \bibinfo{number}{2} (\bibinfo{year}{2013}), \bibinfo{pages}{95--107}.
\newblock


\bibitem[Buttcher et~al\mbox{.}(2016)]%
        {buttcher2016information}
\bibfield{author}{\bibinfo{person}{Stefan Buttcher}, \bibinfo{person}{Charles~LA Clarke}, {and} \bibinfo{person}{Gordon~V Cormack}.} \bibinfo{year}{2016}\natexlab{}.
\newblock \bibinfo{booktitle}{\emph{Information retrieval: Implementing and evaluating search engines}}.
\newblock \bibinfo{publisher}{Mit Press}.
\newblock


\bibitem[Calacci and Pentland(2022)]%
        {calacci2022bargaining}
\bibfield{author}{\bibinfo{person}{Dan Calacci} {and} \bibinfo{person}{Alex Pentland}.} \bibinfo{year}{2022}\natexlab{}.
\newblock \showarticletitle{Bargaining with the Black-Box: Designing and Deploying Worker-Centric Tools to Audit Algorithmic Management}.
\newblock \bibinfo{journal}{\emph{Proceedings of the ACM on Human-Computer Interaction}} \bibinfo{volume}{6}, \bibinfo{number}{CSCW2} (\bibinfo{year}{2022}), \bibinfo{pages}{1--24}.
\newblock


\bibitem[Carr and Kefalas(2009)]%
        {carr2009hollowing}
\bibfield{author}{\bibinfo{person}{Patrick~J Carr} {and} \bibinfo{person}{Maria~J Kefalas}.} \bibinfo{year}{2009}\natexlab{}.
\newblock \bibinfo{booktitle}{\emph{Hollowing out the middle: The rural brain drain and what it means for America}}.
\newblock \bibinfo{publisher}{Beacon press}.
\newblock


\bibitem[Carterette and Jones(2007)]%
        {carterette2007evaluating}
\bibfield{author}{\bibinfo{person}{Ben Carterette} {and} \bibinfo{person}{Rosie Jones}.} \bibinfo{year}{2007}\natexlab{}.
\newblock \showarticletitle{Evaluating search engines by modeling the relationship between relevance and clicks}.
\newblock \bibinfo{journal}{\emph{Advances in neural information processing systems}}  \bibinfo{volume}{20} (\bibinfo{year}{2007}).
\newblock


\bibitem[Casilli(2017)]%
        {casilli2017global}
\bibfield{author}{\bibinfo{person}{Antonio~A Casilli}.} \bibinfo{year}{2017}\natexlab{}.
\newblock \showarticletitle{Global digital culture| Digital labor studies go global: Toward a digital decolonial turn}.
\newblock \bibinfo{journal}{\emph{International Journal of Communication}}  \bibinfo{volume}{11} (\bibinfo{year}{2017}), \bibinfo{pages}{21}.
\newblock


\bibitem[Chambers(2014)]%
        {chambers2014rural}
\bibfield{author}{\bibinfo{person}{Robert Chambers}.} \bibinfo{year}{2014}\natexlab{}.
\newblock \bibinfo{booktitle}{\emph{Rural development: Putting the last first}}.
\newblock \bibinfo{publisher}{Routledge}.
\newblock


\bibitem[Chen(2020)]%
        {chen2020}
\bibfield{author}{\bibinfo{person}{Angela Chen}.} \bibinfo{year}{2020}\natexlab{}.
\newblock \bibinfo{title}{Desperate Venezuelans are making money by training AI for self-driving cars}.
\newblock
\newblock
\urldef\tempurl%
\url{https://www.technologyreview.com/2019/08/22/65375/venezuela-crisis-platform-work-trains-self-driving-car-ai-data/}
\showURL{%
\tempurl}


\bibitem[Chen(2021)]%
        {chen2021desperate}
\bibfield{author}{\bibinfo{person}{Angela Chen}.} \bibinfo{year}{2021}\natexlab{}.
\newblock \bibinfo{title}{Desperate Venezuelans are making money by training AI for self-driving cars}.
\newblock
\newblock


\bibitem[Cheon(2024)]%
        {Cheon2024}
\bibfield{author}{\bibinfo{person}{EunJeong Cheon}.} \bibinfo{year}{2024}\natexlab{}.
\newblock \showarticletitle{Examining Algorithmic Metrics and their Effects through the Lens of Reactivity}. In \bibinfo{booktitle}{\emph{Proceedings of the 2024 ACM Designing Interactive Systems Conference}}. \bibinfo{pages}{3179--3192}.
\newblock


\bibitem[Chiang et~al\mbox{.}(2018)]%
        {chiang2018crowd}
\bibfield{author}{\bibinfo{person}{Chun-Wei Chiang}, \bibinfo{person}{Anna Kasunic}, {and} \bibinfo{person}{Saiph Savage}.} \bibinfo{year}{2018}\natexlab{}.
\newblock \showarticletitle{Crowd coach: Peer coaching for crowd workers' skill growth}.
\newblock \bibinfo{journal}{\emph{Proceedings of the ACM on Human-Computer Interaction}} \bibinfo{volume}{2}, \bibinfo{number}{CSCW} (\bibinfo{year}{2018}), \bibinfo{pages}{1--17}.
\newblock


\bibitem[Christakis and Fowler(2009)]%
        {christakis2009connected}
\bibfield{author}{\bibinfo{person}{Nicholas~A Christakis} {and} \bibinfo{person}{James~H Fowler}.} \bibinfo{year}{2009}\natexlab{}.
\newblock \bibinfo{booktitle}{\emph{Connected: The surprising power of our social networks and how they shape our lives}}.
\newblock \bibinfo{publisher}{Little, Brown Spark}.
\newblock


\bibitem[Christensen et~al\mbox{.}(2011)]%
        {christensen2011research}
\bibfield{author}{\bibinfo{person}{Larry~B Christensen}, \bibinfo{person}{Burke Johnson}, \bibinfo{person}{Lisa~Anne Turner}, {and} \bibinfo{person}{Larry~B Christensen}.} \bibinfo{year}{2011}\natexlab{}.
\newblock \bibinfo{booktitle}{\emph{Research methods, design, and analysis}}.
\newblock \bibinfo{publisher}{Pearson Boston}.
\newblock


\bibitem[Ciborra(1997)]%
        {ciborra1997groupware}
\bibfield{author}{\bibinfo{person}{Claudio~U Ciborra}.} \bibinfo{year}{1997}\natexlab{}.
\newblock \bibinfo{booktitle}{\emph{Groupware and teamwork: Invisible aid or technical hindrance?}}
\newblock \bibinfo{publisher}{John Wiley \& Sons, Inc.}
\newblock


\bibitem[Collective(2023)]%
        {ShromaMonitoring2023}
\bibfield{author}{\bibinfo{person}{Shroma~Worker Collective}.} \bibinfo{year}{2023}\natexlab{}.
\newblock \bibinfo{title}{Monitoring - Shroma}.
\newblock \bibinfo{howpublished}{\url{https://shroma.ge/monitor/?lang=en}}.
\newblock
\newblock
\shownote{Accessed: 2023-12-10}.


\bibitem[Collier(2008)]%
        {collier2008bottom}
\bibfield{author}{\bibinfo{person}{Paul Collier}.} \bibinfo{year}{2008}\natexlab{}.
\newblock \bibinfo{booktitle}{\emph{The bottom billion: Why the poorest countries are failing and what can be done about it}}.
\newblock \bibinfo{publisher}{Oxford University Press, USA}.
\newblock


\bibitem[Conole(2012)]%
        {conole2012designing}
\bibfield{author}{\bibinfo{person}{Gr{\'a}inne Conole}.} \bibinfo{year}{2012}\natexlab{}.
\newblock \bibinfo{booktitle}{\emph{Designing for learning in an open world}}. Vol.~\bibinfo{volume}{4}.
\newblock \bibinfo{publisher}{Springer Science \& Business Media}.
\newblock


\bibitem[Cotte and Ratneshwar(1999)]%
        {cotte1999juggling}
\bibfield{author}{\bibinfo{person}{June Cotte} {and} \bibinfo{person}{Srinivasan Ratneshwar}.} \bibinfo{year}{1999}\natexlab{}.
\newblock \showarticletitle{Juggling and hopping: what does it mean to work polychronically?}
\newblock \bibinfo{journal}{\emph{Journal of Managerial Psychology}} (\bibinfo{year}{1999}).
\newblock


\bibitem[Creators(2023)]%
        {ShiptTransparencyCalculator}
\bibfield{author}{\bibinfo{person}{Shipt Creators}.} \bibinfo{year}{2023}\natexlab{}.
\newblock \bibinfo{title}{Shipt Transparency Calculator}.
\newblock \bibinfo{howpublished}{\url{https://home.coworker.org/shiptcalc/}}.
\newblock
\newblock
\shownote{Accessed: 2023-12-10}.


\bibitem[De~Los~Santos et~al\mbox{.}(2023)]%
        {de2023independiente}
\bibfield{author}{\bibinfo{person}{Maya De~Los~Santos}, \bibinfo{person}{Norma~Elva Ch{\'a}vez}, \bibinfo{person}{Alberto Navarrete}, \bibinfo{person}{Cristina Mart{\'\i}nez~Pinto}, \bibinfo{person}{Luz~Elena Gonz{\'a}lez}, \bibinfo{person}{Tatiana Telles-Calderon}, {and} \bibinfo{person}{Saiph Savage}.} \bibinfo{year}{2023}\natexlab{}.
\newblock \showarticletitle{La Independiente: Designing Ubiquitous Systems for Latin American and Caribbean Women Crowdworkers}. In \bibinfo{booktitle}{\emph{Adjunct Proceedings of the 2023 ACM International Joint Conference on Pervasive and Ubiquitous Computing \& the 2023 ACM International Symposium on Wearable Computing}}. \bibinfo{pages}{404--406}.
\newblock


\bibitem[De~los Santos et~al\mbox{.}(2023)]%
        {DelosSantos2023}
\bibfield{author}{\bibinfo{person}{Maya De~los Santos}, \bibinfo{person}{José~Emilio García}, \bibinfo{person}{Norma~Elva Chávez}, \bibinfo{person}{Alberto Navarrete}, \bibinfo{person}{Cristina Martínez}, \bibinfo{person}{Luz~Elena González}, \bibinfo{person}{Tatiana Telles}, \bibinfo{person}{Paola Ricaurte}, {and} \bibinfo{person}{Saiph Savage}.} \bibinfo{year}{2023}\natexlab{}.
\newblock \showarticletitle{La Independiente: an AI-enhanced Platform Co-Designed with Latin-American Crowd-Workers}.
\newblock \bibinfo{journal}{\emph{Avances en Interacción Humano-Computadora}} \bibinfo{volume}{8}, \bibinfo{number}{1} (\bibinfo{year}{2023}), \bibinfo{pages}{6--10}.
\newblock


\bibitem[De~Stefano(2015)]%
        {de2015rise}
\bibfield{author}{\bibinfo{person}{Valerio De~Stefano}.} \bibinfo{year}{2015}\natexlab{}.
\newblock \showarticletitle{The rise of the just-in-time workforce: On-demand work, crowdwork, and labor protection in the gig-economy}.
\newblock \bibinfo{journal}{\emph{Comp. Lab. L. \& Pol'y J.}}  \bibinfo{volume}{37} (\bibinfo{year}{2015}), \bibinfo{pages}{471}.
\newblock


\bibitem[Difallah et~al\mbox{.}(2018)]%
        {difallah2018demographics}
\bibfield{author}{\bibinfo{person}{Djellel Difallah}, \bibinfo{person}{Elena Filatova}, {and} \bibinfo{person}{Panos Ipeirotis}.} \bibinfo{year}{2018}\natexlab{}.
\newblock \showarticletitle{Demographics and dynamics of mechanical turk workers}. In \bibinfo{booktitle}{\emph{Proceedings of the eleventh ACM international conference on web search and data mining}}. \bibinfo{pages}{135--143}.
\newblock


\bibitem[Do et~al\mbox{.}(2024)]%
        {Do2024}
\bibfield{author}{\bibinfo{person}{Kimberly Do}, \bibinfo{person}{Maya De~Los~Santos}, \bibinfo{person}{Michael Muller}, {and} \bibinfo{person}{Saiph Savage}.} \bibinfo{year}{2024}\natexlab{}.
\newblock \showarticletitle{GigSousveillance: Designing Gig Worker Centric Sousveillance Tools}. In \bibinfo{booktitle}{\emph{Proceedings of the ACM CHI Conference on Human Factors in Computing Systems}}.
\newblock


\bibitem[Duncan(2015)]%
        {duncan2015worlds}
\bibfield{author}{\bibinfo{person}{Cynthia~M Duncan}.} \bibinfo{year}{2015}\natexlab{}.
\newblock \bibinfo{booktitle}{\emph{Worlds apart: Poverty and politics in rural America}}.
\newblock \bibinfo{publisher}{Yale University Press}.
\newblock


\bibitem[Edixhoven et~al\mbox{.}(2021)]%
        {edixhoven2021improving}
\bibfield{author}{\bibinfo{person}{Tom Edixhoven}, \bibinfo{person}{Sihang Qiu}, \bibinfo{person}{Lucie Kuiper}, \bibinfo{person}{Olivier Dikken}, \bibinfo{person}{Gwennan Smitskamp}, {and} \bibinfo{person}{Ujwal Gadiraju}.} \bibinfo{year}{2021}\natexlab{}.
\newblock \showarticletitle{Improving Reactions to Rejection in Crowdsourcing Through Self-Reflection}. In \bibinfo{booktitle}{\emph{13th ACM Web Science Conference 2021}}. \bibinfo{pages}{74--83}.
\newblock


\bibitem[Erickson and Kellogg(2000)]%
        {erickson2000social}
\bibfield{author}{\bibinfo{person}{Thomas Erickson} {and} \bibinfo{person}{Wendy~A Kellogg}.} \bibinfo{year}{2000}\natexlab{}.
\newblock \showarticletitle{Social translucence: an approach to designing systems that support social processes}.
\newblock \bibinfo{journal}{\emph{ACM transactions on computer-human interaction (TOCHI)}} \bibinfo{volume}{7}, \bibinfo{number}{1} (\bibinfo{year}{2000}), \bibinfo{pages}{59--83}.
\newblock


\bibitem[Fiorentino and Tomkowicz(2019)]%
        {fiorentino2019petitions}
\bibfield{author}{\bibinfo{person}{Susan~R Fiorentino} {and} \bibinfo{person}{Sandra~M Tomkowicz}.} \bibinfo{year}{2019}\natexlab{}.
\newblock \showarticletitle{E-Petitions and Protected Concerted Activity: The Millennial Response to Organized Labor?}
\newblock \bibinfo{journal}{\emph{ABAJ Lab. \& Emp. L.}}  \bibinfo{volume}{34} (\bibinfo{year}{2019}), \bibinfo{pages}{71}.
\newblock


\bibitem[Flores-Saviaga et~al\mbox{.}(2020)]%
        {flores2020challenges}
\bibfield{author}{\bibinfo{person}{Claudia Flores-Saviaga}, \bibinfo{person}{Yuwen Li}, \bibinfo{person}{Benjamin Hanrahan}, \bibinfo{person}{Jeffrey Bigham}, {and} \bibinfo{person}{Saiph Savage}.} \bibinfo{year}{2020}\natexlab{}.
\newblock \showarticletitle{The Challenges of Crowd Workers in Rural and Urban America}.
\newblock \bibinfo{journal}{\emph{Proceedings of the AAAI Conference on Human Computation and Crowdsourcing}} \bibinfo{volume}{8}, \bibinfo{number}{1} (\bibinfo{year}{2020}), \bibinfo{pages}{159--162}.
\newblock


\bibitem[Fox et~al\mbox{.}(2020)]%
        {fox2020worker}
\bibfield{author}{\bibinfo{person}{Sarah~E Fox}, \bibinfo{person}{Vera Khovanskaya}, \bibinfo{person}{Clara Crivellaro}, \bibinfo{person}{Niloufar Salehi}, \bibinfo{person}{Lynn Dombrowski}, \bibinfo{person}{Chinmay Kulkarni}, \bibinfo{person}{Lilly Irani}, {and} \bibinfo{person}{Jodi Forlizzi}.} \bibinfo{year}{2020}\natexlab{}.
\newblock \showarticletitle{Worker-centered design: Expanding HCI methods for supporting labor}. In \bibinfo{booktitle}{\emph{Extended abstracts of the 2020 CHI conference on human factors in computing systems}}. \bibinfo{pages}{1--8}.
\newblock


\bibitem[Friedman(2014)]%
        {friedman2014best}
\bibfield{author}{\bibinfo{person}{Ron Friedman}.} \bibinfo{year}{2014}\natexlab{}.
\newblock \bibinfo{booktitle}{\emph{The best place to work: The art and science of creating an extraordinary workplace}}.
\newblock \bibinfo{publisher}{TarcherPerigee}.
\newblock


\bibitem[Fulmer et~al\mbox{.}(2014)]%
        {inbook}
\bibfield{author}{\bibinfo{person}{Ashley Fulmer}, \bibinfo{person}{Brandon Crosby}, {and} \bibinfo{person}{Michele Gelfand}.} \bibinfo{year}{2014}\natexlab{}.
\newblock \bibinfo{booktitle}{\emph{Cross-cultural Perspectives on Time}}.
\newblock
\urldef\tempurl%
\url{https://doi.org/10.4324/9781315798370}
\showDOI{\tempurl}


\bibitem[Gaikwad et~al\mbox{.}(2015)]%
        {gaikwad2015daemo}
\bibfield{author}{\bibinfo{person}{Snehal Gaikwad}, \bibinfo{person}{Durim Morina}, \bibinfo{person}{Rohit Nistala}, \bibinfo{person}{Megha Agarwal}, \bibinfo{person}{Alison Cossette}, \bibinfo{person}{Radhika Bhanu}, \bibinfo{person}{Saiph Savage}, \bibinfo{person}{Vishwajeet Narwal}, \bibinfo{person}{Karan Rajpal}, \bibinfo{person}{Jeff Regino}, {et~al\mbox{.}}} \bibinfo{year}{2015}\natexlab{}.
\newblock \showarticletitle{Daemo: A self-governed crowdsourcing marketplace}. In \bibinfo{booktitle}{\emph{Adjunct proceedings of the 28th annual ACM symposium on user interface software \& technology}}. \bibinfo{pages}{101--102}.
\newblock


\bibitem[Gaikwad et~al\mbox{.}(2016)]%
        {gaikwad2016boomerang}
\bibfield{author}{\bibinfo{person}{Snehalkumar (Neil)~S Gaikwad}, \bibinfo{person}{Durim Morina}, \bibinfo{person}{Adam Ginzberg}, \bibinfo{person}{Catherine Mullings}, \bibinfo{person}{Shirish Goyal}, \bibinfo{person}{Dilrukshi Gamage}, \bibinfo{person}{Christopher Diemert}, \bibinfo{person}{Mathias Burton}, \bibinfo{person}{Sharon Zhou}, \bibinfo{person}{Mark Whiting}, {et~al\mbox{.}}} \bibinfo{year}{2016}\natexlab{}.
\newblock \showarticletitle{Boomerang: Rebounding the consequences of reputation feedback on crowdsourcing platforms}. In \bibinfo{booktitle}{\emph{Proceedings of the 29th Annual Symposium on User Interface Software and Technology}}. \bibinfo{pages}{625--637}.
\newblock


\bibitem[Gaur et~al\mbox{.}(2016)]%
        {gaur2016effects}
\bibfield{author}{\bibinfo{person}{Yashesh Gaur}, \bibinfo{person}{Walter~S Lasecki}, \bibinfo{person}{Florian Metze}, {and} \bibinfo{person}{Jeffrey~P Bigham}.} \bibinfo{year}{2016}\natexlab{}.
\newblock \showarticletitle{The effects of automatic speech recognition quality on human transcription latency}. In \bibinfo{booktitle}{\emph{Proceedings of the 13th International Web for All Conference}}. \bibinfo{pages}{1--8}.
\newblock


\bibitem[Gemo(2024)]%
        {WageTheftIntro2023}
\bibfield{author}{\bibinfo{person}{Shromis Gemo}.} \bibinfo{year}{2024}\natexlab{}.
\newblock \bibinfo{title}{Introduction to Wage Theft}.
\newblock \bibinfo{howpublished}{\url{https://shroma.ge/en/wagetheft-intro-en/}}.
\newblock
\newblock
\shownote{Accessed: 2023-12-10}.


\bibitem[Gong(2009)]%
        {gong2009national}
\bibfield{author}{\bibinfo{person}{Wen Gong}.} \bibinfo{year}{2009}\natexlab{}.
\newblock \showarticletitle{National culture and global diffusion of business-to-consumer e-commerce}.
\newblock \bibinfo{journal}{\emph{Cross cultural management: an international journal}} (\bibinfo{year}{2009}).
\newblock


\bibitem[Gordon et~al\mbox{.}(2022)]%
        {gordon2022jury}
\bibfield{author}{\bibinfo{person}{Mitchell~L Gordon}, \bibinfo{person}{Michelle~S Lam}, \bibinfo{person}{Joon~Sung Park}, \bibinfo{person}{Kayur Patel}, \bibinfo{person}{Jeff Hancock}, \bibinfo{person}{Tatsunori Hashimoto}, {and} \bibinfo{person}{Michael~S Bernstein}.} \bibinfo{year}{2022}\natexlab{}.
\newblock \showarticletitle{Jury learning: Integrating dissenting voices into machine learning models}. In \bibinfo{booktitle}{\emph{Proceedings of the 2022 CHI Conference on Human Factors in Computing Systems}}. \bibinfo{pages}{1--19}.
\newblock


\bibitem[Gray and Suri(2019)]%
        {gray2019ghost}
\bibfield{author}{\bibinfo{person}{Mary~L Gray} {and} \bibinfo{person}{Siddharth Suri}.} \bibinfo{year}{2019}\natexlab{}.
\newblock \bibinfo{booktitle}{\emph{Ghost work: How to stop Silicon Valley from building a new global underclass}}.
\newblock \bibinfo{publisher}{Eamon Dolan Books}.
\newblock


\bibitem[Gregory et~al\mbox{.}(2016)]%
        {gregory2016learning}
\bibfield{author}{\bibinfo{person}{Sue Gregory}, \bibinfo{person}{Mark~JW Lee}, \bibinfo{person}{Barney Dalgarno}, {and} \bibinfo{person}{Belinda Tynan}.} \bibinfo{year}{2016}\natexlab{}.
\newblock \bibinfo{booktitle}{\emph{Learning in virtual worlds: Research and applications}}.
\newblock \bibinfo{publisher}{Athabasca University Press}.
\newblock


\bibitem[Gupta et~al\mbox{.}(2014)]%
        {gupta2014turk}
\bibfield{author}{\bibinfo{person}{Neha Gupta}, \bibinfo{person}{David Martin}, \bibinfo{person}{Benjamin~V Hanrahan}, {and} \bibinfo{person}{Jacki O'Neill}.} \bibinfo{year}{2014}\natexlab{}.
\newblock \showarticletitle{Turk-life in India}. In \bibinfo{booktitle}{\emph{Proceedings of the 18th International Conference on Supporting Group Work}}. \bibinfo{pages}{1--11}.
\newblock


\bibitem[Hall(1971)]%
        {hall1971paradox}
\bibfield{author}{\bibinfo{person}{Edward Hall}.} \bibinfo{year}{1971}\natexlab{}.
\newblock \showarticletitle{The paradox of culture}.
\newblock \bibinfo{journal}{\emph{B. Landis and ES Tauber (Eds.): In the Name of Life. Essays in Honor of Erich Fromm, New York (Holt, Rinehart and Winston) 1970, pp. 218-235.}} (\bibinfo{year}{1971}).
\newblock


\bibitem[Hall(1989)]%
        {hall1989beyond}
\bibfield{author}{\bibinfo{person}{Edward~Twitchell Hall}.} \bibinfo{year}{1989}\natexlab{}.
\newblock \bibinfo{booktitle}{\emph{Beyond culture}}.
\newblock \bibinfo{publisher}{Anchor}.
\newblock


\bibitem[Hall and Hall(1989)]%
        {hall1989dance}
\bibfield{author}{\bibinfo{person}{Edward~T Hall} {and} \bibinfo{person}{Edward~Twitchell Hall}.} \bibinfo{year}{1989}\natexlab{}.
\newblock \bibinfo{booktitle}{\emph{The dance of life: The other dimension of time}}.
\newblock \bibinfo{publisher}{Anchor}.
\newblock


\bibitem[Hall and Hall(1990)]%
        {hall1990understanding}
\bibfield{author}{\bibinfo{person}{Edward~Twitchell Hall} {and} \bibinfo{person}{Mildred~Reed Hall}.} \bibinfo{year}{1990}\natexlab{}.
\newblock \bibinfo{booktitle}{\emph{Understanding cultural differences:[keys to success in West Germany, France, and the United States]}}.
\newblock \bibinfo{publisher}{Intercultural Press}.
\newblock


\bibitem[Hall et~al\mbox{.}(1989)]%
        {hall1989understanding}
\bibfield{author}{\bibinfo{person}{Edward~Twitchell Hall}, \bibinfo{person}{Mildred~Reed Hall}, {et~al\mbox{.}}} \bibinfo{year}{1989}\natexlab{}.
\newblock \bibinfo{booktitle}{\emph{Understanding cultural differences}}.
\newblock \bibinfo{publisher}{Intercultural press}.
\newblock


\bibitem[Hall and Hall(1959)]%
        {hall1959silent}
\bibfield{author}{\bibinfo{person}{Edward~Twitchell Hall} {and} \bibinfo{person}{T Hall}.} \bibinfo{year}{1959}\natexlab{}.
\newblock \bibinfo{booktitle}{\emph{The silent language}}. Vol.~\bibinfo{volume}{948}.
\newblock \bibinfo{publisher}{Anchor books}.
\newblock


\bibitem[Hanrahan et~al\mbox{.}(2020)]%
        {hanrahan2020reciprocal}
\bibfield{author}{\bibinfo{person}{Benjamin~V Hanrahan}, \bibinfo{person}{Ning~F Ma}, \bibinfo{person}{Eber Betanzos}, {and} \bibinfo{person}{Saiph Savage}.} \bibinfo{year}{2020}\natexlab{}.
\newblock \showarticletitle{Reciprocal research: Providing value in design research from the outset in the rural united states}. In \bibinfo{booktitle}{\emph{Proceedings of the 2020 International Conference on Information and Communication Technologies and Development}}. \bibinfo{pages}{1--5}.
\newblock


\bibitem[Hanrahan et~al\mbox{.}(2015)]%
        {Hanrahan2015}
\bibfield{author}{\bibinfo{person}{B.~V. Hanrahan}, \bibinfo{person}{J.~K. Willamowski}, \bibinfo{person}{S. Swaminathan}, {and} \bibinfo{person}{D.~B. Martin}.} \bibinfo{year}{2015}\natexlab{}.
\newblock \showarticletitle{TurkBench: Rendering the market for Turkers}. In \bibinfo{booktitle}{\emph{Proceedings of the 33rd Annual ACM Conference on Human Factors in Computing Systems}}. \bibinfo{pages}{1613--1616}.
\newblock


\bibitem[Hara et~al\mbox{.}(2018)]%
        {hara2018data}
\bibfield{author}{\bibinfo{person}{Kotaro Hara}, \bibinfo{person}{Abigail Adams}, \bibinfo{person}{Kristy Milland}, \bibinfo{person}{Saiph Savage}, \bibinfo{person}{Chris Callison-Burch}, {and} \bibinfo{person}{Jeffrey~P Bigham}.} \bibinfo{year}{2018}\natexlab{}.
\newblock \showarticletitle{A data-driven analysis of workers' earnings on Amazon Mechanical Turk}. In \bibinfo{booktitle}{\emph{Proceedings of the 2018 CHI Conference on Human Factors in Computing Systems}}. \bibinfo{pages}{1--14}.
\newblock


\bibitem[Hara et~al\mbox{.}(2019)]%
        {hara2019worker}
\bibfield{author}{\bibinfo{person}{Kotaro Hara}, \bibinfo{person}{Abigail Adams}, \bibinfo{person}{Kristy Milland}, \bibinfo{person}{Saiph Savage}, \bibinfo{person}{Benjamin~V Hanrahan}, \bibinfo{person}{Jeffrey~P Bigham}, {and} \bibinfo{person}{Chris Callison-Burch}.} \bibinfo{year}{2019}\natexlab{}.
\newblock \showarticletitle{Worker demographics and earnings on amazon mechanical turk: An exploratory analysis}. In \bibinfo{booktitle}{\emph{Extended Abstracts of the 2019 CHI Conference on Human Factors in Computing Systems}}. \bibinfo{pages}{1--6}.
\newblock


\bibitem[Harboe and Huang(2015)]%
        {harboe2015real}
\bibfield{author}{\bibinfo{person}{Gunnar Harboe} {and} \bibinfo{person}{Elaine~M Huang}.} \bibinfo{year}{2015}\natexlab{}.
\newblock \showarticletitle{Real-world affinity diagramming practices: Bridging the paper-digital gap}. In \bibinfo{booktitle}{\emph{Proceedings of the 33rd annual ACM conference on human factors in computing systems}}. \bibinfo{pages}{95--104}.
\newblock


\bibitem[Hardy(2019)]%
        {hardy2019design}
\bibfield{author}{\bibinfo{person}{Jean Hardy}.} \bibinfo{year}{2019}\natexlab{}.
\newblock \showarticletitle{How the design of social technology fails rural America}. In \bibinfo{booktitle}{\emph{Companion Publication of the 2019 on Designing Interactive Systems Conference 2019 Companion}}. \bibinfo{pages}{189--193}.
\newblock


\bibitem[Hardy et~al\mbox{.}(2018)]%
        {hardy2018rural}
\bibfield{author}{\bibinfo{person}{Jean Hardy}, \bibinfo{person}{Dharma Dailey}, \bibinfo{person}{Susan Wyche}, {and} \bibinfo{person}{Norman~Makoto Su}.} \bibinfo{year}{2018}\natexlab{}.
\newblock \showarticletitle{Rural computing: Beyond access and infrastructure}. In \bibinfo{booktitle}{\emph{Companion of the 2018 ACM Conference on Computer Supported Cooperative Work and Social Computing}}. \bibinfo{pages}{463--470}.
\newblock


\bibitem[Hardy et~al\mbox{.}(2019)]%
        {hardy2019designing}
\bibfield{author}{\bibinfo{person}{Jean Hardy}, \bibinfo{person}{Chanda Phelan}, \bibinfo{person}{Morgan Vigil-Hayes}, \bibinfo{person}{Norman~Makoto Su}, \bibinfo{person}{Susan Wyche}, {and} \bibinfo{person}{Phoebe Sengers}.} \bibinfo{year}{2019}\natexlab{}.
\newblock \showarticletitle{Designing from the rural}.
\newblock \bibinfo{journal}{\emph{Interactions}} \bibinfo{volume}{26}, \bibinfo{number}{4} (\bibinfo{year}{2019}), \bibinfo{pages}{37--41}.
\newblock


\bibitem[Hettiachchi et~al\mbox{.}(2021)]%
        {hettiachchi2021challenge}
\bibfield{author}{\bibinfo{person}{Danula Hettiachchi}, \bibinfo{person}{Mike Schaekermann}, \bibinfo{person}{Tristan~J McKinney}, {and} \bibinfo{person}{Matthew Lease}.} \bibinfo{year}{2021}\natexlab{}.
\newblock \showarticletitle{The Challenge of Variable Effort Crowdsourcing and How Visible Gold Can Help}.
\newblock \bibinfo{journal}{\emph{Proceedings of the ACM on Human-Computer Interaction}} \bibinfo{volume}{5}, \bibinfo{number}{CSCW2} (\bibinfo{year}{2021}), \bibinfo{pages}{1--26}.
\newblock


\bibitem[Heylighen(2014)]%
        {heylighen2014nature}
\bibfield{author}{\bibinfo{person}{Ann Heylighen}.} \bibinfo{year}{2014}\natexlab{}.
\newblock \showarticletitle{About the nature of design in universal design}.
\newblock \bibinfo{journal}{\emph{Disability and rehabilitation}} \bibinfo{volume}{36}, \bibinfo{number}{16} (\bibinfo{year}{2014}), \bibinfo{pages}{1360--1368}.
\newblock


\bibitem[Hill et~al\mbox{.}(2004)]%
        {hill2004cross}
\bibfield{author}{\bibinfo{person}{E~Jeffrey Hill}, \bibinfo{person}{Chongming Yang}, \bibinfo{person}{Alan~J Hawkins}, {and} \bibinfo{person}{Maria Ferris}.} \bibinfo{year}{2004}\natexlab{}.
\newblock \showarticletitle{A cross-cultural test of the work-family interface in 48 countries}.
\newblock \bibinfo{journal}{\emph{Journal of Marriage and the Family}} (\bibinfo{year}{2004}), \bibinfo{pages}{1300--1316}.
\newblock


\bibitem[Hmelo-Silver(2013)]%
        {hmelo2013international}
\bibfield{author}{\bibinfo{person}{Cindy~E Hmelo-Silver}.} \bibinfo{year}{2013}\natexlab{}.
\newblock \showarticletitle{The international handbook of collaborative learning}.
\newblock  (\bibinfo{year}{2013}).
\newblock


\bibitem[Hofstede(1984)]%
        {hofstede1984cultural}
\bibfield{author}{\bibinfo{person}{Geert Hofstede}.} \bibinfo{year}{1984}\natexlab{}.
\newblock \showarticletitle{Cultural dimensions in management and planning}.
\newblock \bibinfo{journal}{\emph{Asia Pacific journal of management}}  \bibinfo{volume}{1} (\bibinfo{year}{1984}), \bibinfo{pages}{81--99}.
\newblock


\bibitem[Hofstede(2001)]%
        {hofstede2001culture}
\bibfield{author}{\bibinfo{person}{Geert Hofstede}.} \bibinfo{year}{2001}\natexlab{}.
\newblock \bibinfo{booktitle}{\emph{Culture's consequences: Comparing values, behaviors, institutions and organizations across nations}}.
\newblock \bibinfo{publisher}{Sage publications}.
\newblock


\bibitem[Hofstede et~al\mbox{.}(2005)]%
        {hofstede2005cultures}
\bibfield{author}{\bibinfo{person}{Geert Hofstede}, \bibinfo{person}{Gert~Jan Hofstede}, {and} \bibinfo{person}{Michael Minkov}.} \bibinfo{year}{2005}\natexlab{}.
\newblock \bibinfo{booktitle}{\emph{Cultures and organizations: Software of the mind}}. Vol.~\bibinfo{volume}{2}.
\newblock \bibinfo{publisher}{Mcgraw-hill New York}.
\newblock


\bibitem[Hooker(2003)]%
        {hooker2003working}
\bibfield{author}{\bibinfo{person}{John Hooker}.} \bibinfo{year}{2003}\natexlab{}.
\newblock \bibinfo{booktitle}{\emph{Working across cultures}}.
\newblock \bibinfo{publisher}{Stanford University Press}.
\newblock


\bibitem[Huang et~al\mbox{.}(2010)]%
        {huang2010effects}
\bibfield{author}{\bibinfo{person}{Ding-Long Huang}, \bibinfo{person}{Pei-Luen~Patrick Rau}, \bibinfo{person}{Hui Su}, \bibinfo{person}{Nan Tu}, {and} \bibinfo{person}{Chen Zhao}.} \bibinfo{year}{2010}\natexlab{}.
\newblock \showarticletitle{Effects of communication style and time orientation on notification systems and anti-virus software}.
\newblock \bibinfo{journal}{\emph{Behaviour \& Information Technology}} \bibinfo{volume}{29}, \bibinfo{number}{5} (\bibinfo{year}{2010}), \bibinfo{pages}{483--495}.
\newblock


\bibitem[Huang et~al\mbox{.}(2016)]%
        {huang2016there}
\bibfield{author}{\bibinfo{person}{Ting-Hao~Kenneth Huang}, \bibinfo{person}{Walter~S Lasecki}, \bibinfo{person}{Amos Azaria}, {and} \bibinfo{person}{Jeffrey~P Bigham}.} \bibinfo{year}{2016}\natexlab{}.
\newblock \showarticletitle{" Is There Anything Else I Can Help You With?" Challenges in Deploying an On-Demand Crowd-Powered Conversational Agent}. In \bibinfo{booktitle}{\emph{Fourth AAAI Conference on Human Computation and Crowdsourcing}}.
\newblock


\bibitem[Irani(2015)]%
        {irani2015cultural}
\bibfield{author}{\bibinfo{person}{Lilly Irani}.} \bibinfo{year}{2015}\natexlab{}.
\newblock \showarticletitle{The cultural work of microwork}.
\newblock \bibinfo{journal}{\emph{New media \& society}} \bibinfo{volume}{17}, \bibinfo{number}{5} (\bibinfo{year}{2015}), \bibinfo{pages}{720--739}.
\newblock


\bibitem[Irani et~al\mbox{.}(2010)]%
        {irani2010postcolonial}
\bibfield{author}{\bibinfo{person}{Lilly Irani}, \bibinfo{person}{Janet Vertesi}, \bibinfo{person}{Paul Dourish}, \bibinfo{person}{Kavita Philip}, {and} \bibinfo{person}{Rebecca~E Grinter}.} \bibinfo{year}{2010}\natexlab{}.
\newblock \showarticletitle{Postcolonial computing: a lens on design and development}. In \bibinfo{booktitle}{\emph{Proceedings of the SIGCHI conference on human factors in computing systems}}. \bibinfo{pages}{1311--1320}.
\newblock


\bibitem[Irani and Silberman(2013)]%
        {irani2013turkopticon}
\bibfield{author}{\bibinfo{person}{Lilly~C Irani} {and} \bibinfo{person}{M~Six Silberman}.} \bibinfo{year}{2013}\natexlab{}.
\newblock \showarticletitle{Turkopticon: Interrupting worker invisibility in amazon mechanical turk}. In \bibinfo{booktitle}{\emph{Proceedings of the SIGCHI conference on human factors in computing systems}}. \bibinfo{pages}{611--620}.
\newblock


\bibitem[Janssen et~al\mbox{.}(2019)]%
        {Janssen2019}
\bibfield{author}{\bibinfo{person}{Christian~P. Janssen}, \bibinfo{person}{Shamsi~T. Iqbal}, \bibinfo{person}{Andrew~L. Kun}, {and} \bibinfo{person}{Stella~F. Donker}.} \bibinfo{year}{2019}\natexlab{}.
\newblock \showarticletitle{Interrupted by my car? Implications of interruption and interleaving research for automated vehicles}.
\newblock \bibinfo{journal}{\emph{International Journal of Human-Computer Studies}}  \bibinfo{volume}{130} (\bibinfo{year}{2019}), \bibinfo{pages}{221--233}.
\newblock


\bibitem[Jones(1983)]%
        {jones1983transaction}
\bibfield{author}{\bibinfo{person}{Gareth~R Jones}.} \bibinfo{year}{1983}\natexlab{}.
\newblock \showarticletitle{Transaction costs, property rights, and organizational culture: An exchange perspective}.
\newblock \bibinfo{journal}{\emph{Administrative Science Quarterly}} (\bibinfo{year}{1983}), \bibinfo{pages}{454--467}.
\newblock


\bibitem[Jung and Lease(2012)]%
        {jung2012improving}
\bibfield{author}{\bibinfo{person}{Hyun~Joon Jung} {and} \bibinfo{person}{Matthew Lease}.} \bibinfo{year}{2012}\natexlab{}.
\newblock \showarticletitle{Improving quality of crowdsourced labels via probabilistic matrix factorization}. In \bibinfo{booktitle}{\emph{Workshops at the Twenty-Sixth AAAI Conference on Artificial Intelligence}}.
\newblock


\bibitem[Kapania et~al\mbox{.}(2023)]%
        {kapania2023hunt}
\bibfield{author}{\bibinfo{person}{Shivani Kapania}, \bibinfo{person}{Alex~S Taylor}, {and} \bibinfo{person}{Ding Wang}.} \bibinfo{year}{2023}\natexlab{}.
\newblock \showarticletitle{A hunt for the Snark: Annotator Diversity in Data Practices}. In \bibinfo{booktitle}{\emph{Proceedings of the 2023 CHI Conference on Human Factors in Computing Systems}}. \bibinfo{pages}{1--15}.
\newblock


\bibitem[Kaplan et~al\mbox{.}(2018)]%
        {kaplan2018striving}
\bibfield{author}{\bibinfo{person}{Toni Kaplan}, \bibinfo{person}{Susumu Saito}, \bibinfo{person}{Kotaro Hara}, {and} \bibinfo{person}{Jeffrey~P Bigham}.} \bibinfo{year}{2018}\natexlab{}.
\newblock \showarticletitle{Striving to earn more: a survey of work strategies and tool use among crowd workers}. In \bibinfo{booktitle}{\emph{Sixth AAAI Conference on Human Computation and Crowdsourcing}}.
\newblock


\bibitem[Kasunic et~al\mbox{.}(2019a)]%
        {kasunic2019crowd}
\bibfield{author}{\bibinfo{person}{Anna Kasunic}, \bibinfo{person}{Chun-Wei Chiang}, \bibinfo{person}{Geoff Kaufman}, {and} \bibinfo{person}{Saiph Savage}.} \bibinfo{year}{2019}\natexlab{a}.
\newblock \showarticletitle{Crowd Work on a CV? Understanding How AMT Fits into Turkers' Career Goals and Professional Profiles}.
\newblock \bibinfo{journal}{\emph{arXiv preprint arXiv:1902.05361}} (\bibinfo{year}{2019}).
\newblock


\bibitem[Kasunic et~al\mbox{.}(2019b)]%
        {kasunic2019turker}
\bibfield{author}{\bibinfo{person}{Anna Kasunic}, \bibinfo{person}{Chun-Wei Chiang}, \bibinfo{person}{Geoff Kaufman}, {and} \bibinfo{person}{Saiph Savage}.} \bibinfo{year}{2019}\natexlab{b}.
\newblock \showarticletitle{Turker Tales: Integrating Tangential Play into Crowd Work}. In \bibinfo{booktitle}{\emph{Proceedings of the 2019 on Designing Interactive Systems Conference}}. \bibinfo{pages}{21--34}.
\newblock


\bibitem[Kaufman-Scarborough and Lindquist(1999)]%
        {article}
\bibfield{author}{\bibinfo{person}{Carol Kaufman-Scarborough} {and} \bibinfo{person}{Jay Lindquist}.} \bibinfo{year}{1999}\natexlab{}.
\newblock \showarticletitle{Time Management and Polychronicity: Comparisons, Contrasts, and Insights for the Workplace}.
\newblock \bibinfo{journal}{\emph{Journal of Managerial Psychology}}  \bibinfo{volume}{14} (\bibinfo{date}{06} \bibinfo{year}{1999}), \bibinfo{pages}{288--312}.
\newblock
\urldef\tempurl%
\url{https://doi.org/10.1108/02683949910263819}
\showDOI{\tempurl}


\bibitem[Kaziunas et~al\mbox{.}(2019)]%
        {kaziunas2019precarious}
\bibfield{author}{\bibinfo{person}{Elizabeth Kaziunas}, \bibinfo{person}{Michael~S Klinkman}, {and} \bibinfo{person}{Mark~S Ackerman}.} \bibinfo{year}{2019}\natexlab{}.
\newblock \showarticletitle{Precarious interventions: Designing for ecologies of care}.
\newblock \bibinfo{journal}{\emph{Proceedings of the ACM on Human-Computer Interaction}} \bibinfo{volume}{3}, \bibinfo{number}{CSCW} (\bibinfo{year}{2019}), \bibinfo{pages}{1--27}.
\newblock


\bibitem[Khovanskaya et~al\mbox{.}(2017)]%
        {khovanskaya2017reworking}
\bibfield{author}{\bibinfo{person}{Vera Khovanskaya}, \bibinfo{person}{Phoebe Sengers}, \bibinfo{person}{Melissa Mazmanian}, {and} \bibinfo{person}{Charles Darrah}.} \bibinfo{year}{2017}\natexlab{}.
\newblock \showarticletitle{Reworking the gaps between design and ethnography}. In \bibinfo{booktitle}{\emph{Proceedings of the 2017 CHI Conference on Human Factors in Computing Systems}}. \bibinfo{pages}{5373--5385}.
\newblock


\bibitem[Kim et~al\mbox{.}(2023)]%
        {Kim2023}
\bibfield{author}{\bibinfo{person}{P. Kim}, \bibinfo{person}{E. Cheon}, {and} \bibinfo{person}{S. Sawyer}.} \bibinfo{year}{2023}\natexlab{}.
\newblock \showarticletitle{Online Freelancing on Digital Labor Platforms: A Scoping Review}. In \bibinfo{booktitle}{\emph{Companion Publication of the 2023 Conference on Computer Supported Cooperative Work and Social Computing}}. \bibinfo{pages}{259--266}.
\newblock


\bibitem[Kingsley et~al\mbox{.}(2024)]%
        {Kingsley2024}
\bibfield{author}{\bibinfo{person}{S. Kingsley}, \bibinfo{person}{M.~S. Silberman}, \bibinfo{person}{C. Wang}, \bibinfo{person}{R. Lambeth}, \bibinfo{person}{J. Zhi}, \bibinfo{person}{M. Eslami}, \bibinfo{person}{B. Li}, {and} \bibinfo{person}{J. Bigham}.} \bibinfo{year}{2024}\natexlab{}.
\newblock \showarticletitle{‘Your Duties Are To Sweep A Floor Remotely’: Low Information Quality in Job Advertisements is a Barrier to Low-Income Job-Seekers’ Successful Use of Digital Platforms}. In \bibinfo{booktitle}{\emph{Proceedings of the 3rd Annual Meeting of the Symposium on Human-Computer Interaction for Work}}. \bibinfo{pages}{1--20}.
\newblock


\bibitem[Kingsley et~al\mbox{.}(2015)]%
        {Kingsley2015}
\bibfield{author}{\bibinfo{person}{Sara~Constance Kingsley}, \bibinfo{person}{Mary~L. Gray}, {and} \bibinfo{person}{Siddharth Suri}.} \bibinfo{year}{2015}\natexlab{}.
\newblock \showarticletitle{Accounting for market frictions and power asymmetries in online labor markets}.
\newblock \bibinfo{journal}{\emph{Policy \& Internet}} \bibinfo{volume}{7}, \bibinfo{number}{4} (\bibinfo{year}{2015}), \bibinfo{pages}{383--400}.
\newblock


\bibitem[Kittur et~al\mbox{.}(2013)]%
        {kittur2013future}
\bibfield{author}{\bibinfo{person}{Aniket Kittur}, \bibinfo{person}{Jeffrey~V Nickerson}, \bibinfo{person}{Michael Bernstein}, \bibinfo{person}{Elizabeth Gerber}, \bibinfo{person}{Aaron Shaw}, \bibinfo{person}{John Zimmerman}, \bibinfo{person}{Matt Lease}, {and} \bibinfo{person}{John Horton}.} \bibinfo{year}{2013}\natexlab{}.
\newblock \showarticletitle{The future of crowd work}. In \bibinfo{booktitle}{\emph{Proceedings of the 2013 conference on Computer supported cooperative work}}. \bibinfo{pages}{1301--1318}.
\newblock


\bibitem[Kose(1998)]%
        {kose1998barrier}
\bibfield{author}{\bibinfo{person}{Satoshi Kose}.} \bibinfo{year}{1998}\natexlab{}.
\newblock \showarticletitle{From barrier-free to universal design: an international perspective}.
\newblock \bibinfo{journal}{\emph{Assistive technology}} \bibinfo{volume}{10}, \bibinfo{number}{1} (\bibinfo{year}{1998}), \bibinfo{pages}{44--50}.
\newblock


\bibitem[Kraemer et~al\mbox{.}(2009)]%
        {kraemer2009one}
\bibfield{author}{\bibinfo{person}{Kenneth~L Kraemer}, \bibinfo{person}{Jason Dedrick}, {and} \bibinfo{person}{Prakul Sharma}.} \bibinfo{year}{2009}\natexlab{}.
\newblock \showarticletitle{One laptop per child: vision vs. reality}.
\newblock \bibinfo{journal}{\emph{Commun. ACM}} \bibinfo{volume}{52}, \bibinfo{number}{6} (\bibinfo{year}{2009}), \bibinfo{pages}{66--73}.
\newblock


\bibitem[Kraut and Resnick(2012)]%
        {kraut2012building}
\bibfield{author}{\bibinfo{person}{Robert~E Kraut} {and} \bibinfo{person}{Paul Resnick}.} \bibinfo{year}{2012}\natexlab{}.
\newblock \bibinfo{booktitle}{\emph{Building successful online communities: Evidence-based social design}}.
\newblock \bibinfo{publisher}{Mit Press}.
\newblock


\bibitem[Krishna et~al\mbox{.}(2017)]%
        {krishna2017visual}
\bibfield{author}{\bibinfo{person}{Ranjay Krishna}, \bibinfo{person}{Yuke Zhu}, \bibinfo{person}{Oliver Groth}, \bibinfo{person}{Justin Johnson}, \bibinfo{person}{Kenji Hata}, \bibinfo{person}{Joshua Kravitz}, \bibinfo{person}{Stephanie Chen}, \bibinfo{person}{Yannis Kalantidis}, \bibinfo{person}{Li-Jia Li}, \bibinfo{person}{David~A Shamma}, {et~al\mbox{.}}} \bibinfo{year}{2017}\natexlab{}.
\newblock \showarticletitle{Visual genome: Connecting language and vision using crowdsourced dense image annotations}.
\newblock \bibinfo{journal}{\emph{International journal of computer vision}}  \bibinfo{volume}{123} (\bibinfo{year}{2017}), \bibinfo{pages}{32--73}.
\newblock


\bibitem[Kun and Shaer(2023)]%
        {Kun2023}
\bibfield{author}{\bibinfo{person}{Andrew~L. Kun} {and} \bibinfo{person}{Orit Shaer}.} \bibinfo{year}{2023}\natexlab{}.
\newblock \showarticletitle{Designing an Inclusive and Engaging Hybrid Event: Experiences From CHIWORK}.
\newblock \bibinfo{journal}{\emph{IEEE Pervasive Computing}} \bibinfo{volume}{22}, \bibinfo{number}{3} (\bibinfo{year}{2023}), \bibinfo{pages}{52--57}.
\newblock


\bibitem[Lascau et~al\mbox{.}(2024)]%
        {Lasacau2024}
\bibfield{author}{\bibinfo{person}{L. Lascau}, \bibinfo{person}{D.~P. Brumby}, \bibinfo{person}{S.~J. Gould}, {and} \bibinfo{person}{A.~L. Cox}.} \bibinfo{year}{2024}\natexlab{}.
\newblock \showarticletitle{“Sometimes It’s Like Putting the Track in Front of the Rushing Train”: Having to Be ‘On Call’ for Work Limits the Temporal Flexibility of Crowdworkers}.
\newblock \bibinfo{journal}{\emph{ACM Transactions on Computer-Human Interaction}} \bibinfo{volume}{31}, \bibinfo{number}{2} (\bibinfo{date}{29 Jan} \bibinfo{year}{2024}), \bibinfo{pages}{1--45}.
\newblock


\bibitem[Lascau et~al\mbox{.}(2022)]%
        {lascau2022crowdworkers}
\bibfield{author}{\bibinfo{person}{Laura Lascau}, \bibinfo{person}{Sandy~JJ Gould}, \bibinfo{person}{Duncan~P Brumby}, {and} \bibinfo{person}{Anna~L Cox}.} \bibinfo{year}{2022}\natexlab{}.
\newblock \showarticletitle{Crowdworkers’ Temporal Flexibility is Being Traded for the Convenience of Requesters Through 19 ‘Invisible Mechanisms’ Employed by Crowdworking Platforms: A Comparative Analysis Study of Nine Platforms}. In \bibinfo{booktitle}{\emph{CHI Conference on Human Factors in Computing Systems Extended Abstracts}}. \bibinfo{pages}{1--8}.
\newblock


\bibitem[Lascau et~al\mbox{.}(2019)]%
        {lascau2019monotasking}
\bibfield{author}{\bibinfo{person}{Laura Lascau}, \bibinfo{person}{Sandy~JJ Gould}, \bibinfo{person}{Anna~L Cox}, \bibinfo{person}{Elizaveta Karmannaya}, {and} \bibinfo{person}{Duncan~P Brumby}.} \bibinfo{year}{2019}\natexlab{}.
\newblock \showarticletitle{Monotasking or multitasking: Designing for crowdworkers' preferences}. In \bibinfo{booktitle}{\emph{Proceedings of the 2019 CHI Conference on Human Factors in Computing Systems}}. \bibinfo{pages}{1--14}.
\newblock


\bibitem[Lee(1999)]%
        {lee1999time}
\bibfield{author}{\bibinfo{person}{Heejin Lee}.} \bibinfo{year}{1999}\natexlab{}.
\newblock \showarticletitle{Time and information technology: monochronicity, polychronicity and temporal symmetry}.
\newblock \bibinfo{journal}{\emph{European Journal of Information Systems}} \bibinfo{volume}{8}, \bibinfo{number}{1} (\bibinfo{year}{1999}), \bibinfo{pages}{16--26}.
\newblock


\bibitem[Lee and Carter(2011)]%
        {lee2011global}
\bibfield{author}{\bibinfo{person}{Kiefer Lee} {and} \bibinfo{person}{Steve Carter}.} \bibinfo{year}{2011}\natexlab{}.
\newblock \showarticletitle{Global marketing management}.
\newblock \bibinfo{journal}{\emph{Strategic Direction}} \bibinfo{volume}{27}, \bibinfo{number}{1} (\bibinfo{year}{2011}).
\newblock


\bibitem[Lewis(2010)]%
        {lewis2010cultures}
\bibfield{author}{\bibinfo{person}{Richard Lewis}.} \bibinfo{year}{2010}\natexlab{}.
\newblock \bibinfo{booktitle}{\emph{When cultures collide: Leading across cultures}}.
\newblock \bibinfo{publisher}{Nicholas Brealey International}.
\newblock


\bibitem[Li et~al\mbox{.}(2023)]%
        {li2023improving}
\bibfield{author}{\bibinfo{person}{Xiaoyan Li}, \bibinfo{person}{Naomi Yamashita}, \bibinfo{person}{Wen Duan}, \bibinfo{person}{Yoshinari Shirai}, {and} \bibinfo{person}{Susan~R Fussell}.} \bibinfo{year}{2023}\natexlab{}.
\newblock \showarticletitle{Improving Non-Native Speakers' Participation with an Automatic Agent in Multilingual Groups}.
\newblock \bibinfo{journal}{\emph{Proceedings of the ACM on Human-Computer Interaction}} \bibinfo{volume}{7}, \bibinfo{number}{GROUP} (\bibinfo{year}{2023}), \bibinfo{pages}{1--28}.
\newblock


\bibitem[Lindtner et~al\mbox{.}(2012)]%
        {lindtner2012cultural}
\bibfield{author}{\bibinfo{person}{Silvia Lindtner}, \bibinfo{person}{Ken Anderson}, {and} \bibinfo{person}{Paul Dourish}.} \bibinfo{year}{2012}\natexlab{}.
\newblock \showarticletitle{Cultural appropriation: information technologies as sites of transnational imagination}. In \bibinfo{booktitle}{\emph{Proceedings of the ACM 2012 conference on computer supported cooperative work}}. \bibinfo{pages}{77--86}.
\newblock


\bibitem[Lops et~al\mbox{.}(2019)]%
        {lops2019trends}
\bibfield{author}{\bibinfo{person}{Pasquale Lops}, \bibinfo{person}{Dietmar Jannach}, \bibinfo{person}{Cataldo Musto}, \bibinfo{person}{Toine Bogers}, {and} \bibinfo{person}{Marijn Koolen}.} \bibinfo{year}{2019}\natexlab{}.
\newblock \showarticletitle{Trends in content-based recommendation}.
\newblock \bibinfo{journal}{\emph{User Modeling and User-Adapted Interaction}} \bibinfo{volume}{29}, \bibinfo{number}{2} (\bibinfo{year}{2019}), \bibinfo{pages}{239--249}.
\newblock


\bibitem[Lucero(2015)]%
        {lucero2015using}
\bibfield{author}{\bibinfo{person}{Andr{\'e}s Lucero}.} \bibinfo{year}{2015}\natexlab{}.
\newblock \showarticletitle{Using affinity diagrams to evaluate interactive prototypes}. In \bibinfo{booktitle}{\emph{IFIP conference on human-computer interaction}}. Springer, \bibinfo{pages}{231--248}.
\newblock


\bibitem[Manning et~al\mbox{.}(2008)]%
        {manning2008introduction}
\bibfield{author}{\bibinfo{person}{Christopher~D Manning}, \bibinfo{person}{Prabhakar Raghavan}, {and} \bibinfo{person}{Hinrich Sch{\"u}tze}.} \bibinfo{year}{2008}\natexlab{}.
\newblock \bibinfo{booktitle}{\emph{Introduction to information retrieval}}.
\newblock \bibinfo{publisher}{Cambridge university press}.
\newblock


\bibitem[Manrai and Manrai(1995)]%
        {manrai1995effects}
\bibfield{author}{\bibinfo{person}{Lalita~A Manrai} {and} \bibinfo{person}{Ajay~K Manrai}.} \bibinfo{year}{1995}\natexlab{}.
\newblock \showarticletitle{Effects of cultural-context, gender, and acculturation on perceptions of work versus social/leisure time usage}.
\newblock \bibinfo{journal}{\emph{Journal of Business Research}} \bibinfo{volume}{32}, \bibinfo{number}{2} (\bibinfo{year}{1995}), \bibinfo{pages}{115--128}.
\newblock


\bibitem[Mansouri and Taylor(2024)]%
        {mansouri2024does}
\bibfield{author}{\bibinfo{person}{Masoumeh Mansouri} {and} \bibinfo{person}{Henry Taylor}.} \bibinfo{year}{2024}\natexlab{}.
\newblock \showarticletitle{Does Cultural Robotics Need Culture? Conceptual Fragmentation and the Problems of Merging Culture with Robot Design}.
\newblock \bibinfo{journal}{\emph{International Journal of Social Robotics}} \bibinfo{volume}{16}, \bibinfo{number}{2} (\bibinfo{year}{2024}), \bibinfo{pages}{385--401}.
\newblock


\bibitem[Marcus(2006)]%
        {marcus2006cross}
\bibfield{author}{\bibinfo{person}{Aaron Marcus}.} \bibinfo{year}{2006}\natexlab{}.
\newblock \showarticletitle{Cross-cultural user-experience design}. In \bibinfo{booktitle}{\emph{International Conference on Theory and Application of Diagrams}}. Springer, \bibinfo{pages}{16--24}.
\newblock


\bibitem[Marcus and Gould(2000)]%
        {marcus2000crosscurrents}
\bibfield{author}{\bibinfo{person}{Aaron Marcus} {and} \bibinfo{person}{Emilie~West Gould}.} \bibinfo{year}{2000}\natexlab{}.
\newblock \showarticletitle{Crosscurrents: cultural dimensions and global Web user-interface design}.
\newblock \bibinfo{journal}{\emph{interactions}} \bibinfo{volume}{7}, \bibinfo{number}{4} (\bibinfo{year}{2000}), \bibinfo{pages}{32--46}.
\newblock


\bibitem[Martin et~al\mbox{.}(2014)]%
        {martin2014being}
\bibfield{author}{\bibinfo{person}{David Martin}, \bibinfo{person}{Benjamin~V Hanrahan}, \bibinfo{person}{Jacki O'Neill}, {and} \bibinfo{person}{Neha Gupta}.} \bibinfo{year}{2014}\natexlab{}.
\newblock \showarticletitle{Being a turker}. In \bibinfo{booktitle}{\emph{Proceedings of the 17th ACM conference on Computer supported cooperative work \& social computing}}. \bibinfo{pages}{224--235}.
\newblock


\bibitem[McCrickard(2003)]%
        {mccrickard2003chewar}
\bibfield{author}{\bibinfo{person}{DS McCrickard}.} \bibinfo{year}{2003}\natexlab{}.
\newblock \showarticletitle{Chewar. CM, Somervell, JP, \& Ndiwalana}.
\newblock \bibinfo{journal}{\emph{A." A Model for Notification Systems Evaluation--Assessing User Goals for Multitasking Activity." ACM Transactions on Computer-Human Interaction}} (\bibinfo{year}{2003}), \bibinfo{pages}{312--338}.
\newblock


\bibitem[Meiselwitz(2020)]%
        {meiselwitz2020social}
\bibfield{author}{\bibinfo{person}{Gabriele Meiselwitz}.} \bibinfo{year}{2020}\natexlab{}.
\newblock \bibinfo{booktitle}{\emph{Social Computing and Social Media. Design, Ethics, User Behavior, and Social Network Analysis: 12th International Conference, SCSM 2020, Held as Part of the 22nd HCI International Conference, HCII 2020, Copenhagen, Denmark, July 19--24, 2020, Proceedings, Part I}}. Vol.~\bibinfo{volume}{12194}.
\newblock \bibinfo{publisher}{Springer Nature}.
\newblock


\bibitem[Mella et~al\mbox{.}(2024)]%
        {Mella2024}
\bibfield{author}{\bibinfo{person}{Jonathan Mella}, \bibinfo{person}{Ioanna Iacovides}, {and} \bibinfo{person}{Anna Cox}.} \bibinfo{year}{2024}\natexlab{}.
\newblock \showarticletitle{‘Jumping Out from the Pressure of Work and into the Game’: Curating Immersive Digital Game Experiences for Post-Work Recovery}.
\newblock \bibinfo{journal}{\emph{ACM Games: Research and Practice}} (\bibinfo{year}{2024}).
\newblock


\bibitem[Meredith et~al\mbox{.}(2017)]%
        {meredith2017project}
\bibfield{author}{\bibinfo{person}{Jack~R Meredith}, \bibinfo{person}{Scott~M Shafer}, {and} \bibinfo{person}{Samuel~J Mantel~Jr}.} \bibinfo{year}{2017}\natexlab{}.
\newblock \bibinfo{booktitle}{\emph{Project management: a strategic managerial approach}}.
\newblock \bibinfo{publisher}{John Wiley \& Sons}.
\newblock


\bibitem[Meyer(2014)]%
        {meyer2014culture}
\bibfield{author}{\bibinfo{person}{Erin Meyer}.} \bibinfo{year}{2014}\natexlab{}.
\newblock \bibinfo{booktitle}{\emph{The culture map: Breaking through the invisible boundaries of global business}}.
\newblock \bibinfo{publisher}{Public Affairs}.
\newblock


\bibitem[Monk et~al\mbox{.}(2002)]%
        {monk2002funology}
\bibfield{author}{\bibinfo{person}{Andrew Monk}, \bibinfo{person}{Marc Hassenzahl}, \bibinfo{person}{Mark Blythe}, {and} \bibinfo{person}{Darren Reed}.} \bibinfo{year}{2002}\natexlab{}.
\newblock \showarticletitle{Funology: designing enjoyment}. In \bibinfo{booktitle}{\emph{CHI'02 extended abstracts on human factors in computing systems}}. \bibinfo{pages}{924--925}.
\newblock


\bibitem[Morden(1999)]%
        {morden1999models}
\bibfield{author}{\bibinfo{person}{Tony Morden}.} \bibinfo{year}{1999}\natexlab{}.
\newblock \showarticletitle{Models of national culture--a management review}.
\newblock \bibinfo{journal}{\emph{Cross Cultural Management: An International Journal}} \bibinfo{volume}{6}, \bibinfo{number}{1} (\bibinfo{year}{1999}), \bibinfo{pages}{19--44}.
\newblock


\bibitem[Moreschi et~al\mbox{.}(2020)]%
        {moreschi2020brazilian}
\bibfield{author}{\bibinfo{person}{Bruno Moreschi}, \bibinfo{person}{Gabriel Pereira}, {and} \bibinfo{person}{Fabio~G Cozman}.} \bibinfo{year}{2020}\natexlab{}.
\newblock \showarticletitle{The Brazilian Workers in Amazon Mechanical Turk: Dreams and realities of ghost workers.}
\newblock \bibinfo{journal}{\emph{Contracampo}} \bibinfo{volume}{39}, \bibinfo{number}{1} (\bibinfo{year}{2020}).
\newblock


\bibitem[Nardon and Steers(2009)]%
        {nardon2009culture}
\bibfield{author}{\bibinfo{person}{Luciara Nardon} {and} \bibinfo{person}{Richard~M Steers}.} \bibinfo{year}{2009}\natexlab{}.
\newblock \showarticletitle{The culture theory jungle: Divergence and convergence in models of national culture}.
\newblock \bibinfo{journal}{\emph{Cambridge handbook of culture, organizations, and work}} (\bibinfo{year}{2009}), \bibinfo{pages}{3--22}.
\newblock


\bibitem[Newbold et~al\mbox{.}(2022)]%
        {Newbold2022}
\bibfield{author}{\bibinfo{person}{Joseph~W. Newbold}, \bibinfo{person}{Anna Rudnicka}, \bibinfo{person}{David Cook}, \bibinfo{person}{Marta~E. Cecchinato}, \bibinfo{person}{Sandy~JJ Gould}, {and} \bibinfo{person}{Anna~L. Cox}.} \bibinfo{year}{2022}\natexlab{}.
\newblock \showarticletitle{The new normals of work: a framework for understanding responses to disruptions created by new futures of work}.
\newblock \bibinfo{journal}{\emph{Human–Computer Interaction}} \bibinfo{volume}{37}, \bibinfo{number}{6} (\bibinfo{year}{2022}), \bibinfo{pages}{508--531}.
\newblock


\bibitem[Newlands and Lutz(2021)]%
        {newlands2021crowdwork}
\bibfield{author}{\bibinfo{person}{Gemma Newlands} {and} \bibinfo{person}{Christoph Lutz}.} \bibinfo{year}{2021}\natexlab{}.
\newblock \showarticletitle{Crowdwork and the mobile underclass: Barriers to participation in India and the United States}.
\newblock \bibinfo{journal}{\emph{new media \& society}} \bibinfo{volume}{23}, \bibinfo{number}{6} (\bibinfo{year}{2021}), \bibinfo{pages}{1341--1361}.
\newblock


\bibitem[Norman(2012)]%
        {norman2012does}
\bibfield{author}{\bibinfo{person}{Don Norman}.} \bibinfo{year}{2012}\natexlab{}.
\newblock \showarticletitle{Does culture matter for product design}.
\newblock \bibinfo{journal}{\emph{Core 77 Design magazine \&}} (\bibinfo{year}{2012}).
\newblock


\bibitem[Norman(2013)]%
        {norman2013design}
\bibfield{author}{\bibinfo{person}{Don Norman}.} \bibinfo{year}{2013}\natexlab{}.
\newblock \bibinfo{booktitle}{\emph{The design of everyday things: Revised and expanded edition}}.
\newblock \bibinfo{publisher}{Basic books}.
\newblock


\bibitem[Norman(1991)]%
        {norman1991cognitive}
\bibfield{author}{\bibinfo{person}{Donald~A Norman}.} \bibinfo{year}{1991}\natexlab{}.
\newblock \showarticletitle{Cognitive artifacts}.
\newblock \bibinfo{journal}{\emph{Designing interaction: Psychology at the human-computer interface}} \bibinfo{volume}{1}, \bibinfo{number}{1} (\bibinfo{year}{1991}), \bibinfo{pages}{17--38}.
\newblock


\bibitem[Norman(2016)]%
        {norman2016living}
\bibfield{author}{\bibinfo{person}{Donald~A Norman}.} \bibinfo{year}{2016}\natexlab{}.
\newblock \bibinfo{booktitle}{\emph{Living with complexity}}.
\newblock \bibinfo{publisher}{MIT press}.
\newblock


\bibitem[Platform(2020)]%
        {platform2020wage}
\bibfield{author}{\bibinfo{person}{Georgia Fair~Labor Platform}.} \bibinfo{year}{2020}\natexlab{}.
\newblock \bibinfo{title}{Wage Theft Calculator. Georgia Fair Labor Platform}.
\newblock
\newblock


\bibitem[Poposki and Oswald(2010)]%
        {poposki2010multitasking}
\bibfield{author}{\bibinfo{person}{Elizabeth~M Poposki} {and} \bibinfo{person}{Frederick~L Oswald}.} \bibinfo{year}{2010}\natexlab{}.
\newblock \showarticletitle{The multitasking preference inventory: Toward an improved measure of individual differences in polychronicity}.
\newblock \bibinfo{journal}{\emph{Human Performance}} \bibinfo{volume}{23}, \bibinfo{number}{3} (\bibinfo{year}{2010}), \bibinfo{pages}{247--264}.
\newblock


\bibitem[Posada(2022)]%
        {posada2022coloniality}
\bibfield{author}{\bibinfo{person}{Julian Posada}.} \bibinfo{year}{2022}\natexlab{}.
\newblock \emph{\bibinfo{title}{The Coloniality of Data Work: Power and Inequality in Outsourced Data Production for Machine Learning}}.
\newblock \bibinfo{thesistype}{Ph.\,D. Dissertation}. \bibinfo{school}{University of Toronto (Canada)}.
\newblock


\bibitem[Potocka-Sionek(2022)]%
        {potocka2022crowdwork}
\bibfield{author}{\bibinfo{person}{Nastazja Potocka-Sionek}.} \bibinfo{year}{2022}\natexlab{}.
\newblock \showarticletitle{Crowdwork and global supply chains: Regulating digital piecework}.
\newblock In \bibinfo{booktitle}{\emph{A Research Agenda for the Gig Economy and Society}}. \bibinfo{publisher}{Edward Elgar Publishing}, \bibinfo{pages}{215--234}.
\newblock


\bibitem[Pradhan et~al\mbox{.}(2022)]%
        {pradhan2022search}
\bibfield{author}{\bibinfo{person}{Vivek~Krishna Pradhan}, \bibinfo{person}{Mike Schaekermann}, {and} \bibinfo{person}{Matthew Lease}.} \bibinfo{year}{2022}\natexlab{}.
\newblock \showarticletitle{In Search of Ambiguity: A Three-Stage Workflow Design to Clarify Annotation Guidelines for Crowd Workers}.
\newblock \bibinfo{journal}{\emph{Frontiers in Artificial Intelligence}}  \bibinfo{volume}{5} (\bibinfo{year}{2022}).
\newblock


\bibitem[Prahalad(2008)]%
        {prahalad2008fortune}
\bibfield{author}{\bibinfo{person}{CK Prahalad}.} \bibinfo{year}{2008}\natexlab{}.
\newblock \showarticletitle{The fortune at the bottom of the pyramid: Eradicating poverty through profits}.
\newblock \bibinfo{journal}{\emph{McKinsey briefing notes series}} \bibinfo{volume}{36}, \bibinfo{number}{3} (\bibinfo{year}{2008}), \bibinfo{pages}{52--74}.
\newblock


\bibitem[Qiu et~al\mbox{.}(2021)]%
        {qiu2021using}
\bibfield{author}{\bibinfo{person}{Sihang Qiu}, \bibinfo{person}{Alessandro Bozzon}, \bibinfo{person}{Max~V Birk}, {and} \bibinfo{person}{Ujwal Gadiraju}.} \bibinfo{year}{2021}\natexlab{}.
\newblock \showarticletitle{Using worker avatars to improve microtask crowdsourcing}.
\newblock \bibinfo{journal}{\emph{Proceedings of the ACM on Human-Computer Interaction}} \bibinfo{volume}{5}, \bibinfo{number}{CSCW2} (\bibinfo{year}{2021}), \bibinfo{pages}{1--28}.
\newblock


\bibitem[Quesenbery and Szuc(2011)]%
        {quesenbery2011global}
\bibfield{author}{\bibinfo{person}{Whitney Quesenbery} {and} \bibinfo{person}{Daniel Szuc}.} \bibinfo{year}{2011}\natexlab{}.
\newblock \bibinfo{booktitle}{\emph{Global UX: design and research in a connected world}}.
\newblock \bibinfo{publisher}{Elsevier}.
\newblock


\bibitem[Rahman et~al\mbox{.}(2019)]%
        {rahman2019constructing}
\bibfield{author}{\bibinfo{person}{Md~Mustafizur Rahman}, \bibinfo{person}{Mucahid Kutlu}, {and} \bibinfo{person}{Matthew Lease}.} \bibinfo{year}{2019}\natexlab{}.
\newblock \showarticletitle{Constructing test collections using multi-armed bandits and active learning}. In \bibinfo{booktitle}{\emph{The World Wide Web Conference}}. \bibinfo{pages}{3158--3164}.
\newblock


\bibitem[Rai et~al\mbox{.}(2017)]%
        {rai2017editor}
\bibfield{author}{\bibinfo{person}{Arun Rai}, \bibinfo{person}{Andrew Burton-Jones}, \bibinfo{person}{Hsinchun Chen}, \bibinfo{person}{Alok Gupta}, \bibinfo{person}{Alan~R Hevner}, \bibinfo{person}{Wolfgang Ketter}, \bibinfo{person}{Jeffrey Parsons}, \bibinfo{person}{H~Raghav Rao}, \bibinfo{person}{Sumit Sarkar}, {and} \bibinfo{person}{Youngjin Yoo}.} \bibinfo{year}{2017}\natexlab{}.
\newblock \showarticletitle{Editor’s comments: Diversity of design science research}.
\newblock  (\bibinfo{year}{2017}).
\newblock


\bibitem[Rau et~al\mbox{.}(2012)]%
        {rau2012cross}
\bibfield{author}{\bibinfo{person}{Pei-Luen~Patrick Rau}, \bibinfo{person}{Tom Plocher}, {and} \bibinfo{person}{Yee-Yin Choong}.} \bibinfo{year}{2012}\natexlab{}.
\newblock \bibinfo{booktitle}{\emph{Cross-cultural design for IT products and services}}.
\newblock \bibinfo{publisher}{CRC Press}.
\newblock


\bibitem[Rau(2015)]%
        {rau2015cross}
\bibfield{author}{\bibinfo{person}{PL~Patrick Rau}.} \bibinfo{year}{2015}\natexlab{}.
\newblock \bibinfo{booktitle}{\emph{Cross-Cultural Design Methods, Practice and Impact: 7th International Conference, CCD 2015, Held as Part of HCI International 2015, Los Angeles, CA, USA, August 2-7, 2015, Proceedings, Part I}}. Vol.~\bibinfo{volume}{9180}.
\newblock \bibinfo{publisher}{Springer}.
\newblock


\bibitem[Reinecke(2010)]%
        {reinecke2010culturally}
\bibfield{author}{\bibinfo{person}{Katharina Reinecke}.} \bibinfo{year}{2010}\natexlab{}.
\newblock \emph{\bibinfo{title}{Culturally adaptive user interfaces}}.
\newblock \bibinfo{thesistype}{Ph.\,D. Dissertation}. \bibinfo{school}{University of Zurich}.
\newblock


\bibitem[Reinecke and Bernstein(2011)]%
        {reinecke2011improving}
\bibfield{author}{\bibinfo{person}{Katharina Reinecke} {and} \bibinfo{person}{Abraham Bernstein}.} \bibinfo{year}{2011}\natexlab{}.
\newblock \showarticletitle{Improving performance, perceived usability, and aesthetics with culturally adaptive user interfaces}.
\newblock \bibinfo{journal}{\emph{ACM Transactions on Computer-Human Interaction (TOCHI)}} \bibinfo{volume}{18}, \bibinfo{number}{2} (\bibinfo{year}{2011}), \bibinfo{pages}{1--29}.
\newblock


\bibitem[Reynolds and Fletcher-Janzen(2007)]%
        {reynolds2007encyclopedia}
\bibfield{author}{\bibinfo{person}{Cecil~R Reynolds} {and} \bibinfo{person}{Elaine Fletcher-Janzen}.} \bibinfo{year}{2007}\natexlab{}.
\newblock \bibinfo{booktitle}{\emph{Encyclopedia of Special Education: A Reference for the Education of Children, Adolescents, and Adults with Disabilities and Other Exceptional Individuals, Volume 3}}. Vol.~\bibinfo{volume}{3}.
\newblock \bibinfo{publisher}{John Wiley \& Sons}.
\newblock


\bibitem[Ricaurte(2019)]%
        {Ricaurte2019}
\bibfield{author}{\bibinfo{person}{Paola Ricaurte}.} \bibinfo{year}{2019}\natexlab{}.
\newblock \showarticletitle{Data epistemologies, the coloniality of power, and resistance}.
\newblock \bibinfo{journal}{\emph{Television \& New Media}} \bibinfo{volume}{20}, \bibinfo{number}{4} (\bibinfo{year}{2019}), \bibinfo{pages}{350--365}.
\newblock


\bibitem[Rivera and Lee(2021)]%
        {rivera2021want}
\bibfield{author}{\bibinfo{person}{Veronica~A Rivera} {and} \bibinfo{person}{David~T Lee}.} \bibinfo{year}{2021}\natexlab{}.
\newblock \showarticletitle{I Want to, but First I Need to: Understanding Crowdworkers' Career Goals, Challenges, and Tensions}.
\newblock \bibinfo{journal}{\emph{Proceedings of the ACM on Human-Computer Interaction}} \bibinfo{volume}{5}, \bibinfo{number}{CSCW1} (\bibinfo{year}{2021}), \bibinfo{pages}{1--22}.
\newblock


\bibitem[Roberts(2005)]%
        {roberts2005computer}
\bibfield{author}{\bibinfo{person}{Tim~S Roberts}.} \bibinfo{year}{2005}\natexlab{}.
\newblock \showarticletitle{Computer-supported collaborative learning in higher education}.
\newblock In \bibinfo{booktitle}{\emph{Computer-supported collaborative learning in higher education}}. \bibinfo{publisher}{IGI global}, \bibinfo{pages}{1--18}.
\newblock


\bibitem[Robinson and Giles(1990)]%
        {robinson1990handbook}
\bibfield{author}{\bibinfo{person}{William~Peter Robinson} {and} \bibinfo{person}{Howard Giles}.} \bibinfo{year}{1990}\natexlab{}.
\newblock \bibinfo{booktitle}{\emph{Handbook of language and social psychology}}.
\newblock \bibinfo{publisher}{Wiley Chichester, UK}.
\newblock


\bibitem[Roffarello et~al\mbox{.}(2023)]%
        {Roffarello2023}
\bibfield{author}{\bibinfo{person}{A.~M. Roffarello}, \bibinfo{person}{L. De~Russis}, \bibinfo{person}{D. Lottridge}, {and} \bibinfo{person}{M.~E. Cecchinato}.} \bibinfo{year}{2023}\natexlab{}.
\newblock \showarticletitle{Understanding digital wellbeing within complex technological contexts}.
\newblock \bibinfo{journal}{\emph{International Journal of Human-Computer Studies}}  \bibinfo{volume}{175} (\bibinfo{year}{2023}), \bibinfo{pages}{103034}.
\newblock


\bibitem[Ross et~al\mbox{.}(2010)]%
        {ross2010crowdworkers}
\bibfield{author}{\bibinfo{person}{Joel Ross}, \bibinfo{person}{Lilly Irani}, \bibinfo{person}{M~Six Silberman}, \bibinfo{person}{Andrew Zaldivar}, {and} \bibinfo{person}{Bill Tomlinson}.} \bibinfo{year}{2010}\natexlab{}.
\newblock \showarticletitle{Who are the crowdworkers? Shifting demographics in Mechanical Turk}.
\newblock In \bibinfo{booktitle}{\emph{CHI'10 extended abstracts on Human factors in computing systems}}. \bibinfo{pages}{2863--2872}.
\newblock


\bibitem[Saito et~al\mbox{.}(2019)]%
        {saito2019turkscanner}
\bibfield{author}{\bibinfo{person}{Susumu Saito}, \bibinfo{person}{Chun-Wei Chiang}, \bibinfo{person}{Saiph Savage}, \bibinfo{person}{Teppei Nakano}, \bibinfo{person}{Tetsunori Kobayashi}, {and} \bibinfo{person}{Jeffrey Bigham}.} \bibinfo{year}{2019}\natexlab{}.
\newblock \showarticletitle{Turkscanner: Predicting the hourly wage of microtasks}. In \bibinfo{booktitle}{\emph{The World Wide Web Conference}}. \bibinfo{pages}{3187--3193}.
\newblock


\bibitem[Sannon and Cosley(2019)]%
        {sannon2019privacy}
\bibfield{author}{\bibinfo{person}{Shruti Sannon} {and} \bibinfo{person}{Dan Cosley}.} \bibinfo{year}{2019}\natexlab{}.
\newblock \showarticletitle{Privacy, power, and invisible labor on Amazon Mechanical Turk}. In \bibinfo{booktitle}{\emph{Proceedings of the 2019 CHI conference on human factors in computing systems}}. \bibinfo{pages}{1--12}.
\newblock


\bibitem[Savage et~al\mbox{.}(2020)]%
        {savage2020becoming}
\bibfield{author}{\bibinfo{person}{Saiph Savage}, \bibinfo{person}{Chun~Wei Chiang}, \bibinfo{person}{Susumu Saito}, \bibinfo{person}{Carlos Toxtli}, {and} \bibinfo{person}{Jeffrey Bigham}.} \bibinfo{year}{2020}\natexlab{}.
\newblock \showarticletitle{Becoming the super turker: Increasing wages via a strategy from high earning workers}. In \bibinfo{booktitle}{\emph{Proceedings of The Web Conference 2020}}. \bibinfo{pages}{1241--1252}.
\newblock


\bibitem[Savage and Garcia-Murillo(2024)]%
        {Savage2024}
\bibfield{author}{\bibinfo{person}{Saiph Savage} {and} \bibinfo{person}{Martha Garcia-Murillo}.} \bibinfo{year}{2024}\natexlab{}.
\newblock \showarticletitle{Tools for crowdworkers coding data for AI}.
\newblock In \bibinfo{booktitle}{\emph{Handbook of Artificial Intelligence at Work}}, \bibfield{editor}{\bibinfo{person}{Joy Xiang} {and} \bibinfo{person}{Peter Lee}} (Eds.). \bibinfo{publisher}{Edward Elgar Publishing}, \bibinfo{pages}{76--94}.
\newblock


\bibitem[Savage et~al\mbox{.}(2021)]%
        {savage2021research}
\bibfield{author}{\bibinfo{person}{S Savage}, \bibinfo{person}{C Toxtli}, {and} \bibinfo{person}{E Betanzos}.} \bibinfo{year}{2021}\natexlab{}.
\newblock \showarticletitle{Research Methods to Study \& Empower Crowd Workers}.
\newblock \bibinfo{journal}{\emph{Oxford University Press}} (\bibinfo{year}{2021}).
\newblock


\bibitem[Sayago(2023)]%
        {sayago2023cultures}
\bibfield{author}{\bibinfo{person}{Sergio Sayago}.} \bibinfo{year}{2023}\natexlab{}.
\newblock \bibinfo{booktitle}{\emph{Cultures in human-computer interaction}}.
\newblock \bibinfo{publisher}{Springer Nature}.
\newblock


\bibitem[Schwanda~Sosik et~al\mbox{.}(2012)]%
        {schwanda2012see}
\bibfield{author}{\bibinfo{person}{Victoria Schwanda~Sosik}, \bibinfo{person}{Xuan Zhao}, {and} \bibinfo{person}{Dan Cosley}.} \bibinfo{year}{2012}\natexlab{}.
\newblock \showarticletitle{See friendship, sort of: How conversation and digital traces might support reflection on friendships}. In \bibinfo{booktitle}{\emph{Proceedings of the ACM 2012 Conference on Computer Supported Cooperative Work}}. \bibinfo{pages}{1145--1154}.
\newblock


\bibitem[Seering et~al\mbox{.}(2018)]%
        {Seering2018}
\bibfield{author}{\bibinfo{person}{Joseph Seering}, \bibinfo{person}{Juan~Pablo Flores}, \bibinfo{person}{Saiph Savage}, {and} \bibinfo{person}{Jessica Hammer}.} \bibinfo{year}{2018}\natexlab{}.
\newblock \showarticletitle{The social roles of bots: evaluating impact of bots on discussions in online communities}.
\newblock \bibinfo{journal}{\emph{Proceedings of the ACM on Human-Computer Interaction}} \bibinfo{volume}{2}, \bibinfo{number}{CSCW} (\bibinfo{year}{2018}), \bibinfo{pages}{1--29}.
\newblock


\bibitem[Seering et~al\mbox{.}(2019)]%
        {Seering2019}
\bibfield{author}{\bibinfo{person}{Joseph Seering}, \bibinfo{person}{Ole~J. Mengshoel}, \bibinfo{person}{Quinn Liao}, {and} \bibinfo{person}{Geoffrey Kaufman}.} \bibinfo{year}{2019}\natexlab{}.
\newblock \showarticletitle{Moderator engagement and community development in the age of algorithms}.
\newblock \bibinfo{journal}{\emph{New Media \& Society}} \bibinfo{volume}{21}, \bibinfo{number}{7} (\bibinfo{year}{2019}), \bibinfo{pages}{1417--1443}.
\newblock


\bibitem[Setlock et~al\mbox{.}(2004)]%
        {setlock2004taking}
\bibfield{author}{\bibinfo{person}{Leslie~D Setlock}, \bibinfo{person}{Susan~R Fussell}, {and} \bibinfo{person}{Christine Neuwirth}.} \bibinfo{year}{2004}\natexlab{}.
\newblock \showarticletitle{Taking it out of context: collaborating within and across cultures in face-to-face settings and via instant messaging}. In \bibinfo{booktitle}{\emph{Proceedings of the 2004 ACM conference on Computer supported cooperative work}}. \bibinfo{pages}{604--613}.
\newblock


\bibitem[Shaer et~al\mbox{.}(2024)]%
        {Shaer2024}
\bibfield{author}{\bibinfo{person}{Orit Shaer}, \bibinfo{person}{Angelora Cooper}, \bibinfo{person}{Osnat Mokryn}, \bibinfo{person}{Andrew~L. Kun}, {and} \bibinfo{person}{Hagit Ben~Shoshan}.} \bibinfo{year}{2024}\natexlab{}.
\newblock \showarticletitle{AI-Augmented Brainwriting: Investigating the use of LLMs in group ideation}. In \bibinfo{booktitle}{\emph{Proceedings of the CHI Conference on Human Factors in Computing Systems}}. \bibinfo{pages}{1--17}.
\newblock


\bibitem[Sikkema and Niyekawa(1987)]%
        {sikkema1987design}
\bibfield{author}{\bibinfo{person}{Mildred Sikkema} {and} \bibinfo{person}{Agnes Niyekawa}.} \bibinfo{year}{1987}\natexlab{}.
\newblock \bibinfo{booktitle}{\emph{Design for Cross-Cultural Learning.}}
\newblock \bibinfo{publisher}{ERIC}.
\newblock


\bibitem[Silberman(2015)]%
        {silberman2015operating}
\bibfield{author}{\bibinfo{person}{M Silberman}.} \bibinfo{year}{2015}\natexlab{}.
\newblock \showarticletitle{Operating an employer reputation system: Lessons from Turkopticon, 2008-2015}.
\newblock \bibinfo{journal}{\emph{Comp. Lab. L. \& Pol'y J.}}  \bibinfo{volume}{37} (\bibinfo{year}{2015}), \bibinfo{pages}{505}.
\newblock


\bibitem[Silberman et~al\mbox{.}(2018)]%
        {silberman2018responsible}
\bibfield{author}{\bibinfo{person}{M~Six Silberman}, \bibinfo{person}{Bill Tomlinson}, \bibinfo{person}{Rochelle LaPlante}, \bibinfo{person}{Joel Ross}, \bibinfo{person}{Lilly Irani}, {and} \bibinfo{person}{Andrew Zaldivar}.} \bibinfo{year}{2018}\natexlab{}.
\newblock \showarticletitle{Responsible research with crowds: pay crowdworkers at least minimum wage}.
\newblock \bibinfo{journal}{\emph{Commun. ACM}} \bibinfo{volume}{61}, \bibinfo{number}{3} (\bibinfo{year}{2018}), \bibinfo{pages}{39--41}.
\newblock


\bibitem[Slack and Wise(2005)]%
        {slack2005culture}
\bibfield{author}{\bibinfo{person}{Jennifer~Daryl Slack} {and} \bibinfo{person}{John~Macgregor Wise}.} \bibinfo{year}{2005}\natexlab{}.
\newblock \bibinfo{booktitle}{\emph{Culture+ technology: A primer}}.
\newblock \bibinfo{publisher}{Peter Lang}.
\newblock


\bibitem[Stahl(2006)]%
        {stahl2006group}
\bibfield{author}{\bibinfo{person}{Gerry Stahl}.} \bibinfo{year}{2006}\natexlab{}.
\newblock \bibinfo{booktitle}{\emph{Group cognition: Computer support for building collaborative knowledge (acting with technology)}}.
\newblock \bibinfo{publisher}{The MIT Press}.
\newblock


\bibitem[Steinfeld and Maisel(2012)]%
        {steinfeld2012universal}
\bibfield{author}{\bibinfo{person}{Edward Steinfeld} {and} \bibinfo{person}{Jordana Maisel}.} \bibinfo{year}{2012}\natexlab{}.
\newblock \bibinfo{booktitle}{\emph{Universal design: Creating inclusive environments}}.
\newblock \bibinfo{publisher}{John Wiley \& Sons}.
\newblock


\bibitem[Stiglitz and Pike(2004)]%
        {stiglitz2004globalization}
\bibfield{author}{\bibinfo{person}{Joseph Stiglitz} {and} \bibinfo{person}{Robert~M Pike}.} \bibinfo{year}{2004}\natexlab{}.
\newblock \showarticletitle{Globalization and its Discontents}.
\newblock  (\bibinfo{year}{2004}).
\newblock


\bibitem[Story(2001)]%
        {story2001principles}
\bibfield{author}{\bibinfo{person}{Molly~Follette Story}.} \bibinfo{year}{2001}\natexlab{}.
\newblock \showarticletitle{Principles of universal design}.
\newblock \bibinfo{journal}{\emph{Universal design handbook}} (\bibinfo{year}{2001}).
\newblock


\bibitem[Sun(2012)]%
        {sun2012cross}
\bibfield{author}{\bibinfo{person}{Huatong Sun}.} \bibinfo{year}{2012}\natexlab{}.
\newblock \bibinfo{booktitle}{\emph{Cross-cultural technology design: Creating culture-sensitive technology for local users}}.
\newblock \bibinfo{publisher}{OUP USA}.
\newblock


\bibitem[Suzuki et~al\mbox{.}(2016)]%
        {suzuki2016atelier}
\bibfield{author}{\bibinfo{person}{Ryo Suzuki}, \bibinfo{person}{Niloufar Salehi}, \bibinfo{person}{Michelle~S Lam}, \bibinfo{person}{Juan~C Marroquin}, {and} \bibinfo{person}{Michael~S Bernstein}.} \bibinfo{year}{2016}\natexlab{}.
\newblock \showarticletitle{Atelier: Repurposing expert crowdsourcing tasks as micro-internships}. In \bibinfo{booktitle}{\emph{Proceedings of the 2016 CHI conference on human factors in computing systems}}. \bibinfo{pages}{2645--2656}.
\newblock


\bibitem[Toxtli et~al\mbox{.}(2020)]%
        {toxtli2020reputation}
\bibfield{author}{\bibinfo{person}{Carlos Toxtli}, \bibinfo{person}{Angela Richmond-Fuller}, {and} \bibinfo{person}{Saiph Savage}.} \bibinfo{year}{2020}\natexlab{}.
\newblock \showarticletitle{Reputation agent: Prompting fair reviews in gig markets}. In \bibinfo{booktitle}{\emph{Proceedings of The Web Conference 2020}}. \bibinfo{pages}{1228--1240}.
\newblock


\bibitem[Toxtli et~al\mbox{.}(2021)]%
        {toxtli2021quantifying}
\bibfield{author}{\bibinfo{person}{Carlos Toxtli}, \bibinfo{person}{Siddharth Suri}, {and} \bibinfo{person}{Saiph Savage}.} \bibinfo{year}{2021}\natexlab{}.
\newblock \showarticletitle{Quantifying the invisible labor in crowd work}.
\newblock \bibinfo{journal}{\emph{Proceedings of the ACM on Human-Computer Interaction}} \bibinfo{volume}{5}, \bibinfo{number}{CSCW2} (\bibinfo{year}{2021}), \bibinfo{pages}{1--26}.
\newblock


\bibitem[Trompenaars and Hampden-Turner(2004)]%
        {trompenaars2004managing}
\bibfield{author}{\bibinfo{person}{Fons Trompenaars} {and} \bibinfo{person}{Charles Hampden-Turner}.} \bibinfo{year}{2004}\natexlab{}.
\newblock \bibinfo{booktitle}{\emph{Managing people across cultures}}.
\newblock \bibinfo{publisher}{Capstone Chichester}.
\newblock


\bibitem[Trompenaars and Hampden-Turner(2011)]%
        {trompenaars2011riding}
\bibfield{author}{\bibinfo{person}{Fons Trompenaars} {and} \bibinfo{person}{Charles Hampden-Turner}.} \bibinfo{year}{2011}\natexlab{}.
\newblock \bibinfo{booktitle}{\emph{Riding the waves of culture: Understanding diversity in global business}}.
\newblock \bibinfo{publisher}{Nicholas Brealey International}.
\newblock


\bibitem[Upwork(2022)]%
        {upwork_2022}
\bibfield{author}{\bibinfo{person}{Upwork}.} \bibinfo{year}{2022}\natexlab{}.
\newblock \bibinfo{title}{Use your work diary - upwork customer service \& support}.
\newblock
\newblock
\urldef\tempurl%
\url{https://support.upwork.com/hc/en-us/articles/211068518-Use-Your-Work-Diary}
\showURL{%
\tempurl}


\bibitem[Ursu and Ciortescu(2021)]%
        {ursu2021exploring}
\bibfield{author}{\bibinfo{person}{Oana Ursu} {and} \bibinfo{person}{Elena Ciortescu}.} \bibinfo{year}{2021}\natexlab{}.
\newblock \showarticletitle{Exploring Cultural Patterns in Business Communication. Insights from Europe and Asia}.
\newblock \bibinfo{journal}{\emph{CES Working Papers}} \bibinfo{volume}{13}, \bibinfo{number}{2} (\bibinfo{year}{2021}), \bibinfo{pages}{149--158}.
\newblock


\bibitem[Varanasi et~al\mbox{.}(2022)]%
        {varanasi2022feeling}
\bibfield{author}{\bibinfo{person}{Rama~Adithya Varanasi}, \bibinfo{person}{Divya Siddarth}, \bibinfo{person}{Vivek Seshadri}, \bibinfo{person}{Kalika Bali}, {and} \bibinfo{person}{Aditya Vashistha}.} \bibinfo{year}{2022}\natexlab{}.
\newblock \showarticletitle{Feeling Proud, Feeling Embarrassed: Experiences of Low-income Women with Crowd Work}. In \bibinfo{booktitle}{\emph{Proceedings of the 2022 CHI Conference on Human Factors in Computing Systems}}. \bibinfo{pages}{1--18}.
\newblock


\bibitem[Voorhees et~al\mbox{.}(2005)]%
        {voorhees2005trec}
\bibfield{author}{\bibinfo{person}{Ellen~M Voorhees}, \bibinfo{person}{Donna~K Harman}, {et~al\mbox{.}}} \bibinfo{year}{2005}\natexlab{}.
\newblock \bibinfo{booktitle}{\emph{TREC: Experiment and evaluation in information retrieval}}. Vol.~\bibinfo{volume}{63}.
\newblock \bibinfo{publisher}{Citeseer}.
\newblock


\bibitem[Wallerstein(2020)]%
        {wallerstein2020world}
\bibfield{author}{\bibinfo{person}{Immanuel Wallerstein}.} \bibinfo{year}{2020}\natexlab{}.
\newblock \bibinfo{booktitle}{\emph{World-systems analysis: An introduction}}.
\newblock \bibinfo{publisher}{duke university Press}.
\newblock


\bibitem[Wang et~al\mbox{.}(2022)]%
        {10.1145/3491102.3502121}
\bibfield{author}{\bibinfo{person}{Ding Wang}, \bibinfo{person}{Shantanu Prabhat}, {and} \bibinfo{person}{Nithya Sambasivan}.} \bibinfo{year}{2022}\natexlab{}.
\newblock \showarticletitle{Whose AI Dream? In Search of the Aspiration in Data Annotation.}. In \bibinfo{booktitle}{\emph{Proceedings of the 2022 CHI Conference on Human Factors in Computing Systems}} (New Orleans, LA, USA) \emph{(\bibinfo{series}{CHI '22})}. \bibinfo{publisher}{Association for Computing Machinery}, \bibinfo{address}{New York, NY, USA}, Article \bibinfo{articleno}{582}, \bibinfo{numpages}{16}~pages.
\newblock
\showISBNx{9781450391573}
\urldef\tempurl%
\url{https://doi.org/10.1145/3491102.3502121}
\showDOI{\tempurl}


\bibitem[Wang et~al\mbox{.}(2017)]%
        {wang2017community}
\bibfield{author}{\bibinfo{person}{Xinyi Wang}, \bibinfo{person}{Haiyi Zhu}, \bibinfo{person}{Yangyun Li}, \bibinfo{person}{Yu Cui}, {and} \bibinfo{person}{Joseph Konstan}.} \bibinfo{year}{2017}\natexlab{}.
\newblock \showarticletitle{A community rather than a union: Understanding self-organization phenomenon on Mturk and how it impacts Turkers and requesters}. In \bibinfo{booktitle}{\emph{Proceedings of the 2017 CHI conference extended abstracts on human factors in computing systems}}. \bibinfo{pages}{2210--2216}.
\newblock


\bibitem[Wijenayake et~al\mbox{.}(2023)]%
        {wijenayake2023combining}
\bibfield{author}{\bibinfo{person}{Senuri Wijenayake}, \bibinfo{person}{Danula Hettiachchi}, {and} \bibinfo{person}{Jorge Goncalves}.} \bibinfo{year}{2023}\natexlab{}.
\newblock \showarticletitle{Combining Worker Factors for Heterogeneous Crowd Task Assignment}. In \bibinfo{booktitle}{\emph{Proceedings of the ACM Web Conference 2023}}. \bibinfo{pages}{3794--3805}.
\newblock


\bibitem[Williams et~al\mbox{.}(2019)]%
        {williams2019perpetual}
\bibfield{author}{\bibinfo{person}{Alex~C Williams}, \bibinfo{person}{Gloria Mark}, \bibinfo{person}{Kristy Milland}, \bibinfo{person}{Edward Lank}, {and} \bibinfo{person}{Edith Law}.} \bibinfo{year}{2019}\natexlab{}.
\newblock \showarticletitle{The Perpetual Work Life of Crowdworkers: How Tooling Practices Increase Fragmentation in Crowdwork}.
\newblock \bibinfo{journal}{\emph{Proceedings of the ACM on Human-Computer Interaction}} \bibinfo{volume}{3}, \bibinfo{number}{CSCW} (\bibinfo{year}{2019}), \bibinfo{pages}{1--28}.
\newblock


\bibitem[Xie(2023)]%
        {xie2023understanding}
\bibfield{author}{\bibinfo{person}{Haoyu Xie}.} \bibinfo{year}{2023}\natexlab{}.
\newblock \emph{\bibinfo{title}{Understanding the Use of HIT Catchers and Crowd Knowledge Sharing: A Case Study of Crowdworkers on Amazon Mechanical Turk}}.
\newblock \bibinfo{thesistype}{Ph.\,D. Dissertation}. \bibinfo{school}{University of Sheffield}.
\newblock


\bibitem[Yaaqoubi(2020)]%
        {yaaqoubi2020practitioners}
\bibfield{author}{\bibinfo{person}{Judith Yaaqoubi}.} \bibinfo{year}{2020}\natexlab{}.
\newblock \bibinfo{booktitle}{\emph{Practitioners’ Views on Cultural Adaptation of Web-based Products}}.
\newblock \bibinfo{publisher}{University of Washington}.
\newblock


\bibitem[Yao et~al\mbox{.}(2021)]%
        {yao2021together}
\bibfield{author}{\bibinfo{person}{Zheng Yao}, \bibinfo{person}{Silas Weden}, \bibinfo{person}{Lea Emerlyn}, \bibinfo{person}{Haiyi Zhu}, {and} \bibinfo{person}{Robert~E Kraut}.} \bibinfo{year}{2021}\natexlab{}.
\newblock \showarticletitle{Together but alone: Atomization and peer support among gig workers}.
\newblock \bibinfo{journal}{\emph{Proceedings of the ACM on Human-Computer Interaction}} \bibinfo{volume}{5}, \bibinfo{number}{CSCW2} (\bibinfo{year}{2021}), \bibinfo{pages}{1--29}.
\newblock


\bibitem[Yuan et~al\mbox{.}(2013)]%
        {yuan2013understanding}
\bibfield{author}{\bibinfo{person}{Chien~Wen Yuan}, \bibinfo{person}{Leslie~D Setlock}, \bibinfo{person}{Dan Cosley}, {and} \bibinfo{person}{Susan~R Fussell}.} \bibinfo{year}{2013}\natexlab{}.
\newblock \showarticletitle{Understanding informal communication in multilingual contexts}. In \bibinfo{booktitle}{\emph{Proceedings of the 2013 conference on Computer supported cooperative work}}. \bibinfo{pages}{909--922}.
\newblock


\bibitem[Zichermann(2020)]%
        {zichermann2020gamification}
\bibfield{author}{\bibinfo{person}{G Zichermann}.} \bibinfo{year}{2020}\natexlab{}.
\newblock \showarticletitle{Gamification at Work: Designing Engaging Business Software}.
\newblock \bibinfo{journal}{\emph{Retrieved from The Interaction Design Foundation: https://www. interactiondesign. org/literature/book/gamification-at-work-designing-engaging-businesssoftware}} (\bibinfo{year}{2020}).
\newblock


\bibitem[Zyskowski and Milland(2018)]%
        {zyskowski2018crowded}
\bibfield{author}{\bibinfo{person}{Kathryn Zyskowski} {and} \bibinfo{person}{Kristy Milland}.} \bibinfo{year}{2018}\natexlab{}.
\newblock \showarticletitle{A crowded future: Working against abstraction on Turker Nation}.
\newblock \bibinfo{journal}{\emph{Catalyst: Feminism, Theory, Technoscience}} \bibinfo{volume}{4}, \bibinfo{number}{2} (\bibinfo{year}{2018}), \bibinfo{pages}{1--30}.
\newblock


\end{thebibliography}

\begin{appendices}
\section{}
\blu{Our Appendix provides information on: 1) CultureFit's recommender system that powers its notification system's task feeds; 2) the surveys conducted before and after the study.}

\subsection{\blu{CultureFit's Recommendation Algorithm}}
\blu{CultureFit utilizes a content-based recommendation algorithm framed as a regression model, which operates through a feed-forward back-propagation neural network implemented via TensorFlow.js\footnote{https://www.tensorflow.org/js} \cite{lops2019trends}. This model is designed to predict the likelihood that a task will be initiated or completed by a crowdworker, taking into account the specific characteristics of individual tasks or task batches. The input features for the model include payment per task, payment per batch of tasks, the number of tasks previously completed by the worker from the requester who is posting the task, task acceptance rate (a metric indicating the percentage of tasks accepted by workers, which can reflect task attractiveness or suitability), task duration, one-hot-encoded task category, and the task type (regular, training, or exam).}

\blu{Given the relatively sparse data available per worker, we implemented a continuous learning approach, enabling dynamic updates to the model as tasks are completed. The model is re-trained locally each time a worker completes a new type of task, allowing it to adapt in real-time to the evolving actions and preferences of the worker. This strategy enhances the model’s accuracy over time and tailors task recommendations more effectively to individual workers.}

\blu{Our continuous learning framework employs regression techniques to consistently update and refine predictions based on incoming data. Regression is particularly well-suited for generating continuous outputs, such as the probability of task completion in this context. This output precision facilitates meticulous computation of ranking metrics, ensuring that the tasks most likely to be completed by the worker are prioritized.} \blu{The model uses the Adam optimizer and Mean Squared Error (MSE) as the loss function, with a learning rate of $0.001$ and a training period spanning 1,000 epochs. CultureFit maintains a consistent application of this algorithm across various cultural contexts—both monochronic and polychronic—ensuring uniformity in task recommendation practices.}

\subsubsection{\blu{General Evaluation of CultureFit's Recommendation Algorithm}}
\blu{We evaluated the performance of the recommendation algorithm integrated into our tool. For this purpose, we computed the mean average precision (MAP) at k (mAP@k) per participant to understand the relevance of the recommendations. CultureFit's recommendation algorithm achieved overall MAP scores of 0.68 at mAP@3 and 0.52 at mAP@5. For monochronic workers, the mAP@3 score was 0.67 and the mAP@5 score was 0.52. For polychronic workers, the mAP@3 score reached 0.69, while the mAP@5 was 0.51. Ultimately, this means that the first two task recommendation items were likely to be completed by both polychronic and monochronic workers.}

\blu{Note that we decided to use MAP to evaluate the recommender system we integrated into CultureFit due to:} \begin{itemize}
    \item[(a)]\textbf{\blu{MAP Evaluates Precision and Ranking of Recommendations}}: \blu{MAP is commonly used for evaluating the results of search engines \cite{buttcher2016information,manning2008introduction,voorhees2005trec}, particularly because it can measure the precision of top-ranked items and assesses how well these rankings align with user preferences \cite{baeza1999modern,carterette2007evaluating}. For our tool, designed to minimize workers' search time, accurately measuring the relevance of top tasks is important—if these tasks are not relevant, we do not effectively reduce workers' search time. Therefore, MAP's sensitivity to the order of task presentation was a key reason for selecting it to evaluate the recommendation component of CultureFit. Additionally, since CultureFit employs regression techniques to refine predictions and produce a ranked output of task completion probabilities, MAP is particularly suitable. It inherently assesses the quality of these rankings by evaluating the precision at various list depths \cite{buttcher2016information,baeza1999modern}.}
    
    \item[(b)] \textbf{\blu{Monitors Adaptation:}} \blu{For CultureFit's recommendation algorithm, we integrated a continuous learning model with regression techniques, which evolves by learning from new data. On the other hand, MAP can be recalculated periodically to monitor any improvements or declines in the precision of recommendations over time. Therefore, in our context, MAP proved to be a useful tool for assessing the effectiveness of the evolving recommendation algorithm integrated into CultureFit.}

    \item[(c)] \textbf{\blu{Feedback Integration:}} \blu{MAP can also help to quantify how well a recommendation system integrates user feedback to refine its recommendations by measuring the precision of the recommendation list at various cutoff points. This approach can provide clear insights into how a recommender system adapts to user feedback. For CultureFit, MAP is particularly useful in evaluating how the recommendation system incorporates feedback from both completed and uncompleted tasks into its model updates. This capability is important  for ensuring that the recommendations accurately reflect observed worker behaviors and preferences.}
    
\end{itemize}
\blu{Overall, MAP effectively measures whether the most relevant recommendations are presented first, factoring in user satisfaction and engagement within dynamic learning environments, such as those where CultureFit's algorithm operates, making it a suitable method for assessing our tool's success.}\\

\subsubsection{\blu{Comparing CultureFit's Recommendation Algorithm to Baselines}}
\blu{To further validate the effectiveness of CultureFit's recommendation algorithm, we compared its predictive performance against two baselines: a simpler, less resource-intensive algorithm, and a version of our model that is updated less frequently.}

\blu{The less resource-intensive baseline, which uses a heuristic approach of recommending tasks based on the most frequently completed tasks across all workers, achieved a mean average precision at k (mAP@k) of 0.64 at mAP@3 and 0.49 at mAP@5. This indicates that although simpler, this model underperforms in predicting task relevance compared to CultureFit's more sophisticated model, which integrates continuous learning. Recall that CultureFit's recommendation algorithm achieved overall MAP scores of 0.68 at mAP@3 and 0.52 at mAP@5.}


\blu{We also evaluated a version of our recommendation algorithm that is retrained only once every three days, as opposed to immnediate retraining for the main model. This less frequently updated model scored 0.66 at mAP@3 and 0.47 at mAP@5, indicating a decrease in predictive accuracy over time, likely due to stale data. However it is important to note that this model did outperform the heuristic baseline.} \blu{These comparisons underscore that although CultureFit's more advanced model demands additional resources and frequent updates, it does deliver more relevant task recommendations. There is always a trade-off between resource consumption and predictive performance. Nonetheless, we hope that these results will guide researchers in future decisions regarding the optimization of update frequencies and algorithm complexity when deploying AI-enhanced tools for workers across various scenarios.}

\subsubsection{\blu{Evaluating the Benefits of Continuous Learning in CultureFit's Recommendation Algorithm: An Ablation Study}}
We also aimed to evaluate the effectiveness of the continuous learning component within CultureFit's recommendation module. For this purpose, we conducted an ablation study on CultureFit's recommendation algorithm, comparing its performance with and without retraining (i.e., the continuous learning component). We assessed the MAP at various levels of k for both configurations. The results showed that the model with continuous retraining—CultureFit's standard model—achieved MAP scores of 0.68 at mAP@3 and 0.52 at mAP@5, as previously mentioned. This outperformed the non-retraining version, which scored 0.49 at mAP@3 and 0.36 at mAP@5. This highlights the role of the continuous learning feature in adapting to new data and enhancing task prediction accuracy. Furthermore, these findings underscore the benefits of dynamic model updates for optimizing performance in real-time task prediction scenarios within crowd work tools.


\subsection{\blu{Surveys}}
\blu{In our study, crowdworkers were required to complete a pre-survey before gaining access to our tool and a post-survey upon concluding the study. Both surveys were administered to participants in both the control group and those using the CultureFit system.}
\subsubsection{\blu{Pre-survey}}
\blu{In the following we present the pre-survey we gave to participants.} 

\begin{longtable}{| p{.50\textwidth} | p{.50\textwidth} |} 
\hline
\textbf{\blu{Question}} & \textbf{\blu{Options}} \\ \hline
{\bf \blu{EXPERIENCES WITH CROWD WORK}}&  \\ \hline
\blu{How long have you been working on Toloka?} & \blu{Less than a month; Between 1 and 3 months; Between 3 and 6 months; Between 6 months and 1 year; Between 1 and 2 years; More than 2 and 3 years; More than 3 years} \\ \hline

\blu{How often do you work in Toloka?} & \blu{Everyday; From five to six days a week; From three to four days a week; Once or twice a week; Less than once a week} \\ \hline

\blu{How fast can you find tasks to work on in Toloka?} & \blu{1) Very Slowly - It takes me a long time to find tasks I can work on; 2) Slowly - It often takes some time to find tasks; 3) Moderately - I can find tasks in a reasonable amount of time; 4) Quickly - I usually find tasks to work on quickly; 5)Very Quickly - I can find tasks to work on immediately.} \\ \hline

\blu{How often do you get the chance to use your strengths when working on Toloka?} & \blu{1. No Extent - I do not use my strengths at all when working on Toloka; 2. Slight Extent - I use my strengths to a slight extent when working on Toloka; 3. Moderate Extent - I use my strengths to a moderate extent when working on Toloka; 4. Great Extent - I use my strengths to a great extent when working on Toloka; 5. Very Great Extent - I use my strengths to a very great extent when working on Toloka.} \\ \hline

\blu{How much do you enjoy planning your tasks on Toloka?} & \blu{1) Not at all - I do not enjoy planning my tasks at all; 2) Slightly - I enjoy planning my tasks a little; 3) Moderately - I somewhat enjoy planning my tasks; 4) Quite a bit - I enjoy planning my tasks quite a lot; 5) Extremely - I enjoy planning my tasks immensely.} \\ \hline

\blu{How much do you like having someone oversee the work you do?} & \blu{1) Not at all - I do not like having someone oversee my work at all; 2) Slightly - I slightly dislike having someone oversee my work; 3) Neutral - I am neutral about having someone oversee my work; 4) Somewhat - I somewhat like having someone oversee my work; 5) Very much - I very much like having someone oversee my work.} \\ \hline

\blu{How often do you use Toloka on your Desktop?} & \blu{1) Never - I never use Toloka on a desktop; 2) Rarely - I rarely use Toloka on a desktop; 3) Sometimes - I sometimes use Toloka on a desktop.
4) Often - I often use Toloka on a desktop; 5) Always - I always use Toloka on a desktop.} \\ \hline

\blu{How often do you use Toloka on your Smartphone?} & \blu{1) Never - I never use Toloka on a smartphone; 2) Rarely - I rarely use Toloka on a smartphone; 3) Sometimes - I sometimes use Toloka on a smartphone; 4) Often - I often use Toloka on a smartphone; 4) Always - I always use Toloka on a smartphone.} \\ \hline

\blu{What web browsers do you use? Check all that apply.} & \blu{Chrome; Firefox; Opera; Yandex; Brave; Other} \\ \hline

\blu{Name a few of the tools you use to help you do work on Toloka?} & \blu{Open-ended} \\ \hline

\blu{Which forums do you visit to discuss topics related to Toloka, including tasks and requesters?} & \blu{Open-ended} \\ \hline

\blu{What other crowdsourcing or Gig markets have you worked on previously? Check all that apply.} & \blu{Amazon Mechanical Turk; LiveOps; Samasource; Galaxy Zoo; Prolific; Upwork; Other} \\ \hline

\blu{Since when have you been working on other Crowdsourcing or Gig Markets?} & \blu{Never; Less than a month; Between 1 and 3 months; Between 3 and 6 months; Between 6 months and 1 year; Between 1 and 2 years; More than 2 and 3 years; More than 3 years} \\ \hline
{\bf \blu{MPI QUESTIONS}} &  \\ \hline
\blu{I prefer to work on several projects in a day, rather than completing one project and then switching to another.} & \blu{1) Not at all; 2) Slightly; 3) Neutral; 4) Somewhat; 5) Very much.} \\ \hline

\blu{I would like to work in a job where I was constantly shifting from one task to another, like a receptionist or an air traffic controller.} & \blu{1) Not at all; 2) Slightly; 3) Neutral; 4) Somewhat; 5) Very much.} \\ \hline

\blu{I lose interest in what I am doing if I have to focus on the same task for long periods of time, without thinking about or doing something else.} & \blu{1) Not at all; 2) Slightly; 3) Neutral; 4) Somewhat; 5) Very much.} \\ \hline

\blu{When doing a number of assignments, I like to switch back and forth between them rather than do one at a time.} & \blu{1) Not at all; 2) Slightly; 3) Neutral; 4) Somewhat; 5) Very much.} \\ \hline

\blu{To see if you're still paying attention, please select the choice that says neutral.} & \blu{Strongly Disagree 1	2	3	4	5 Strongly Agree} \\ \hline

\blu{I like to finish one task completely before focusing on anything else.} & \blu{1) Not at all; 2) Slightly; 3) Neutral; 4) Somewhat; 5) Very much.} \\ \hline

\blu{It makes me uncomfortable when I am not able to finish one task completely before focusing on another task.} & \blu{1) Not at all; 2) Slightly; 3) Neutral; 4) Somewhat; 5) Very much.} \\ \hline

\blu{I am much more engaged in what I am doing if I am able to switch between several different tasks.} & \blu{1) Not at all; 2) Slightly; 3) Neutral; 4) Somewhat; 5) Very much.} \\ \hline

\blu{I do not like having to shift my attention between multiple tasks.} & \blu{1) Not at all; 2) Slightly; 3) Neutral; 4) Somewhat; 5) Very much.} \\ \hline

\blu{I would rather switch back and forth between several projects than concentrate my efforts on just one.} & \blu{1) Not at all; 2) Slightly; 3) Neutral; 4) Somewhat; 5) Very much.} \\ \hline

\blu{I would prefer to work in an environment where I can finish one task before starting the next.} & \blu{1) Not at all; 2) Slightly; 3) Neutral; 4) Somewhat; 5) Very much.} \\ \hline

\blu{I do not like when I have to stop in the middle of a task to work on something else.} & \blu{1) Not at all; 2) Slightly; 3) Neutral; 4) Somewhat; 5) Very much.} \\ \hline

\blu{When I have a task to complete, I like to break it up by switching to other tasks intermittently.} & \blu{1) Not at all; 2) Slightly; 3) Neutral; 4) Somewhat; 5) Very much.} \\ \hline

\blu{I have a ``one-track'' mind.} & \blu{1) Not at all; 2) Slightly; 3) Neutral; 4) Somewhat; 5) Very much.} \\ \hline

\blu{I prefer not to be interrupted when working on a task.} & \blu{1) Not at all; 2) Slightly; 3) Neutral; 4) Somewhat; 5) Very much.} \\ \hline

{\bf \blu{BACKGROUND AND DEMOGRAPHICS}}&  \\ \hline
\blu{Please select your current country of residence from the list below.} & \blu{List of Countries} \\ \hline
\blu{Which country have you lived in for the majority of your life? Please specify below.} & \blu{List of Countries}\\ \hline
\blu{Have you lived in countries other than your current residence? If so, please list them:}& \blu{Open-ended}\\ \hline
\blu{Please state what is your educational background?} & \blu{No schooling completed; Elementary school; Some high school, no diploma; High school graduate, diploma or the equivalent (for example: GED); Some college credit, no degree; Trade/technical/vocational training; Associate degree; Bachelor’s degree; Master’s degree; Professional degree; Doctorate degree }\\ \hline
\blu{Please state what is your gender:} & \blu{Male; Female; Non-binary; Prefer not to say} \\ \hline
\blu{Please state what is your age:} & \blu{18-24 years old; 25-34 years old; 35-44 years old; 45-54 years old; 55-64 years old; 65- 74 years old; 75 years or older} \\ \hline

\end{longtable}

\newpage
\subsubsection{\blu{Post-survey}}
\blu{In the following we present the post-survey we gave to participants.} 
\begin{table}[H]
\begin{tabular}{ | m{9cm} | m{5cm} | }
\hline
\textbf{\blu{Question}} & \textbf{\blu{Options}} \\ \hline              
\blu{How much has your work schedule on Toloka changed this week compared to previous weeks?} & \blu{Not at all 1	2	3	4	5 Very Much}\\ \hline
\blu{How much did you get to try out new tasks on Toloka this week?} & \blu{Not at all 1	2	3	4	5 Very Much} \\ \hline
\blu{How often did you get to use your strengths while working on Toloka this week?}  & \blu{1. No Extent - I did not use my strengths at all when working on Toloka this week; 2. Slight Extent - I used my strengths to a slight extent when working on Toloka this week; 3. Moderate Extent - I used my strengths to a moderate extent when working on Toloka this week; 4. Great Extent - I used my strengths to a great extent when working on Toloka this week; 5. Very Great Extent - I used my strengths to a very great extent when working on Toloka this week.}\\ \hline
\blu{How different did your experience feel while working on Toloka this week?} & \blu{Not different at all 1	2	3	4	5 Very different}\\ \hline
\blu{What did you like most about the plugin, and why?} & \blu{Open-ended} \\ \hline
\blu{What aspect of the plugin did you like the least, and why?} & \blu{Open-ended} \\ \hline
\blu{If you could magically change one thing about the plugin by adding or removing something to it, what would it be and why?}& \blu{Open-ended} \\ \hline
\blu{To see if you're still paying attention, please select the choice that says strongly disagree.} & \blu{Strongly disagree 1 2 3 4 5 Strongly agree}\\ \hline
\blu{Do you have any final thoughts or comments? Feel free to share also any questions you might have for us, or anything else you'd like to discuss.} & \blu{Open-ended} \\ \hline
\end{tabular}
\end{table}

\end{appendices}

\end{document}